Freeze Casting – A Review of Processing, Microstructure and Properties via the Open Data Repository, FreezeCasting.net

*Kristen L. Scotti*[1] and *David C.* Dunand[*1]

[1]Department of Materials Science and Engineering, Northwestern University, Evanston, IL 60208, USA

## Abstract

Freeze-casting produces materials with complex, three-dimensional pore structures which may be tuned during the solidification process. The range of potential applications of freeze-cast materials is vast, and includes: structural materials, biomaterials, filtration membranes, pharmaceuticals, and foodstuffs. Fabrication of materials with application-specific microstructures is possible via freeze casting, however, the templating process is highly complex and the underlying principles are only partially understood. Here, we report the creation of a freeze-casting experimental data repository, which contains data extracted from ~800 different freeze-casting papers (as of August 2017). These data pertain to variables that link processing conditions to microstructural characteristics, and finally, mechanical properties. The aim of this work is to facilitate broad dissemination of relevant data to freeze-casting researchers, promote better informed experimental design, and encourage modeling efforts that relate processing conditions to microstructure formation and material properties. An initial, systematic analysis of these data is provided and key processing-structure-property relationships posited in the freeze-casting literature are discussed and tested against the database. Tools for data visualization and exploration available through the web interface are also provided.

**Keywords:** ice-templating, solidification, porous materials, mechanical properties, data repository

## 1. Introduction

Freeze-casting is a solidification technique for fabricating porous materials. During solidification of a suspension or solution, walls are templated due to the rejection of particles and/or solute by a solidifying fluid. Pore structures, which are created after post-solidification solid removal, replicate the morphology of the solidified fluid. Freeze-casting gained much of its initial attention in the early 2000’s as a processing route for polymer [1-4] and ceramic [5] biomaterials. Concurrent investigations of freeze-cast materials for other ceramic-based applications, including filtration membranes [6-8] and fuel cell electrodes [9-16], began to reveal the true versatility of the technique. Freeze-casting of metals [17], pharmaceuticals [18-24], and foodstuffs [25] was demonstrated within the same decade. From the early 2000’s to 2017, the number and diversity of potential applications for these materials grew considerably; applications investigated include substrates for supercapacitors [26-32], photocatalysis [33-43], liquid chromatography [44], sensors (e.g. pressure [45-47], biological [48] and gas [49-51]), and batteries [52-62]; biomaterials [63-105] remain the most extensively investigated application. Dense composite materials may also be fabricated by infiltrating freeze-cast skeletons with a secondary phase. Thus far, ceramic/metal [52, 106-119], ceramic/polymer [120-138], metal/polymer [139-143], and metal/metal [144, 145] composites have been demonstrated.

[*] Corresponding author. Tel.: +1 847 491 5370. Fax: +1 847 467 6573
E-mail address: dunand@northwestern.edu (D.C. Dunand)

The flexibility of the freeze-casting process is an advantage of the technique. The volume fraction, size, shape, and orientation of porosity templated during the process may be tuned by changing suspension characteristics (*e.g.* fluid type [146], additives [147], particle fraction [148]) as well as solidification conditions (*e.g.* solidification technique [149], cold plate temperature [150], mold design [151, 152] and substrate material [153]). However, this freedom also adds significant complexity when attempting to understand the underlying principles that govern the microstructure templated during the process. By the end of 2016, ~900 papers (Fig. 1) had been published on the freeze-casting technique, most of them since 2006; over 800 of these papers represent independent experimental investigations. Each study presents—to some degree—its own set of parameters, so that drawing conclusions from any one paper is nearly impossible and systematic review of processing-structure-property relationships is difficult. Although we have a basic understanding of the underlying principles that govern microstructure formation during solidification, new parameters thought to influence these principles are still being identified (e.g., surface wetting of the freezing substrate [152] and filler material [154], coarsening [155], mold design [151], and gravity [156]); new models are also being applied to describe resulting, anisotropic properties [157].

Here, we expand upon the initial meta-analysis of Deville *et al.* published in 2016 [158], with the creation of a freeze-casting experimental data repository. As of May 2017, the repository contains data extracted from over 800 different freeze-casting papers (consisting of over 6,000 unique experimental specimens which total over 10,000 samples). These data pertain to variables that link processing conditions to microstructural characteristics, and finally, mechanical and physical properties. The aim of the database (and this paper) is to facilitate broad dissemination of relevant data to freeze-casting researchers and users, promote better informed experimental design, and encourage modeling efforts that relate processing conditions to microstructure formation and material properties. To that end, the freeze-casting database is provided free to the public through a web interface (http://www.freezecasting.net), under a Creative Commons ShareAlike license.

As an introduction to the FreezeCasting.net database, we summarize microstructural characteristics observed in freeze-cast materials and test key processing-structure-property relationships posited in the freeze-casting literature against the FreezeCasting.net database. General freeze-casting principles are reviewed briefly in the following; these are discussed in greater detail elsewhere (basic freeze-casting principles [3, 159-168], processing-structure-properties relationships [5, 160, 167, 169-173], and broader relevance of the technique [174, 175]). In comparison to previous reviews, we also provide more granular data analysis by making distinctions between (i) solidified, unsintered and sintered samples, (ii) anisotropic and isotropic structures, and (iii) ceramics, metals, and polymers. We also include ~500 additional papers since the latest meta-analysis. All data figures were created using a Python/Jupyter notebook and data contained within the FreezeCasting.net database, as of August 1, 2017. The corresponding notebook and database are available for download from the FreezeCasting.net website (http://www.freezecasting.net/downloads; Creative Commons ShareAlike license).

## 1.1. Basic principles

### *1.1.1. Anisotropic vs. isotropic freezing*

Freeze-casting solidification techniques can broadly be categorized as "anisotropic" or "isotropic." Anisotropic, usually "unidirectional", techniques are more extensively investigated than isotropic techniques. This process is depicted in Fig. 2. First, (a) a suspension of particles or a solution containing dissolved polymer is placed into a mold. Typically, the mold is composed of thermally insulating sides (*e.g.*, Teflon) and a thermally conductive base (*e.g.*, copper). The base of the mold is cooled, promoting (b) nucleation and directional propagation of a solidification front. At first, the solidification velocity is very high and suspended particles (and/or solute) are initially engulfed by the solidification front, resulting in a dense particle layer at the base of the sample [3, 176, 177]. The latent heat released by this early freezing event slows the solidification velocity and particles are subsequently pushed by the solidification front rather than engulfed, forming a (c) particle/solute accumulation region ahead of the front [178, 179]. As solidification progresses, rejected particles and/or solute are concentrated within interdendritic space, where solidification of interstitial fluid eventually occurs as well. Once (d) solidification is complete, the solidified fluid is (e) removed via sublimation, and in the case of ceramic and metal material processing, (f) the resulting scaffold is sintered to densify particle-packed walls. The pore morphology in the final material is a near replica of the morphology of the solidified fluid, and shows both a vertically-aligned, elongated structure and horizontal orientational alignment within colonies.

For anisotropic freeze-casting, solidification typically takes place vertically, from the bottom to the top (hereinafter, referred to as "unidirectional solidification"), as depicted in Fig. 2. Demonstrated solidification heights for this configuration range from 100 μm [180], using a combined doctor-blade/freeze-casting technique, to 9 cm [181] via traditional methods as described above. Sample areas as large as 38 x 15 cm [182] have been demonstrated via unidirectional freeze-tape-casting [182-188]. Anisotropic radial freeze-casting is also employed. In this case, the top and base of the mold are composed of thermally insulating material and, either the sides, or a rod placed in the center, of the mold is thermally conductive. When cooled, solidification is initiated at either the perimeter [9, 189, 190] or center [69, 191] of the suspension, and the resulting front propagates horizontally, in a radial manner.

Isotropic freeze-casting techniques were demonstrated as early as 1954 [192] and are still used today for the production of non-aligned porosity [193]. This process is depicted in Fig. 3. A suspension contained within a thermally insulating mold is (a) placed in a freezer, and (b) solid nucleation occurs randomly throughout the suspension. After nucleation, (c) solid crystal growth occurs at random orientations; a preferential growth direction is not observed. After (d) sublimation and (e) sintering, the resulting pore structures are nearly isotropic [194-196]. Depending on the isotropic technique employed, pore structures range from closed, equiaxed cells (when isotropic freezing and space-holder techniques are combined [197]) to open, reticular [198] networks. In some cases, thermally conductive molds are used and freezing progresses directionally from the bottom as well as radially from the sides [199]. Although some microstructural directionality is often observed in the resulting material (especially radially and at the bottom), overall, the structure is nearly isotropic. Isotropic freeze-casting techniques are most commonly used for freeze-gelation; in such cases, the precursor suspension or solution undergoes a gelling or cross-linking stage prior to freezing [200-203]. Unidirectional freeze-gelation techniques have also been demonstrated [204-216].

#### *1.1.2. Pore structures*

The microstructures of freeze-cast materials are qualitatively described in terms of their pore structure, with the most commonly reported anisotropic pore structures being lamellar (60%), dendritic (20%), and honeycomb (19%). Anisotropic columnar [217-220], needle-like [221, 222], and fishbone [223] morphologies have also been demonstrated. For isotropic structures, cellular (92%) and equiaxed (7%) structures predominate. A schematic representation of these pore structures is shown in Fig. 4, though as discussed in Section 3.1, considerable variation within these structures is observed. Nevertheless, for the purposes of drawing processing-structure-property connections, pore structures are categorized as shown in Fig. 4 and defined as follows: Anisotropic lamellar structures (Fig. 4-a) exhibit plate-like walls which align perpendicular to the freezing direction. Lamellar walls often exhibit dendritic features ranging from surface roughness [147] to dendritic arm stubs [150]. As shown in Fig. 4(b), dendritic structures exhibit primary trunks ("primary arms") and tree-like branching, commonly referred to as secondary arms. When anisotropic freeze-casting techniques are employed, honeycomb structures are elongated and tubular [224], as shown in Fig. 4(c), whereas prismatic cells [225] result when non-directional, freeze-gel-casting techniques are used (Fig. 4-d). These prismatic cells are relatively uncommon; thus, we refer to the elongated honeycomb structures shown in Fig. 4(c) as "honeycomb" throughout. Lastly, near-isotropic structures of closed, equiaxed cells [197] and open, reticular-cellular networks [226] are shown in Fig. 4(e) and (f), respectively.

Quantitatively, freeze-cast microstructures are most commonly described in terms of total porosity, structure wavelength, and pore and wall sizes. Structure wavelength, $\lambda$ is defined as the average sum of the width of one pore plus its adjacent wall [150]. This parameter, traditionally used to describe primary dendrite spacing in metal alloys [227], only applies to anisotropic freeze-cast materials and is most commonly used to describe lamellar structures. For dendritic structures, primary [228] and secondary dendrite spacing [229] (the distance between secondary arms) are less often reported. For anisotropic microstructures, pore and wall size is typically described in terms of width (the length of the minor axis), whereas for isotropic structures, diameter is used.

### *1.1.3. Characterization methods*

Microstructural characteristics of freeze-cast materials are primarily determined during solidification. The morphological evolution of the solidification front is highly dynamic; processing-structure relationships are complex and incompletely understood. Our current understanding is largely influenced by experimental investigations of sintered samples. Microstructural characteristics of sintered samples are most often quantified using two-dimensional images obtained using optical and scanning electron microscopy techniques. Two main issues exist in attempting to discern processing-structure relationships using these data. First, data obtained from two-dimensional images may not reflect the true, three-dimensional structure of these materials. Second, microstructures templated during the solidification process are often modified as a result of subsequent sublimation [230] and sintering [231] steps.

Fife *et al*. [232] compared measurements of pore width obtained using three-dimensional X-ray tomography reconstructions and two-dimensional optical micrographs of unidirectionally freeze-casted titanium materials. Data corresponding to mean pore width obtained from optical micrographs were 5-45% higher than corresponding values of inverse surface area per unit volume (related to the two-dimensional measurement of pore width) obtained from three-dimensional reconstructions. In addition to providing a more accurate representation of feature sizes, three-dimensional reconstructions allow for quantification of parameters such as interconnectivity, percolation, surface curvature and tortuosity [228, 233-235]. These are important parameters for many freeze-cast material applications (*e.g.*, filters; fuel and solar cell electrodes), but are rarely reported, as they are intrinsically three-dimensional properties and often inaccessible to two-dimensional measurements.

The first attempts to address confounding effects of post-solidification processing on freeze-cast microstructures focused on utilizing green (sublimated, but unsintered) samples infiltrated with epoxy [236-238]. Although this approach mitigates any sintering shrinkage effects, microstructural effects of sublimation (such as local particle rearrangement as well as warping or cracking of the green body) remain unaccounted for. It is also unknown whether, or to what extent, infiltrating these fragile compacts with epoxy might introduce additional changes. Later attempts to account for post-solidification processing effects utilized rates of linear sintering shrinkage as a back-estimate of pre-sintering microstructural feature sizes [239]. This approach assumes that sintering shrinkage is near-isotropic; for anisotropic pore structures, this is not necessarily the case [240]. Deville *et al.* [241-244] published a series of papers which provided the first experimental data obtained during *in-situ* investigation of the freeze-casting solidification process. These and subsequent *in-situ* investigations [206, 245-261] have provided better insight into the principles that underlie the templating process during solidification.

Undoubtedly, *in-situ* investigations of the solidification process, combined with 3D reconstruction comparisons of frozen, not-yet-sublimated freeze-cast samples and sintered samples, hold great promise in clarifying processing-structure-property relationships. Few investigations of this nature have been published [178, 262], therefore, we attempt to clarify these relationships using data obtained for a variety of samples at various processing steps. This effort is facilitated by categorical designations within the FreezeCasting.net database where microstructural and mechanical data correspond to that obtained from (i) *in-situ* investigations, (ii) solidified, not-yet-sublimated samples, (iii) sublimated, not-yet-sintered samples, and (iv) sintered samples, are designated as “solidification”, “frozen”, “green”, and “sintered”, respectively. These conventions are utilized throughout.

## 2. Methods

### 2.1. Systematic literature review

A systematic literature search was conducted to identify papers published between the years 1998 and 2017. Two strategies were employed to identify relevant literature. First, computer searches were performed on Google Scholar and Web of Science utilizing the search terms: “freeze-casting” OR “freeze-cast” OR “ice-templating” OR “ice-template” OR “ice segregation induced self-assembly” OR “unidirectional freeze-drying” OR “freeze-gelation” OR “gel-casting” OR “phase separation method.” Second, sources cited in relevant literature obtained from the first step were manually checked; relevant literature was added to the database and associated references were also checked and added when appropriate.

Overall, these searches yielded over 2,000 keyword-related studies. Studies were then excluded for any of the following reasons: (i) the technique did not involve a solidification step (e.g., “gel-casting,” without freezing), (ii) the second phase was not sacrificial (e.g., traditional metal matrix composites), or (iii) the primary materials produced were fibers and/or powders, rather than porous scaffolds. Additionally, data pertaining to papers that: (i) were not available in the English language (33 papers), (ii) did not contain data within the categories of interest (typically, freeze-casting application papers reporting functional property data only; categories of interest are defined in the following section), (iii) consisted of solely numerical (12 papers) and/or theoretical analyses (12 papers), or (iv) review articles (43 papers), were added to a general literature table only and data were not extracted. In the latter case, this was to prevent over-representation of data subsets.

## 2.2. Data extraction

Data were manually extracted from included papers, preferentially from text and tables; data were extracted from plots which were digitized using Plot Digitizer (http://plotdigitizer.sourceforge.net) when values were not provided in text or tables. Unit conversions were performed as needed. In particular, solid loadings were converted to volume percent (vol.%) when provided in weight percent (wt.%) using, in order of preference: (1) density of solid as provided in paper, (2) supplier data, or (3) reasonable estimates of density using other literature sources.

## 2.3. FreezeCasting.net database

The FreezeCasting.net database was constructed using structured query language (SQL) and the free, open-source relational database system, MySQL. However, the database may be used with any SQL-compliant database system and additional file formats are available for download on the FreezeCasting.net website (including: JSON, XML, and CSV).

### *2.3.1. Database architecture*

A modified relational model was utilized in database design; the database schema consists of a group of 'primary', 'secondary', 'property', and supplementary table(s). The primary group serves as the backbone of the database. This group includes the tables named: (1) 'authors', (2) 'papers', and (3) 'samples'. The secondary group consists of tables which hold all freeze-casting experimental data; these tables are categorized by processing steps or characterization types. Lastly, property tables provide property data for particles, fluids, molds, and bulk materials. As of May 2017, the database contains one supplemental table ('susp_stab'), which contains experimental data relevant to suspension stabilization. This hierarchical approach circumvents limitations of a preset number of columns in any particular table and provides flexibility to add new variables of interest as well as supplemental tables without altering the overall database schema.

It is expected that the majority of database users will work with the data using the web interface. Therefore, priority was placed on efficiency over normalization during database design. Although the majority of tables within the freezecasting.net database are in third normal form (in terms of database normalization), exceptions, which serve to reduce required join operations, can be found within secondary tables. A simplified depiction of the database schema is shown in Fig. 5; a full schema may be viewed/downloaded from the FreezeCasting.net website (http://www.freezecasting.net/downloads).

#### *2.3.1.1. Primary tables*

Primary tables are shown in blue in Fig. 5. The 'papers' table contains relevant citation information for each paper in the database. Each paper is classified by several categorical designations. First, freeze-casting papers are distinguished from general suspension stabilization papers by an 'fc' descriptor in the 'category' field (as opposed to 'susp' for suspension stabilization papers). Papers which provide suspension stabilization data in the context of freeze-casting work receive the hierarchical, 'fc' designation. Freeze-casting papers are further categorized in the 'subcategory' field as experimental investigations ('exp'), review articles ('review'), theoretical ('theory'), and numerical ('num') analyses. Lastly, papers published in peer-reviewed journals can be distinguished from non-peer-reviewed in the 'peer_review' field, and open-access and publications holding Creative Commons licenses are designated within the 'usage' field. Each paper is linked to the corresponding author's contact information through the foreign key (in terms of database design, a foreign key is used to connect a row in one table with a corresponding row in a different table), 'author_ID,' which is the primary key for the 'authors' table.

In most cases, individual papers report results that are obtained under varying processing conditions, e.g., microstructural and/or mechanical properties obtained after sintering at different temperatures while all other processing conditions (solid loading, cold plate temperature, etc.) are held constant. For each of these cases, a unique ID number is assigned in the 'samples' table ('sample_id'). The primary key for the papers table ('paper_id') serves as a foreign key in the 'samples' table and links citation information to each sample. In turn, the sample ID acts as a foreign key for each of the secondary tables such that row data may be appropriately linked.

##### *2.3.1.2. Secondary tables*

Processing-structure-property variables are grouped into secondary tables based mainly on processing steps. These include, for example: suspension preparation, including suspension constituents (*e.g*., particle, fluid, and additive types and concentrations), as well as suspension characterization (*e.g*., viscosity, zeta potential, pH), solidification (*e.g*., technique, cooling rate, cold plate temperature), sublimation (*e.g.,* time, temperature, pressure), sintering (*e.g*., sintering time, temperature, and heating/cooling rates), characterization (microstructure; *e.g*., pore and wall width), and properties (mechanical; *e.g*., compressive strength). All secondary tables are shown in green in Fig. 5.

### ***2.3.2. User interface***

The FreezeCasting.net data repository is provided to the public through a web interface (http://www.freezecasting.net) to facilitate data exchange among researchers of the freeze-casting community. The web interface offers three main features: (1) basic background information about the freeze-casting technique, (2) downloadable formats of the database, and (3) an interface for data exploration.

##### *2.3.2.1. Data exploration and visualization*

An interactive plotting application, developed using the JavaScript library, Data Driven Documents (D3.js) [263], is provided on the FreezeCasting.net website (http://freezecasting.net/plotting.html). A subset of 24 database variables (including those related to suspension characteristics, solidification conditions, microstructure, and mechanical properties) were selected for use in this application. Using D3.js, these data are bound to a Document Object Model (DOM), and, upon user input, the document may be manipulated and transformed based on the dataset that is bound to the DOM. The user may select any two of the 24 available variables to plot on the axes and filter results based on material group, material type, and/or fluid type; each selection or filtration initiates a plot transformation. Additional transformation options include coloring points based on fluid or material group, toggling between linear and logarithmic axis scales, performing basic least square linear regressions, and enabling "data point removal," where plot points can be discarded by dragging the points off the plot canvas. Citation data pertaining to each plot point are accessible via tooltip on mouse hover; within the tooltip, an html link to the corresponding paper is also provided. Lastly, tables may be generated based on filtered plot data and exported in CSV format.

An interactive table, featuring the same database subset as utilized in the interactive plotting application, is also provided on the FreezeCasting.net website (http://freezecasting.net/data.html). This application was developed using the DataTables [264, 265] plug-in for the JQuery [266], JavaScript library. The plug-in supports server-side processing and, similar to the interactive plotting application, allows users to select variables for display in the table and to export results.

### 2.4. Statistical analysis

Python was used for all statistical analyses; corresponding code is provided in the form of a Jupyter notebook [267] using the IPython kernel [268] as a supplemental material. The Python library, SciPy was used for linear [269] and non-linear [270] least squares curve fitting. Data are expressed as mean, $\bar{x} \pm$ the standard deviation and median ($\tilde{x}$). Logarithmic transforms were performed on $x$ and $y$ arrays prior to performing curve fitting and/or regression analysis for power law relationships. Paired $t$-tests were used to compare means of two groups; one-way ANOVA and post hoc, paired t-tests with Bonferroni corrections were used to compare means between more than two groups. A probability value of $p < 0.05$ was used to determine statistical significance. Pearson correlation coefficients, hereinafter referred to as correlation coefficients, or $r$ were obtained using SciPy.

## 3. Processing-microstructure relationships

### 3.1. Templating pore structures using fluids and additives

For freeze-cast materials, pore structure is largely determined by the morphology of the solidified fluid. Although most freeze-cast materials are fabricated using aqueous suspensions, unique pore structures have been demonstrated using non-aqueous fluids as well (most commonly, camphene [271-273] and *tert*-butyl alcohol (TBA) [274, 275]). Suspension additives, such as dispersants and binders, are typically added to obtain well-dispersed suspensions and increase green strength of sublimated specimens, respectively. However, recent work aims to utilize these and other additives to tailor pore structures by manipulating freezing kinetics [229], the crystalline structure of the solidified fluid [276], and/or modifying particle redistribution behavior [277].

#### *3.1.1. Aqueous processing*

The choice of water as the fluid for freeze-casting is often attributed to its environmental friendliness, low cost, and bio-inertness. An important, but often unnoted advantage of aqueous processing routes is the large variety of attainable pore structures. When anisotropic freeze-casting techniques are employed, the predominant pore structure is lamellar. These plate-like structures predominate as a result of the anisotropic growth kinetics of hexagonal ice [278]; growth along the crystallographic $c$-axis (perpendicular to a vertically induced temperature gradient) is 43-200% slower than that of the $a$-axis [279, 280]. Thus, growth of dendritic features (*i.e.*, 'side-branching' or 'secondary arms'), which occur along the $c$-axis, are restricted for hexagonal ice. As a result, these structures tend to exhibit high aspect ratios (defined as the ratio of the length of the major pore axis to that of the minor). The mean aspect ratio calculated from literature values for lamellar structures is 5.5±3.2 ($N$=212, [141, 217-219, 281-299]). High aspect ratios are advantageous in applications such as batteries or heat exchangers, where open, aligned channels facilitate fluid flow [300, 301]. However, for load-bearing applications, increased interconnectivity between solid features is often desired as it improves the material's compressive response (*e.g.* by preventing internal wall buckling [147, 295, 302]). Wall interconnectivity can be increased by inducing a morphological transition from lamellar to dendritic ice structures during solidification. This is achievable by modifying suspension characteristics (*e.g.* including additives such as binders [229, 303] or cryoprotectants [304] or increasing particle volume fraction [305]) and/or solidification conditions (*e.g.* increasing solidification velocity [3, 223, 306, 307]).

A schematic representation of the morphological transition of ice is shown in Fig. 6. In Fig. 6(a), a lamellar morphology is depicted and lamellae faces along the crystallographic *a*-axis are perfectly flat. In reality, some degree of dendritic features is almost always present. These typically present in the form of dendritic arm "stubs" which may appear on only one side of the lamellae [150, 308]. The onset of dendritic side branching is depicted in Fig. 6(b); these "secondary arms" increase in both length and area density (number of arms per unit area). In Fig. 6(c), the ice morphology has completely transitioned from lamellar to dendritic. As side branching increases, the diameter of the primary branch decreases. This reflects a decrease in the magnitude of anisotropic growth between the *a*- and *c*-axes. Although the morphological structure of ice is still dendritic in Fig. 6(d), the resulting porous structure is likely to be cellular because of the highly interconnected ice network that has formed.

##### *3.1.1.1. Binders*

Microstructural changes resulting from a morphological transition during solidification from lamellar to dendritic are shown in Fig. 7(a-d) for sintered $Al_2O_3$ [309]. Cross-sections are taken parallel to the solidification direction; dark regions represent pores and the light regions are sintered $Al_2O_3$ walls. Here, the dendritic transition is induced through the inclusion of increasing concentrations of a binder, polyvinyl alcohol (PVA), to aqueous suspensions consisting of 25 vol.% $Al_2O_3$. Fig. 7(a) shows the microstructure obtained when no binder is added; the pore morphology is lamellar and remnants of one-sided dendritic features are observed within the walls. Microstructures obtained with the inclusion of 5, 15, and 20 wt.% PVA (with respect to $Al_2O_3$) are shown in Fig. 7 (b) - (d), respectively. Secondary branching progressively increases as the weight fraction of PVA increases and pore width decrease from ~25 μm at 0 wt.% PVA (Fig. 7-a) to ~9 μm at 20 wt.% PVA (Fig. 7-d). Similar results were obtained when adding increasing amounts of PVA to aqueous suspensions of 50 wt.% hydroxyapatite (HAP) [310], 23.5 vol.% $Al_2O_3$ [311], and 75 wt.% yttria-stabilized zirconia (YSZ) [312] powders, as well as for polymers solutions (*e.g.* collagen and PVA [313]).

Fig. 7(e-i) shows a similar transition for tungsten disulfide ($WS_2$) powder, where dendritic morphologies are observed with the inclusion of increasing weight fractions of gelatin as a binder [314, 315]. Again, cross-sections are taken parallel to the solidification direction, however, the templated $WS_2$ is sublimated but not sintered. The lamellar microstructure shown in Fig. 7(e) is obtained with the inclusion of 1 wt.% gelatin (with respect to powder); the binder content increases to 2, 3, 4, and 5 wt.% in Fig. 7 (f) -(i). As gelatin increases from 1 to 2 wt.%, wall roughness increases (Fig. 7(f)); this indicates that secondary dendritic arms are likely beginning to develop. Dendritic morphologies are observed in Fig. 7(g) and (h) and cellular

morphologies are observed with the inclusion of 5 wt.% gelatin (Fig. 8(i)). Similarly, dendritic morphologies were reported in sintered hydroxyapatite (HAP) materials with the addition of 2-5 wt.% gelatin (with respect to powder) to aqueous suspensions containing 50 wt.% HAP; at 6 wt.% gelatin, a cellular morphology was also observed [316].

The transition from lamellar to dendritic morphologies due to the inclusion of binder is commonly attributed to the increase in the suspension viscosity [311]. As suspension viscosity increases, particle pushing by the advancing solidification front is inhibited due to increased viscous drag. As a result, an increasing number of particles are engulfed by the growing solid. Depletion flocculation may also play a role [317-319], which would also increase the number of particles engulfed (since the size of aggregates increases). This explanation is more likely for microstructural bridging between solid walls than a complete transformation to a dendritic ice morphology. When large amounts of binder are added, constitutional undercooling must also be considered. Although most binders are soluble in water, they have limited solubility in ice. Thus, binder is rejected at the solid/liquid interface, its accumulation depresses the local equilibrium liquidus temperature which, in turn, decreases the degree of undercooling and promotes coarsening of secondary arms [309]. Similar lamellar to dendritic transitions were reported when adding increasing amounts of PVA to aqueous suspensions with 50 wt.% hydroxyapatite (HAP) [310] and 75 wt.% yttria-stabilized zirconia (YSZ) [312], as well as for $ZrO_2$ with the addition of 30 wt.%, (with respect to dry particle mass) [320] of $Na_2SiO_3$ binder.

Sucrose was also observed to trigger a dendritic to cellular transition [147]; dendritic and cellular structures were observed after anisotropic solidification of $Al_2O_3$ suspended in aqueous solutions of citric acid containing 5 and 10 wt.% sucrose (with respect to dry powder mass), respectively, as well with 8 wt.% trehalose, another sugar containing two glucose molecules. The addition of sucrose introduces an interfacial instability within the solidification front which promotes side branching during directional solidification [306]. Similar structures were obtained by Nguyen *et al*. [321] for pharmaceutical tablets containing 20 wt.% sucrose (with respect to water) and for cocoa tablets freeze-casted from aqueous suspensions containing up to 20 wt.% sugar alcohol binders, including sucrose, isomalt and xylitol. The authors attribute the cellular pattern to a decrease in the supersaturation of aqueous solution as a result of high binder content and a concomitant decrease in the ice crystal growth rate, which resulted in smaller, spherical shaped ice crystals.

#### *3.1.1.2. Simple alcohols*

The utilization of binary fluid systems consisting of water and simple alcohols (*e.g.* isopropyl alcohol (IPA), ethanol, *n*-propanol) offers increased flexibility over the microstructural interconnectivity, which can be tailored based on the employed alcohol concentration. For example, the highest aspect ratio (defined above as the ratio of the length of the major pore axis to that of the minor) from the literature comes from Naleway *et al.* [292] who demonstrated increased aspect ratios using a binary fluid system consisting of IPA and water and a solid loading of 10 vol.% $ZrO_2$. The maximum aspect ratio of 20, corresponding to a pore width of 30 μm, was achieved using an IPA concentration of 5 vol.% (with respect to total fluid volume). Comparatively, the authors report an aspect ratio of ~5 (pore width = 10 μm) without IPA. The increase in pore width and aspect ratios due to inclusion of IPA is attributed to the formation of clathrate hydrates during the freezing process [292, 295] and is shown to be maximized at 5 vol.% IPA. Above 5 vol.% IPA, pore width and aspect ratios decrease until 20 vol.% IPA is reached, at which point, values similar to those obtained without IPA are reached [292, 295]. Fig.

8 shows similar results reported by Porter *et al*. [295] for 10 vol.% $TiO_2$. These microstructural cross-sections are taken parallel to the solidification direction; Fig. 8(a) shows the sintered $TiO_2$ material obtained without IPA, and (b) - (d) show microstructures obtained with the addition of 1, 5, and 30 vol.% IPA (with respect to fluid volume). Pore aspect ratios are shown to be highest (at ~10) with the inclusion of 5 vol.% IPA (Fig. 8-c), representing a three-fold increase from the aspect ratio of ~3 reported without IPA (Fig. 8-a); the pore widths are also observably much larger in Fig. 8(c) as compared to 8(a). However, with the addition of 30 vol.% IPA (Fig. 8-d), microstructures more closely resemble those obtained without IPA.

The increase in both pore width and aspect ratio at relatively low fractions of IPA followed by a subsequent decrease with increasing concentrations of IPA may be attributable to Ostwald ripening (coarsening), where larger ice crystals increase in size at the expense of smaller ones. The rate at which this occurs is strongly influenced by the type and concentration of solute; relatively low concentrations of solute can increase the coarsening rate considerably [322]. However, at increasing concentrations of solute (in this case, IPA), mass transfer is increasingly restricted and Ostwald ripening is impeded. This behavior can be exploited to increase interconnectivity of the resulting solid microstructure. Similar behaviors are observed for other alcohols, including ethanol [147, 290, 323], *n*-propanol [290], and *n*-butanol [290]. For 10 vol.% $ZrO_2$, aspect ratio enhancement is reported to peak at 10, 5, and 3 vol.%, respectively [290] for ethanol, *n*-propanol, and *n*-butanol. A decrease in pore aspect ratios (*i.e.*, increased interconnectivity) is reported when ethanol is increased to 15-30 vol.% for $Al_2O_3$ [324] and 5-10 vol.% for Ti-6Al-4V [141]. Similarly, at 10-30 vol.% *n*-propanol in water, decreased aspect ratios are observed for 10 vol.% $Al_2O_3$ [324].

*3.1.1.3. Glycerol*

Interconnectedness between lamellar walls may be also be increased with the inclusion of glycerol as a suspension additive [284]. Microstructures consisting of lamellar walls joined by bridges that form nearly perpendicular to, and join, the walls have been demonstrated using aqueous suspensions containing 5 wt.% glycerol (with respect to water) [284]. The number density of these bridges increases as the glycerol concentration increases [220, 285, 302, 304]. This transition is shown in Fig. 9, where the sintered lamellar structure shown in Fig. 9(a) were fabricated via anisotropic solidification of aqueous suspensions of 10 vol.% HAP particles [285]. In Fig. 9(b), the concentration of glycerol in water increases to 20 wt.% glycerol. Similar to the effect of PVA, the corresponding pore widths to decrease from 20-30 μm without glycerol to 1-10 μm with glycerol. At high solid fractions (60 vol.% $Al_2O_3$), the addition of 20 wt.% glycerol (with respect to water) reportedly results in a cellular microstructure [325].

Although glycerol is commonly employed in freeze-cast studies, the underlying mechanisms governing its effect on suspension rheology and ice morphology during solidification are not clearly understood. Solutions of glycerol and water exhibit higher viscosity than water alone [326, 327] because glycerol is much more viscous than water, (~930 mPa·s [304] vs. ~0.9 mPa·s for water). Thus, it has been suggested that the addition of glycerol increases the suspension viscosity and restricts the diffusion of water molecules to the ice lattice [220]. Indeed, in some cases, the addition of glycerol increases suspension viscosity [302, 328]. However, decreased suspension viscosities have also been reported [304, 326, 329], especially at higher particle volume fractions [304]. In each of these cases, the suspended particle is $Al_2O_3$ and the glycerol contents are comparable (ranging from 10-30 wt.% with respect to water). It has also been suggested that glycerol may interact with nonionic dispersants and reduce suspension viscosity by forming micelles around particles [325, 330] or by untangling dispersant chains [326]. Lending support to this hypothesis, decreased zeta potential values have been reported with the addition of glycerol for suspensions where decreased viscosity is also observed [325], whereas for increased viscosity, zeta potential remains approximately the same [302]. However, glycerol is reported to decrease pore width regardless of whether the suspension viscosity increased or decreased as a result of its addition. This suggests that glycerol's cryoprotectant properties are a dominating factor in the reduction ice crystal size, rather than suspension viscosity.

It should also be noted that the addition of glycerol, both depresses the freezing point of the suspension and decreases the volumetric expansion of water during freezing. At 10 wt.% glycerol in water, the volumetric expansion of ice decreases from 9 to 7.4% [302], which is not sufficient, by itself, for the reduction of pore sizes described above. However, inclusion of glycerol might reduce the tendency for materials to develop ice-lens defects. These defects have an appearance similar to that of frost heaves [331-335]. In sintered freeze-cast materials, ice lenses are present as cracks perpendicular to the direction of solidification. Although not reported specifically, ice lens defects appear to be present in Fig. 9(b) from ref. [336]. However, there is no indication of ice lenses in any of the microstructure images after glycerol was added (*e.g.*, Fig. 10 from ref. [336]). Decreased cracking has been reported with the inclusion of glycerol [337] using isotropic freezing techniques.

##### *3.1.1.4. Unique pore structures*

Fig. 10(a) shows honeycomb structures achieved with the addition of a zirconium acetate complex (ZRA) to an aqueous suspension of yttria-stabilized zirconia (YSZ) [276, 299, 338]. The versatility of this technique for modifying pore morphology has also been demonstrated for $Al_2O_3$ [339, 340], SiC [276], and Teflon [276] freeze-cast structures, where particles are suspended in an aqueous solutions containing 18 g/L of ZRA. Honeycomb structures may also be obtained via freeze gel-casting techniques [341, 342]; the structure shown in Fig. 10(b) was obtained for 5 wt.% gelatin scaffolds [281]. In this case, it is hypothesized that prior to freezing, ice molecules from a discontinuous network throughout the gelatinous medium; thus, resulting ice nuclei are segregated by the continuous, gelatin network. During solidification, growth of secondary arms are restricted by the same gelatin network, resulting in columnar honeycomb structures [342].

The equiaxed pore structures shown in Fig. 10(c) and (d) were both achieved by employing isotropic freezing techniques. In Fig. 10(c), near spherical pores were obtained by subjecting silica suspensions to a tumbling step prior to solidification. Air entrapped during the tumbling step, and rejected during solidification, was responsible for the equiaxed macropores [197]. Similar structures can be obtained by adding polystyrene as a temporary space-holder, as shown for the mullite foam depicted in Fig. 10(d) [343], or PMMA as a space-holder [344]. In these cases, the space-holder templates the macroporosity, whereas the solidified fluid templates the microporosity. In contrast, a combined unidirectional solidification and space-holder technique was demonstrated for the fabrication of hydroxyapatite (HAP) materials [345, 346]. In this case, macroporosity is templated by both the solidified fluid (water) and the pore forming agent, 5 vol.% hydrogen peroxide ($H_2O_2$); lamellar walls are separated by lamellar channels or by lamellar arrays of spherical pores. At 9 vol.% $H_2O_2$, the lamellar pattern was lost and a cellular structure resulted.

#### ***3.1.2. Non-aqueous processing***

An overview of microstructures obtained using non-aqueous fluids and anisotropic freezing techniques is shown in Fig. 11. Upon solidification, camphene forms isotropic, cubic crystals; as growth of secondary arms is prevalent, the resulting pore structure is dendritic. This is illustrated in Fig. 11(a) for sintered $Al_2O_3$ obtained after anisotropic solidification of 50 vol.% $Al_2O_3$ suspended in camphene [347]. The predominant structure for cyclohexane is also dendritic as shown in Fig. 11(b) for sintered silicon oxycarbide (SiOC) [348]. This material was solidified from a precursor solution of 20 wt.% preceramic polymer dissolved in cyclohexane. Lastly, the dendritic structure shown in Fig. 11(c) is a polyurethane material templated using dioxane [180]; the precursor was also a solution.

An overview of honeycomb structures obtained via non-aqueous processing routes are shown in Fig. 11(d-f). Fig. 11(d) is a polystyrene material obtained using 30 wt.% polystyrene dissolved in a fluid system of 7-10 wt.% polyethylene glycol in dioxane [349]. A similar structure was obtained by freeze gel-casting using a fluid system consisting of dodecanol-27 wt.% dioxane-23 wt.% triproplyene gycol-17 wt.% cyclohexanol and dissolved monomers, glycidyl methacrylate (GMA) and ethylene glycol dimethacrylate (EDMA) [350]. TBA is, after water, the second most commonly utilized fluid in the freeze-casting literature. When solidified unidirectionally, the predominant structure is described as tubular, or elongated honeycombs; this is shown in Fig. 11(e) for sintered $Al_2O_3$-$ZrO_2$ [351], the precursor suspension contained 10 vol.% particles suspended in TBA. Dimethyl sulfoxide (DMSO) was utilized to obtain the honeycomb structure shown in Fig. 11(f) for a poly (L-lactic acid; PLLA) material [352]; PLLA was dissolved in DMSO prior to solidification.

## 3.2. Suspension stabilization

The first step in the freeze-casting process is to create a stable suspension. Particle-particle interactions influence both particle packing behavior as well as the morphology of the solidified fluid [317, 353]; unstable suspensions promote defect formation [277, 354], especially those which may result from particle engulfment (*e.g.*, ice lenses [158]). For suspensions containing charged particles, the stabilization process can be non-trivial and often involves measuring zeta potential ($\zeta$) and viscosity for suspensions at various pH levels and at varying concentrations of dispersants as well as other additives. The suspension stabilization table is included with the FreezeCasting.net database to facilitate these efforts. As of August 1, 2017, the table contains over 2,000 samples of viscosity and/or zeta potential data for suspensions containing 20 different particle types.

An illustration of $\zeta$ data plots is provided in Fig. 12. Zeta potential quantifies the electrostatic repulsion between particles within a suspension; at a higher magnitude of $\zeta$, suspensions exhibit greater stability. Therefore, a suspension can be stable at both positive and negative values of $\zeta$. Data were filtered to obtain only values for aqueous suspensions of aluminum oxide ($Al_2O_3$). Filtered data of $\zeta$ are plotted against suspension pH in Fig. 12 [304, 353-367]. For these data, $\zeta$ is positive at low pH values and decreases as the pH of the suspension increases. At the isoelectric point (IEP) where $\zeta = 0$, the suspension is very unstable (shown as a dashed red line in Fig. 8). A stable suspension is broadly defined as one that exhibits a $\zeta$ magnitude of at least 30 mV [368]. Using this metric ($\zeta \geq \pm 30$ mV), an aqueous suspension of $Al_2O_3$ may be stable in the pH range of 4-12, depending on the concentration and nature of the dispersant with respect to particle size and volume fraction.

Dispersants are added to aqueous suspensions of charged particles (*e.g.*, $Al_2O_3$) to protect against flocculation and coagulation. Data points in Fig. 12(a) are mapped to their relative concentrations of dispersant vs. zeta potential in Fig. 12(b). The most commonly utilized dispersants for aqueous $Al_2O_3$ suspensions contain a carboxylic acid functional group (*e.g.* Darvan dispersants) [369-371]. In this case, as the pH of the suspension is increased, greater dissociation of the functional group is realized; repulsion between particles is achieved because particle surfaces are negatively charged due to the adsorption of dispersant molecules and ζ of the suspension decreases. However, if the suspension pH is raised too high, the viscosity of the suspension increases drastically due to the formation of a hydroxide layer on the surface of the particles [372]. Too much dispersant is also a problem, because after particle adsorption is saturated, excess dispersant induces depletion flocculation [353]; this behavior is observed in Fig. 12(b) where data points corresponding to 3 and 4 wt.% dispersant (orange and green points) result in suspensions exhibiting lower magnitudes of ζ as compared to suspensions containing 0.5 – 2 wt.% (purple, blue, and teal points).

Ultimately, stabilization procedures for freeze-casting suspensions is dependent on the chosen particle, fluid, and any additives. Nevertheless, the suspension stabilization table may be utilized to guide stabilization efforts. In such cases, the database supports data filtering which include fluid, particle, and additive type and relative concentrations, particle size(s) and additive role(s). Investigations detailing the suspension stabilization process for freeze-casting studies can be found in refs. [317, 326, 337, 353, 354, 356, 373-378].

### 3.3. Controlling porosity via solid fraction

In general, total porosity obtained after solidification, sublimation, and any sintering treatments is less than the total volume fraction of fluid employed in the initial suspension. Fig. 13 shows total porosity plotted against solid volume fraction in the suspension for all corresponding data in the FreezeCasting.net database, irrespective of material type. Data points that describe sintered samples are colored based on relative point-density; red points indicate high density regions of the plot (many points) and purple points represent low density regions of the plot (few points), whereas data points describing sublimated, unsintered samples are shown in white. Consistent with experimental investigations [379-388], a linear relationship between porosity ($\phi_p$) and initial suspension solid fraction ($\phi_s$) is observed:

$$\phi_p = a \cdot \phi_s + b, \qquad \text{Eq. 1.}$$

where $\phi_p$ is the total porosity, $\phi_s$ is the volume fraction of solid (particles or dissolved polymer for polymer materials) in the suspension, *a* is the slope, and *b* is the intercept.

The solid black line in Fig. 13 is the regression line obtained by fitting all data to Eq. 1 ($\phi_p$= -1.0±0.3 · $\phi_s$ + 86%±6%; *N*=2,855, $R^2$ = 0.01, $r$ =-0.12, $p$ <0.0001). The very small value of the coefficient of determination, $R^2$, indicates that, when all data is considered, only ~1% of variation in $\phi_p$ is explained by Eq. 1. A similar value of $R^2$ was obtained when considering only sintered materials (Table 1; short-dashed line in Fig. 13). This, coupled with similar coefficient values obtained for Eq. 1 (Table 1) reflects the overrepresentation of sintered samples. Indeed, ~82% of these data describe sintered samples. When only data obtained from sublimated, unsintered samples are considered, the fit to Eq. 1 improves considerably and ~40% of variance in $\phi_p$ is predicted by Eq. 1 (Table 1; long-dashed line in Fig. 13). The intercept of the regression line obtained for sintered samples ($b$=82%±10%) is ~11% lower than that obtained for unsintered samples ($b$=92%±3%). The lower intercept value for sintered samples is likely attributable to volumetric shrinkage during sintering. Similarly, variations in sintering conditions for any particular material (*e.g.*, sintering time and temperature) will promote variation in volumetric sintering shrinkage and result in an increased scattering of data.

The relationship between $\phi_s$ and $\phi_p$ is categorized by fluid type in Fig. 14. Data describing sintered samples are shown in blue for (a) water, red for (b) camphene, and sublimated, unsintered for (c) *tert*-butyl alcohol (TBA). Sublimated, unsintered samples are differentiated in Fig. 14(a) as white circles. In Fig 14(b), two data points describe frozen (unsublimated) samples; these are shown as white stars. All TBA data points in Fig. 14(c) were obtained from sintered samples. Solid black lines in 14(a-c) are fitted regression lines for Eq. 1. Shaded regions about each solid line represent the 95% confidence interval for the regression line; the dashed lines represent the 95% prediction interval; *i.e*., 95% of future observations are expected to lie within the boundaries of the dashed lines. Independent fittings for green and sintered samples are performed in Fig. 14(a); regression lines are shown as long- and short-dashed lines, respectively.

Increased values of $R^2$ are obtained by fitting water (Fig. 14-a; $R^2$ =0.43, $N$=2,040, $p$ <0.0001) and camphene (Fig. 14-b; $R^2$ =0.37, $N$=290, $p$ <0.0001) data to Eq. 1 in comparison to that obtained for fitting data from Fig. 13. Statistical significance was not achieved for fitting TBA data to Eq. 1. For water (Fig. 14-a), the absolute value of the correlation coefficient ($r$) increases from 0.58 to 0.66 for sintered ($N$=1,582, $p$<0.0001) and green ($N$=458, $p$<0.0001) samples, respectively, and a significant difference between the means is detected ($t$=21, $p$<0.0001). This indicates greater linearity in the relationship between $\phi_s$ and $\phi_p$ for data obtained from green samples in comparison to sintered samples.

These data are further filtered in Fig. 15 by considering only aqueous suspensions and categorizing by material group. Metals are shown in Fig. 15(a), ceramics in 15(b) and polymers in 15(c). For polymers, a cluster of data points is observed at low initial solid fractions/high porosity. This is because the range of attainable porosity that is controlled by solid fraction is comparatively narrower for polymers than for ceramics or metals. Polymer freeze-cast materials are typically obtained using polymer solutions as opposed to particle suspensions. Thus, for any given polymeric material, the lower bound of attainable $\phi_p$ is governed by the solubility limit of the polymer in the fluid. For example, the solubility limit of cellulose is ~5 vol.% in water; the range of attainable porosity has been reported to be ~90-95% after sublimation of ice [389].

The 95% prediction range, shown as dashed lines about the respective regression lines, is larger for ceramics in Fig. 15(b) than for metals (a) or polymers (c). This is likely due to the larger number of data points for ceramics in comparison to the other systems ($N$=1,502 for aqueous ceramic suspensions, and $N$=104 and 204 for metals and polymers, respectively). These data correspond to 111 different ceramic material types, whereas for metals and polymers, only 10 and 27 materials are represented, respectively. Better predictive power can be obtained by further defining the input system. For example, if we consider only sintered titanium dioxide ($TiO_2$) materials derived from aqueous suspensions [38, 187, 295, 390-393], we obtain a regression model for Eq. 1 that accounts for ~93% of variance in $\phi_p$ as a function of $\phi_s$ (Table 1); the resulting coefficients of $a$ =-1.7±0.2 and $b$ =99.1%±0.7%, are in reasonable agreement with the relationship described by Chen *et al.* [379] for sintered freeze-cast materials derived from aqueous suspensions of 2 μm $TiO_2$ particles ($\phi_p$= -1.4· $\phi_s$+129%). Coefficients for Eq. 1 are provided in Table 1 for data described in Figs. 13-15, as well as for ceramic, metal, and polymer systems utilizing camphene and TBA as the fluid.

In Fig. 15, the slope of the regression line is highest for metals (-1.6±0.3) and lowest for polymers (-0.8±0.2; Table 1). At any given solid fraction ($\phi_s$), the resulting, average porosity ($\phi_p$) is highest for polymers and lowest for metals; the magnitude of this difference increases as solid fraction increases. A statistically significant difference among mean values of $\phi_p$ for ceramics, metals, and polymers is found (one-way ANOVA, $F$ = 6.3, $p$<0.01). Post hoc, unequal variance *t*-tests with Bonferroni corrections indicate statistically significant differences between each of these groups (ceramics/metals: $t$=3.0, $p$<0.01, ceramics/polymers: $t$=9, $p$<0.0001, and metals/polymers: $t$=19, $p$<0.0001). However, statistically significant differences are not found when categorizing by fluid type within material groups; specifically, neither for ceramics employing water, camphene, or TBA as the fluid, nor for metals employing water or camphene. These results are consistent with those found by Naviroj *et al.* [394] for the fabrication of silicon oxycarbide (SiOC) materials using fluids of cyclohexane, camphene, and TBA, where it was found that fluid type had little influence on final porosity. Processing shrinkage, which is typically limited to the drying or sublimation stage for polymers, is a likely explanation for the decreasing slopes of the regression lines in Fig. 15. Although not observed here, solidification of aqueous suspensions is reported to result in higher values of porosity in sublimated, not yet sintered silicate materials, compared to that fabricated using TBA; an effect attributed to the volumetric expansion of water (~10%) during freezing [395].

Fig. 16 shows the distribution of sintering shrinkage as observed for freeze-cast ceramics (blue), metals (red) and polymers (green). Sintering shrinkage is measured as a change in diameter, height, and/or volume; these are shown individually in Fig. 16. For ceramics and metals, sintering shrinkage can be quite significant; average change in diameter, height, and volume are higher for metals than ceramics in all categories, and lowest for polymers (Table 2). A statistically significance difference among means for diametric, linear, and volumetric shrinkage is found between metals, ceramics, and polymers (one-way ANOVA, $F$ =66, 41, and 80, $p$ <0.0001 for all); values of volumetric shrinkage tend to be higher than diametric and linear shrinkage for all material groups (which is expected, given the geometric relationship). As stated, polymer shrinkage is typically limited to the sublimation stage, which explains why polymers have the lowest values of shrinkage. Wall densification for ceramics is more difficult to achieve than for metals; residual or "microporosity" within cell walls will reduce shrinkage rates for all categories (diameter, height or volume) as compared to metals and result in higher levels of total porosity (total porosity includes both open, macro- and closed, micro-porosity).

## 3.4. Tuning microstructures during unidirectional solidification

### *3.4.1. Solidification velocity*

For unidirectional freezing, an empirical power law dependence of structure wavelength λ (defined above as the average width of one pore plus the width of its adjacent wall) on solidification velocity, *v*, is well demonstrated in the freeze-casting literature [150, 156, 228, 295, 396]:

$$\lambda = C_1 \cdot v^k \qquad \text{Eq. 2.}$$

where $C_1$ is a constant and the exponent, $k$ typically varies from -0.03 to -1.3 [156]. Based on the above equation, $\lambda$ decreases as solidification velocity increases; the greater the magnitude of $k$, the greater the dependency of $\lambda$ on $v$. Researchers have also attempted to extend the relationship shown in Eq. 2. to describe a dependency of pore [239] and wall [397] width on $v$. Fig. 17 shows (a) $\lambda$, (b) pore width, and (c) wall width plotted against $v$. Only lamellar structures (obtained via aqueous suspensions) are considered because, in context of Eq. 2, the temperature gradient (not commonly reported) must also be taken into account for dendritic structures [228]. Values of $C_1$ and $k$ from Eq. 2 are obtained by curve fitting using ordinary least squares regression after performing logarithmic transforms on $x$ and $y$ arrays. For each plot in Fig. 17, solid black lines are obtained by fitting all data to Eq. 3, whereas blue, green, and red lines are obtained by fitting ceramics, polymers, and metals, respectively. Regression lines are only shown if statistically significant fits were obtained.

For the dependency of $\lambda$ on $v$ shown in Fig. 17(a), values of $C_1$ and $k$ are calculated as 184±52 μm and -0.5±0.1 (Table 3; black line in Fig. 17). For ceramics, $k$ decreases slightly (-0.6±0.1) and $C_1$ remains the same ($C_1$ =184±52 μm). When only data obtained from sintered ceramic samples are considered, Eq. 2 only predicts ~7% of variance in $\lambda$. Conversely, ~58% of variance in $\lambda$ is predicted by Eq. 2 for green samples (Table 3). The absolute value of the k increases from $k$=0.9±0.2 for green ceramics to $k$=0.3±0.2 for sintered ceramics; *i.e.*, the dependency of $\lambda$ on $v$ is observably higher when describing data obtained from green, rather than sintered, ceramic samples. This suggests that sintering may reduce the dependency of $\lambda$ on $v$; although, sintered data were extracted from 18 different papers, whereas unsintered data were obtained from three papers. A combination of measurement technique variances and sintering conditions could contribute to the high data variability observed for sintered ceramic samples. Studies comparing measurements of $\lambda$ obtained from unsintered and sintered samples are needed to clarify this issue.

The dependency of pore width on $v$ is shown in Fig. 17(b). Values for $C_1$ and $k$ from Eq. 3 (with respect to pore width) are calculated to be 80±30 μm and -0.4±0.2 (Table 3). An excellent fit is obtained for Eq. 2 when considering data obtained during *in-situ* solidification of ceramics (blue stars in Fig. 17-b; [243, 398]), where Eq. 2 predicts 96% of variance in pore width; albeit, the number of samples is small ($N$=6). For polymers, ~88% variance in pore width is predicted by Eq. 2; only ~12% variance in pore width is predicted by Eq. 2 for ceramics (Table 3).

Lastly, the relationship between wall width and $v$ is shown in Fig. 17(c). Values for $C_1$ and $k$ from Eq. 2 (with respect to wall width) are calculated as 19±10 μm and -0.2±0.2 (Table 3). Microstructural parameter fits to Eq. 2 are statistically weaker for wall width than for $\lambda$ or pore width, for all material groups and sample types. Moreover, data obtained from *in-situ* investigations during solidification of aqueous $Al_2O_3$ suspensions (blue stars in Fig. 17-c [398]) suggest that wall width may *increase* with increasing solidification velocity; that is, the value of $k$ from Eq. 2 would have the opposite sign when describing a dependency of wall width on $v$ than it has for describing pore width on $v$. A possible explanation for this observation is that, as solidification velocity increases, there is a higher propensity for particles to accumulate at the solidification interface; a higher volume fraction of particles within, or depth of, the particle accumulation region may result in increased wall width since more particles are available for incorporation. However, for any given solidification velocity, the volume fraction of particles within, and the thickness of, the accumulation region also depend on the nature of the suspension, *e.g.,* binder concentration [317]. Eq. 2 may not be sufficient for describing the relationship between wall width and *v* as some parameters are missing (*e.g.* suspension viscosity, particle size, thermal gradient); this issue is further complicated when considering the effect of sintering.

Considering all materials, magnitudes of *k* calculated from Eq. 2 are shown to decrease from 0.5±0.1 for a dependency of λ on *v* to 0.4±0.2 and 0.2±0.2 for a dependency of pore width and wall width, respectively, on *v*. As stated, higher *k* magnitudes indicate a greater dependency of the microstructural parameter on *v*. For example, consider the effect of a change in velocity from 10 to 100 μm/s on the microstructural parameters of λ, pore width, and wall width. By the fitting parameters calculated for Eq. 3, the width of λ will decrease by 68% with the increase in *v* from 10 to 100 μm/s. By contrast, pore and wall and pore width will decrease by 60 and 38%, respectively. This illustrates that the dependency of microstructural parameters on *v* is higher for λ than pore or wall width. Moreover, the dependency of pore and wall width on *v* are not equivalent. Likewise, the coefficient of determination is greatest when describing a dependency of λ on *v* (*adj.* $R^2$=0.29; $p$<0.0001), lower for pore width (*adj.* $R^2$=0.23; $p$<0.0001) and lowest for wall width (*adj.* $R^2$=0.04; $p$<0.05).

In most cases, microstructural parameters are quantified using sintered samples and rates of sintering shrinkage are typically reported in terms of macro-dimensions prior to and after sintering. In these cases, sintering shrinkage may result in an unequal reduction of pore and wall widths, which may depend on the sintering conditions. For example, Li *et al.* [390] reported an increase in macropore volume from 41 to 57% after sintering lamellar, $TiO_2$ at 1000 vs. 900ºC for 3 h. The authors attributed the increased macroporosity to a decrease in microporosity within cell walls (from 46 to 23%, respectively). Interestingly, microstructural images indicated a high density of bridges connecting adjacent walls. This interconnectivity often results in decreased pore sizes, as connected walls draw adjacent walls towards each other during sintering; increased rates of diametric shrinkage are reported in comparison to linear shrinkage rates [239, 240]. When the sintering temperature was increased to 1473 K for 3 h, both macroporosity and microporosity decreased. By its nature, λ, which takes into account both pore width and wall width, may be less affected by the angle of the micrograph (when measurements are obtained from cross-sections taken parallel to the solidification direction) and/or disparate rates of sintering shrinkage. Alternatively, at increased sintering times and/or temperatures, walls may merge together forming thicker struts. For example, increases in mean wall width have been reported when increasing the sintering temperature from 1323 to 1523 K for lamellar $SnO_2$ [399] as well as when increasing the sintering time from 1 to 4 h at 1473 K for lamellar $TiO_2$ [400]. If this occurs, the sintered microstructure will deviate considerably from that which was templated during solidification and Eq. 2 is largely irrelevant. These cases reemphasize the need for *in-situ* investigation during solidification and corresponding comparisons to sintered microstructures.

### ***3.4.2. Cooling techniques for unidirectional solidification***

#### *3.4.2.1. Constant substrate temperature*

In a typical anisotropic freeze-casting set-up, a mold containing a suspension is placed onto a freezing substrate and the substrate is cooled using refrigerants (*e.g.,* liquid nitrogen) or thermoelectric cooling devices [17, 141, 148, 401]. In this "one-sided cooling" set-up, solidification velocity is controlled—to some extent—by the temperature of the cold plate.

Controlling solidification velocity is a key parameter for fabricating materials with predefined microstructural features sizes. The most widely used, and simplest, cooling technique involves setting the freezing substrate to a constant temperature over the course of an experiment. Subsequent experiments are then conducted at lower or higher temperatures to increase or decrease average solidification velocity, respectively. When a constant temperature is applied via one-sided cooling, the position of the ice front advances at a rate that is inversely proportional to the square root of time [402]; *i.e.*, solidification velocity is fastest at the base of the sample and slows as the solidification interface advances away from the freezing substrate. Therefore, one-sided cooling with a constant substrate temperature offers some control over average solidification velocity, but not over actual velocity measured at any given height of the sample. Gradient microstructures are produced where pore width increases from the base of the sample to the top, a consequence of an increase in the average diameter of dendrites as they progress ever more slowly away from the cold source.

Fig. 18 shows pore width measurements at various heights for samples fabricated using a constant substrate temperature [191, 229, 403-406]. Data points are colored to correspond to the same substrate temperature ($T_c$) and linked when these data pertain to mean values from the same sample set. Data obtained from samples of aqueous suspensions are shown as circles and camphene suspensions are shown as triangles. We are unable to detect a relationship between the magnitude of the gradient, nor the height difference for which pore sizes are measured, in relation to each other, nor to $T_c$. The average increase between pore widths measured at the top of the sample compared to the bottom is 236%, with a very large standard deviation of 182%. An inability to account for the magnitude of the gradient in pore size is likely a result of the paucity of data and inconsistencies among the cooling stages employed. Nevertheless, it is clear that constant cooling techniques result in gradient structures. These structures may be useful in applications such as tissue engineering and/or biomedical implants [407]. Gradient structures have also been enhanced for the production of fuel cell electrodes by utilizing suspensions containing a range of particle sizes and subjecting suspensions to a sedimentation stage followed by directional solidification at a constant temperature [408].

*3.4.2.2. Linear cooling*

Cooling rates are defined in the freeze-casting literature as the rate at which the freezing substrate (on which the crucible containing the suspension is placed) is brought down to the final, lowest temperature. Cooling rates are typically employed to reduce the gradient in microstructure feature size from the top of the sample compared to that at the base. Linear cooling rates, which represent a quasi-static approximation to a Stefan problem [237, 307], are most commonly employed. Stefan problems are two-phase moving boundary problems that assume one-directional heat flow (in the case of unidirectional solidification of aqueous suspensions, heat flow is through the ice), and solidification velocity is largely determined based on the temperature of the freezing substrate, height of the solidified fluid layer, and thermal properties of the suspension (in both liquid and solid state) [409].

Fig. 19 shows pore width measurements at various heights for samples fabricated using unidirectional freezing techniques at various linear cooling rates. Data points are colored to correspond to the same cooling rates and linked when these data pertain to the same sample set. Data obtained from samples of aqueous suspensions are shown as circles; camphene suspensions are shown as triangles. Fig. 19(a) shows data pertaining to cooling rates ≤5 K/min [307, 403, 410], and data pertaining to cooling rates ≥5 K/min [307, 403, 410] are shown in 19(b). Lastly,

Fig. 19(c) shows results obtained using double-sided cooling techniques [411], which allow greater control over the thermal gradient during solidification (a “cold plate” is placed at both the base and top of the suspension, with the cold plate at the base being held at a lower temperature than the top). The average increase between pore widths measured at the top of the sample compared to the bottom is 88%±49%, which is considerably lower than that obtained using a constant cooling temperature, especially considering the mean height difference corresponding to these values is higher than it was for constant cooling (20±6 mm vs. 11±4 mm, respectively). Nevertheless, considerable gradients are still observed, though they can be somewhat reduced with the use of double-sided cooling (Fig. 19-c). Although it is difficult to draw any conclusions from Fig. 19, deviations between desired and actual velocities often are highest at high velocities and with increasing sample height. This is evidenced in Fig. 19(b) for the 10 K/min cooling rate (blue), where pore width is relatively constant at sample heights of ~5 and ~16 mm, but increases sharply at a sample height of ~28 mm.

The cooling rate is often taken as analogous to the solidification velocity and researchers have utilized Eq. 2 to describe a dependency of microstructural parameters on the cooling rate [282, 283, 286, 412-415], wherein an attempt to control velocity is made by adjusting the cooling rate. However, the direct measurement of velocity is more generalizable to other studies. As observed by the gradient microstructures shown in Fig. 19, constant velocity is hardly attainable over the height of the sample via linear cooling rates. Among other factors, the size of microstructural features is highly dependent upon the position in the sample at which they were quantified.

Fig. 20 shows literature values of (a) λ, (b) pore width, and (c) wall width plotted as a function of the applied cooling rate; ceramics are shown in blue, polymers in green, metals in red, and ceramic/polymer composites are shown in purple. For purposes of comparison to the relationships between solidification velocity and microstructural parameters (as shown in Fig. 17), only lamellar structures derived from aqueous suspensions are considered. Coefficient values obtained by fitting these data to Eq. 2 are provided in Table 4.

For the dependency of λ on cooling rate shown in Fig. 20(a), values of $C_1$ and *k* from Eq. 2 are calculated to be 49±13 μm and -0.2±0.1 (black line in Fig. 20-a; Table 4). These values are lower and higher than those calculated for a dependency of λ on *v* ($C_1$=184±52 μm; k=-0.5±0.1) and the value of $R^2$ is higher when considering the dependency of λ on *v* ($R^2$=0.29). The dependency of pore width on *v* is shown in Fig. 20(b). Although statistically significant fits are obtained for Eq. 3, values of $R^2$ are very low (Table 4), this is observably inferred by the large range of demonstrated pore width at any particular cooling rate. Lastly, the relationship between wall width and *v* is shown in Fig. 17(c). Values of $R^2$ are slightly higher in terms of wall width than that obtained for pore width ($R^2$=0.17 vs. 0.13; $p$<0.0001 for both).

Only eleven papers in the FreezeCasting.net database report both solidification velocity and cooling rate [307, 396, 398, 411, 416-422]. Among these studies, the correlation coefficient describing a linear relationship between cooling rate and solidification velocity is $r$=0.65 ($N$=82, $p$<0.0001), which is relatively weak under an assumed causal relationship. However, cooling rate, as defined as the rate in which the freezing substrate is cooled, is fundamentally different from the rate at which a suspension is cooled. For the former, any experimental factor that influences the evacuation of latent heat will influence the relationship between cooling rate and velocity (*e.g.*, thermal conductivity of the mold and its base, shape of the mold, thermal resistance between the cold plate and the mold base, thermal conductivity of, and the presence of, insulation around the mold). Also, at the onset of freezing, the temperature of the freezing substrate may be significantly lower for suspensions that take longer to nucleate solid crystals; lower cold plate temperatures result in higher solidification velocities. Often, when cooling rates are reported, neither the solidification time, nor the final substrate temperature are reported. Some notable exceptions are found in the literature where cooling rate of the suspension is directly measured using thermocouples embedded in the mold wall. These temperatures are compared to that which is imposed on the freezing substrate, *e.g.*, refs. [307, 403, 421].

Hunger *et al*. [307] defined a "local cooling rate" for which the cooling rate of the suspension (rather than the freezing substrate) was calculated at various heights based on temperatures obtained from thermocouples embedded in the mold wall and the relative heights of each respective thermocouple. The authors then related the local cooling rate to local solidification velocity. When possible, solidification velocity should be reported in addition to the cooling rate. Reporting solidification velocity at various height within a directionally solidified sample and directly relating $v$ with spatially-resolved microstructural parameters holds much more utility than reporting solidification velocity and microstructural parameters averaged over the height of the sample. Investigations reporting velocity and microstructural parameters from *in situ* investigations [178, 241-243, 280, 317, 398] are especially valuable as they limit confounding effects of subsequent processing steps, *e.g.*, sintering shrinkage.

*3.4.2.3. Achieving constant freezing velocity*

Advanced unidirectional cooling techniques aim to achieve constant freezing front velocity throughout the solidification event, such that feature sizes are relatively constant over the height of the sample. These techniques include: immersion, adjusting velocity by increasing or decreasing an air gap between the freezing substrate and the base of the suspension [296] as well as applying exponential [239] and parabolic  cooling functions. The immersion technique, which involves dipping a vial (typically glass) containing suspension or polymer solution into a cold bath at a constant rate, is the most commonly employed technique for achieving constant velocity [1, 212, 223, 341, 342, 350, 423-440]. In this case, the "dipping rate", analogous to "pulling velocity" using the Bridgman technique is taken as the solidification velocity. However, imposed and actual velocity are not necessarily analogous, especially at high imposed velocities and when solidifying larger samples [437, 441]. More importantly, the technique likely introduces a radial temperature gradient [442-444], thus in most cases, the thermal gradient is "bidirectional" vs. "unidirectional". Indeed, radial heterogeneity has been observed using the immersion technique [436, 437, 445], and the technique has been purposely employed to achieve radial microstructures [446, 447].

Pore width at corresponding heights for samples solidified using the gap method, exponential and parabolic cooling functions are shown in Fig. 21(a-c), respectively. The air gap method is demonstratively more effective at higher (32 μm/s) than lower (14 μm/s) freezing velocities, which presents some limitation on attainable feature sizes. The inherent trial-and-error nature of the technique for correlating gap distance to solidification velocity also make it a less desirable approach in comparison to utilizing exponential or parabolic cooling rates.

The exponential cooling function is a unique solution to a two-phase, constant velocity Stefan problem [176] and has been demonstrated for achieving constant velocities of 10-50 μm/s [176, 239, 448]. The highest demonstrated sample heights (~2 cm) for any of the constant velocity approaches, were demonstrated via exponential cooling by Flauder *et al*. [239]. In this work, aqueous suspensions of 20 vol.% $\beta$-TCP (tricalcium phosphate) were solidified unidirectionally at ~ 20 μm/s, resulting in pore widths of ~18 μm (Fig. 21-b). Exponential cooling was not found to be effective for maintaining constant velocity for polymer solutions [389]. The authors attributed this to the increasing concentration of solute at the interface over time and the resulting dynamic nature of the freezing temperature. Indeed, a limitation to the solution to the Stefan problem is that it assumes pure materials. Consequently, achieving constant velocity is less likely to be effective via this method when utilizing suspensions containing high concentrations of solute additives.

Parabolic cooling functions hold some utility over the exponential cooling function in that the heat equation is solved numerically. Thus, aspects of directional solidification that violate Stefan problem assumptions may be taken into account; *e.g.*, convective heat transfer in the fluid. Although parabolic cooling functions have been utilized in a few studies [178, 348, 394, 449], only one study reports microstructural sizes over the height of the sample. Fig. 21(c) shows results reported by Zhao *et al*. [449], where constant velocity (~10 μm/s) resulting in a homogenous distribution of pore width of 20 μm over the course of a 1.3 mm sample height was achieved.

### ***3.4.3. Undercooling***

Experimental values of freezing point depression for the solidification of freeze-casting suspensions tend to exceed those which would be explained by solute effects due to the addition of various suspension additives. This causes some complication in the utilization of cooling functions, because the starting temperature of the suspension should be within good agreement of the freezing temperature of the suspension. Peppin *et al*. [450-452] derived an expression for the thermodynamic freezing temperature $T_f(\phi_s)$, for suspensions of hard spheres:

$$T_f = T_m\left[1 - \frac{\Pi(\phi_s)}{\rho_f L_f}\right] = \frac{T_m}{1 + m \cdot Z(\phi_s)} \qquad \text{Eq. 3.}$$

where $T_m$ is the melting point of the fluid, $\Pi(\phi_s)$ is the osmotic pressure of the suspension, $\rho_f$ is the density of the fluid, $L_f$ is the latent heat of fusion of the fluid, $m = k_B T_m / v_p \rho_f L_f$, $k_B$ is Boltzmann's constant, and $v_p$ is the volume of a particle ($v_p = 1/6\,\pi \cdot d_p^3$, where $d_p$ is the particle diameter). When $\Pi(\phi_s)$ is unknown, the dimensionless compressibility factor, $Z(\phi_s)$ is utilized to approximate $\Pi(\phi_s)$ for $0 < \phi_s < 0.55$:

$$Z(\phi) = \frac{1+(4-\frac{1}{\phi_b})\phi_s+(10-\frac{4}{\phi_b})\phi_s^2+(18-\frac{10}{\phi_b})\phi_s^3+(\frac{3}{\phi_b^5}-\frac{18}{\phi_b})\phi_s^4}{1-\phi_s/\phi_b}, \qquad \text{Eq. 4.}$$

where $\phi_s$ is the volume fraction of particles in the suspension and $\phi_b$ is the breakthrough concentration (defined as the particle volume fraction at which the osmotic pressure of the suspension exceeds the capillary pressure necessary to allow ice to invade the pore space, taken as $\phi_b$=0.64 [453]).

Fig. 22(a) shows $\phi_s$ plotted against experimental values of $T_f$ obtained from freeze-casting studies employing aqueous suspensions of dissolved polymer [404, 408, 413, 420] and suspended particles [25, 220, 224, 311, 454-456]. Data describing polymer solutions are provided for comparison to particle suspensions only; these data are depicted as gray circles. Data points corresponding to particle suspensions are color-coded based on relative additive concentration with respect to particle content; points corresponding to the highest additive concentration (50 wt.% sucrose with respect to powder [25, 455]) are shown in red. The melting point of water is taken as 273.15 K and is shown as a purple horizontal line in Fig. 22(a). The freezing point of a 20 wt.% aqueous solution of sucrose (which corresponds to data obtained from the highest additive content suspensions) is estimated using $\Delta T = i \cdot m \cdot K_f$, where $i$ is the van't Hoff factor (taken as unity for sucrose, which does not dissociate in water), $m$ is the molarity (0.58 M for an aqueous solution of 20 wt.% sucrose), and $K_f$ is the molal freezing point depression constant of water (1.86 K·kg·mol). The calculated freezing point of ~272 K is shown in Fig. 22(a), as a red horizontal line. Predicted freezing temperature curves for particle suspensions were calculated using Eq. 3 and considering a base fluid of water (shown in purple) as well as an aqueous solution of 20 wt.% sucrose (red); two particle sizes are considered, $d_p$ = 2 and 5 nm (as marked).

The selected particle sizes are significantly smaller than any of the particles utilized in the studies plotted in Fig. 22(a) ($d_p$=0.3 to 150 μm). However, when particle sizes larger than 5 nm are utilized, the prediction curves are barely distinguishable from the horizontal lines and the predicted freezing temperatures are systematically higher than the experimental observations shown here. By Eq. 3, greater freezing depression is expected for smaller particle sizes. Peppin *et al*. [450] found good agreement with experimental values of $T_f$ for aqueous suspensions of bentonite particles ($d_p$= 1 μm ) and those predicted by Eq. 3. However, the researchers utilized experimental values of $\Pi(\phi_s)$ for aqueous bentonite suspensions to obtain fitting parameters for Eq. 4. Experimental values are not presently available for the systems studied here.

A dependency of $\phi_s$ on $T_f$, as predicted by Eq. 3, is not confirmed by experimental data plotted in Fig. 22(a), although a group of data points appear to be following the prediction curve for $d_p$=2 nm. In addition to a lack of experimental values of $\Pi(\phi_s)$ for these data, there are several possible explanations for the scatter observed in Fig. 22(a). First, there is some ambiguity in how freezing temperature is reported in the freeze-casting literature wherein nucleation temperature is sometimes taken as the freezing temperature. These values can be properly interpreted when cooling curves are provided; however, they are rarely provided. Another issue exists in the tendency to consider the melting temperature of water as the freezing temperature of the system without considering any depression resulting from the inclusion of suspension additives; this can be rather significant [454] and varies depending on the nature of the additives. Lastly, reported values of $T_f$ are typically obtained immediately after nucleation, whereas Eq. 3 predicts freezing point depression resulting from a layer of concentrated particles. Thus, its potential use for predicting freezing point depression dynamically (*i.e.*, over the course of solidification as the concentration of particles accumulated ahead of the interfaces changes) holds great utility in being used in conjunction with parabolic cooling rates for attaining constant velocity. However, experimental data reporting the freezing point (at the interface) over the course of solidification is needed for empirical verification.

Undercooling (ΔT), defined as the difference between the freezing temperature and the solid nucleation temperature has been shown to effect pore geometry [454, 456, 457], structure wavelength [456], and the height of the transition zone (typically corresponding to a cellular region between the aforementioned "dense layer" and ordered, lamellar growth) [454]. It is also less sensitive to the ambiguities noted above because its calculation requires that measurements of the actual freezing temperature and nucleation temperature be distinguished. These data are plotted as a function of $\phi_s$ in Fig. 22(b). We find a strong, linear relationship between $\phi_s$ and $\Delta T$ ($N$=37, $r$ =-0.9, $R^2$=0.8, $p$<0.0001). These results may offer some indirect support for Eq. 3, because nucleation has been reported to occur at higher temperatures with increased particle volume fractions [411], which is attributed to particles acting as nucleation sites. Then, if the freezing point of the suspension remained constant across solid fractions, we would expect undercooling to decrease at higher particle fractions, which is the opposite of what we observe here. A more focused statistical analysis (e.g., factor, principal component and/or cluster analysis) is beyond the scope of this review but may be beneficial for elucidating the appropriate explanatory variables and/or uncovering confounding variables, thus, clarifying the relationships explored here (especially if these techniques are used to analyze a greater number of experimental data).

### 3.5. Summary of attainable microstructures

Box plots showing the range of attainable porosity for anisotropic and isotropic structures are shown in Fig. 23, where anisotropic structures, including lamellar (blue), dendritic (red), and honeycomb (green) are shown on the left and isotropic structures are shown on the right (grey background) including cellular (purple), and equiaxed (orange). Darker shades of color represent data describing sintered samples whereas lighter shades describe green (unsintered) samples. For anisotropic materials, mean values of porosity ($\phi_p$) are 65±14, 66±18, and 82±18% for sintered lamellar ($N$=1,116), dendritic ($N$=398), and honeycomb ($N$=334) structures, respectively (Table 5). Increase values of $\phi_p$ are observed when considering green samples (80±18, 80±15, and 87±17% for lamellar ($N$=206), dendritic ($N$=74), and honeycomb ($N$=101) structures, respectively). For lamellar, dendritic, and honeycomb structures, the minimum demonstrated porosities (22, 15, and 28%, respectively) should be considered an approximate lower bound for attainable porosity, as many studies have shown that the corresponding microstructural patterns are lost at lower values of porosity and, if a pattern remains, it is cellular, *e.g.*, [304, 379]. This is reflected in the large range of porosity observed for the isotropic-cellular structures shown in Fig. 23 (purple). Summary statistics for isotropic structures are provided in Table 5.

Box plots showing distributions of pore and wall widths for anisotropic materials are shown in Fig. 24, where (a) is pore width and (b) is wall width for lamellar (blue), dendritic (red), and honeycomb (green) structures. These data are highly skewed, with the majority of pore and wall width observations being below 100 and 30 μm, respectively, for both sintered and green samples. In Fig. 24 (c) and (d), only pore and wall width below these values, are considered. For lamellar structures, the mean ($\bar{x}$) pore width for green and sintered samples are, $\bar{x}$=73±90 and 44±85 μm, respectively, representing an approximate 40% decrease in mean pore width when obtained from sintered ($N$=653) vs. green ($N$=222) samples. In contrast, mean wall width for green and sintered samples are $\bar{x}$=24±47 and 17±24 μm, representing a decrease of ~29% for sintered ($N$=404) vs. green ($N$=70) values. As mentioned earlier, in some cases, wall width can increase during sintering, which may be partially responsible for this difference. However, the standard deviations are large in all cases and the number of data points available for green samples is considerably less than that of sintered. Summary statistics for pore and wall width for anisotropic and isotropic pore structures are provided in Table 6.

The mean pore diameter observed for sintered cellular structures is 42±87 μm (Table 6). For cellular structures, the standard deviation is very large and over 25% of reported pore diameters are less than 5 μm, while the maximum observed pore diameter is 580 μm [226]. This extreme variation can be explained by the drastically different solidification techniques that can be used to create cellular pore structures. For example, a large range of pore sizes may be obtained using a combined sponge-replication and freezing technique, wherein the solidified fluid templates microporosity in the final material and a sponge acts as a template for the macroporosity; pore diameters range from 13 [458] to 580 μm [226]. In contrast, mean pore diameter for cellular structures fabricated using traditional isotropic freezing techniques is often less than 5 μm [187].

Fig. 25 shows the porosity dependence of (a) pore and (b) wall width for all materials, irrespective of pore structure. Here, color indicates the relative density of data points for any particular region of the plot; red data points are high density regions, whereas purple points indicate low density regions. The range of attainable pore width is weakly dependent on total porosity ($N$=1,110, $r$ =0.19, *adj*. $R^2$ =0.04, $p$ <0.0001). For example, at a porosity of 60%, pore widths range from <1 to ~400 μm. Above, we found that solid loading (volume fraction of particles in the suspension) is one of the main predictor variables for total porosity, whereas solidification conditions (especially solidification velocity) is a better predictor of pore width. In both cases, shrinkage resulting from sublimation and/or sintering steps can influence these relationships. Therefore, it is not surprising that we do not find a strong relationship between porosity and pore width. Rather, Fig. 25(a) indicates that a wide range of pore widths are attainable at a given porosity, especially within the porosity range of 60-80%. Wall width shows a moderate dependence on overall porosity (Fig. 25 (b), $N$=379, $r$=-0.46, *adj*. $R^2$ =0.21, $p$ <0.0001) as well as (c) pore width ($N$=541, $r$ =0.46, *adj*. $R^2$ =0.21, $p$ <0.0001). For solidified samples (not yet subjected to sublimation and/or sintering shrinkage), minimum wall thickness is theoretically limited to the diameter of the employed particle. Submicron wall sizes are most often demonstrated for polymer materials where the “particles” are dissolved in the fluid [459].

## 4. Mechanical properties

As of August 1, 2017, the FreezeCasting.net database contains mechanical property data, including compressive strength, flexural strength, and elastic modulus (usually compressive stiffness), for over 2,000 freeze-casting samples. In most cases, these data correspond to porous materials. However, data reported for dense, composite materials (typically, where a second phase is infiltrated or impregnated into a freeze-casted skeleton [460]) are also collected. Compressive strength is the mechanical property that is most often investigated; we review these data here. For anisotropic pore structures, compressive strength is typically reported for loads applied parallel to the wall direction. In the cases where compressive strength is reported for loading perpendicular to the wall direction (in addition to parallel loading), anisotropic compressive strength properties, where increased strengths for parallel loading, are reported [186, 308, 461-465]. Increased flexural [186] strength, and elastic moduli [193] are also reported for parallel loading in comparison to perpendicular loading. Fig. 26 shows the mean ratio of compressive strength taken parallel and perpendicular to the wall direction ($\sigma^*_{||}/\sigma^*_{\perp}$) for metals (red), ceramics (blue), and polymers (green). The mean ratio is lowest for metals ($\sigma^*_{||}/\sigma^*_{\perp}$=2.4±0.9) and highest for ceramics ($\sigma^*_{||}/\sigma^*_{\perp}$=4.4±2.5). Despite these unique, anisotropic properties, modeling efforts for anisotropic freeze-cast materials have focused mostly on composite materials, especially metal/ceramic composites [466-486] with some focus on ceramic/polymer [130].

### 4.1. Compressive strength

The Gibson and Ashby [487] micromechanical models for porous materials are commonly utilized in freeze-casting studies to compare measured and predicted values of compressive strength; several models exist, the most commonly employed for freeze-casting studies are the open-cell [285, 339, 381, 399, 412, 488] and honeycomb [342, 399, 462, 489] models. Both models take the general form [487]:

$$\frac{\sigma_c^*}{\sigma^0} = C_2 \cdot \left(\frac{\rho^*}{\rho^0}\right)^{\beta_1}, \qquad \text{Eq. 5.}$$

where $\sigma_c^*$ is the strength of the porous material, $\sigma^0$ is the strength of the bulk (non-porous) material, $\rho^*/\rho^0$ is the relative density ($\rho^*/\rho^0 = 1 - \phi_p$), and $C_2$ is a proportionality constant. The value of $\beta_1$ varies according to the micromechanical model employed. For the open-cell model, $\beta_1$=3/2. For the honeycomb model, $\beta_1$=1 for the axial loading direction (out-of-plane) and $\beta_1$=2 for radial and tangential (in-plane) loading directions. The out-of-plane honeycomb model is appropriate for anisotropic freeze-cast materials loaded perpendicular to the wall direction [490]. For anisotropic structures, only values corresponding to parallel loads are considered here, thus, $\beta_1$=1. Crushing failure is predicted for ceramics; thus, the modulus of rupture (flexural strength) of the bulk material ($\sigma_f^0$) is utilized as the scale factor. By contrast, metals are predicted to fail by plastic deformation and thus yield strength of the bulk material ($\sigma_c^0$) is used.

Gibson and Ashby [487] provide empirical reference values for $C_2$, which depend on the material type and model employed. Theoretically, $C_2$ depends on pore geometry because $\rho^*/\rho^0$ is related to wall ($t$) and pore ($L$) thickness by Eq. 6.

$$\frac{\rho^*}{\rho^0} = C_2 \cdot \left(\frac{t}{L}\right)^{\beta_1} \qquad \text{Eq. 6.}$$

Thus, it is often more beneficial to calculate $C_2$ when $t$ and $L$ are known, or by fitting experimental data to Eq. 5 when they are unknown [491]. As the combination of $t$ and $L$ are unreported in many freeze-casting studies, we take the latter approach here. Values of $C_2$ for open-cell and honeycomb models are obtained by curve fitting using least squares linear regression after performing logarithmic transforms on relative density and normalized compressive strength. Compressive strength of green composites is sometimes reported (*e.g.*, ref. [492]), however, only sintered samples are considered here.

#### *4.1.1. Metals*

Fig. 27 shows experimental values of normalized compressive strength for freeze cast metals plotted against their respective relative densities; bulk material compressive strength values utilized are provided in Table 7. All plotted values correspond to anisotropic structures. Circle shaped data points represent materials fabricated using aqueous particle suspensions, all of which exhibit lamellar pore structures; these include: Ti [17, 157, 489, 493, 494], Ti-6Al-4V [406], Ti-5wt.%W [157, 494], Cu [462, 495], Fe [496], and W [497]. All materials employing camphene (diamonds) as the fluid exhibited dendritic microstructures; these are: Ti [498-500] and Ti-6Al-4V [501]. The solid lines shown in Fig. 27 are obtained by fitting linear regression equations where $\rho^*/\rho^0$<0.5 and >0.4 (as marked); similarly, the dashed lines represent fits for the open-cell and honeycomb models (as marked). Fitting parameters and regression statistics are provided in Table 7.

Considering cases where $\rho^*/\rho^0$<0.5, the value of $\beta_1$ obtained using the regression model for Eq. 5 is 1.7±0.5 (Table 7), which is reasonably close to the open-cell model prediction of, $\beta_1$=1.5. Honeycomb model predictions of $\sigma_c^*/\sigma_c^0$ are higher than the majority of experimental values until $\rho^*/\rho^0$≈0.4. For $\rho^*/\rho^0$>0.4, the value of $\beta_1$=1.1±0.4 obtained by regression (Table 7) is in better agreement with the honeycomb model ($\beta_1$=1). Similarly, a comparison of *adj.* $R^2$ obtained for the open-cell and honeycomb models indicate that the open-cell model is a better fit for these data when $\rho^*/\rho^0$<0.5 ($R^2$=0.60 for the open-cell model, where $C_2$=0.63±0.07 vs. $R^2$=0.49 for the honeycomb model, where $C_2$=0.38±0.05), whereas the honeycomb model is a better fit when $\rho^*/\rho^0$>0.5 ($R^2$=0.30 for the honeycomb model, where $C_2$=0.50±0.06 vs. $R^2$=0.27 for the open-cell model, where $C_2$=0.70±0.09). Dependencies of fluid type and/or pore structure are not observed.

When the entire range of $\rho^*/\rho^0$ shown in Fig. 27 is considered, the open-cell model predicts ~67% of variance in $\sigma_c^*/\sigma_c^0$, which is approximately equivalent to the regression model (Table 7). For comparison, only ~57% variance in $\sigma_c^*/\sigma_c^0$ is predictable by the honeycomb model (Table 7). We obtain a value of $C_2$ for the open-cell model of 0.64±0.06, which is higher than the empirical value provided by Gibson and Ashby [487] of $C_2$=0.3. As discussed above, the anisotropic microstructure of unidirectional freeze-casted materials results in higher values of compressive strength when the walls of the material are oriented parallel to the loading direction as opposed to perpendicular [462]. These anisotropic strength properties likely contribute to the higher the value of $C_2$ (the intercept) calculated here. Incidentally, the two-fold increase of $C_2$ relative to empirical values is in relatively good agreement with the ratio of enhanced compressive strength parallel vs. perpendicular to the walls calculate above for metals ($\sigma_{||}^*/\sigma_{\perp}^*$=2.4±0.9).

For many metals such as titanium and iron, compressive strength is sensitive to the presence of impurities (*e.g.*, oxygen or carbon) introduced during the sintering step, the presence of which reduces the ductility of the material and increases the yield strength [17]. When impurity data is provided, bulk values of compressive strength can be assigned such that impurity content is accounted for [489] (as they were here). However, these data are often unreported. An inability to account for impurities may be also be increasing the intercept values calculated here.

#### *4.1.2. Ceramics*

Fig. 28 shows experimental values of normalized compressive strength for freeze-cast ceramics plotted against their respective relative densities bulk material compressive strength values utilized are provided in Table 8. Anisotropic (parallel loading) and isotropic materials are shown as blue circles and yellow triangles, respectively. The solid black regression line in Fig. 28 was obtained by fitting all material data to Eq. 5; the corresponding open-cell and honeycomb (out-of-plane) fits are shown as long and short-dashed lines, respectively. The solid blue regression line was obtained by fitting data describing anisotropic materials only; the yellow line describes isotropic data. Coefficient values for all fits are provided in Table 8.

Linear regression coefficients for fitting all data to Eq. 5 are $C_2$=0.69±0.10 and $\beta_1$=2.2±0.1 (Table 8). The value of $\beta_1$ obtained via regression is in closer agreement with the open-cell model ($\beta_1$=1.5) than the honeycomb model ($\beta_1$=1). Values of $\beta_1$ obtained by fitting experimental data for ceramics have been shown to vary significantly; $\beta_1 \geq 2$ is not uncommon [502, 503]. The honeycomb model offers an inferior fit to these data, wherein ~53% of variance in $\sigma_c^*/\sigma_f^0$ is predicted by the open-cell model and the honeycomb model only accounts for ~41% (Table 8). By comparison, 59% of variance in $\sigma_c^*/\sigma_f^0$ is predictable by $\rho^*/\rho^0$ using the linear regression model ($\sigma_c^*/\sigma_f^0 \propto (\rho^*/\rho^0)^{2.2\pm0.1}$; Table 8). For the open-cell model, $C_2$ is calculated to be 0.33±0.02. This value is higher than the recommended, $C_2$=0.2 based on empirical values for brittle materials [503], but within the range of values utilized in anisotropic freeze-casting studies, $C_2$=0.2 [285, 412, 488] to 0.65 [293].

In Fig. 28, linear regression fits for anisotropic (blue) and isotropic (yellow) intersect at $\rho^*/\rho^0 \approx 0.18$. Within the region of $\rho^*/\rho^0 > 0.18$ (pink shaded region of Fig. 28), anisotropic ceramic materials are observed to exhibit, on average, a higher compressive strength to density ratio relative to isotropic structures. As discussed earlier, this is typical behavior for anisotropic freeze-cast materials when compressive loads are applied parallel to the wall direction, as they are here. However, isotropic structures are shown to outperform anisotropic structures when $\rho^*/\rho^0 < 0.18$. An interesting divergence about the blue regression line is observed among anisotropic materials (blue circles) within this region. This divergence may be indicative of two distinct failure modes, wherein anisotropic data points above the regression line (following the honeycomb model line) represent crushing failure, while those below may represent internal wall buckling failure.

Failure by Euler buckling under compressive loads is possible for the individual cell walls of anisotropic ceramics (which may buckle cooperatively within a colony of parallel walls) with low relative densities and relatively low elastic moduli [293, 295]. Over 40% of anisotropic data points that lie below the blue regression line, and within the region of $\rho^*/\rho^0 < 0.18$, have a bulk material elastic modulus of <100 GPa (*e.g.,* hydroxyapatite [240, 286, 504, 505]; calcium silicate [506]), whereas all data above the regression line have bulk material moduli >100 GPa. If data within this region represent buckling failure, we would expect microstructures to exhibit thin walls and high aspect ratios, both of which are most often, unreported. Thus, we are unable to make any conclusions here. Nonetheless, we would expect this behavior to subside with increasing relative density due to a resulting increase in cell wall thickness and interconnectedness. After the point of intersection between the isotropic and anisotropic regression lines, materials having values of bulk moduli <100 GPa are observed above and below the regression line in near equal proportions. Within the entire region of $\rho^*/\rho^0 > 0.18$, ~31% of data points represent materials with bulk moduli <100 GPa; 55% of these data lie above the blue regression line.

The micromechanical models explored here predict strength based on defect-free materials with fully densified walls, thus predictions should be considered an upper bound for attainable strength. Indeed, it is not uncommon for predicted strengths to exceed experimental values, especially where $\rho^*/\rho^0 < 0.5$, which is consistent with the relationship between predicted and experimental values shown in Fig. 28. Ceramics—especially difficult-to-densify ceramics—rarely exhibit fully densified walls; microporosity within cell walls leads to lower values of strength than predicted by these models [307, 338]. However, wall density is seldom reported [492]. The significant scatter observed in Fig. 28 is also common for bulk ceramics; this effect is often attributed to a corresponding scatter in defect population size, since failure during compressive testing is attributed to growth and concentration of these defects [507, 508].

#### *4.1.2.1. Pore structure*

Ceramic materials are grouped by pore structure in Fig. 29, where anisotropic materials are shown in (a-c); including, (a) lamellar, (b) dendritic, and (c) honeycomb. Isotropic structures are shown in (d-e); those are, (d) cellular and (e) equiaxed. The solid black regression lines in Fig. 29 were obtained by fitting data to Eq. 5; the corresponding open-cell and honeycomb (out-of-plane) fits are shown as long and short-dashed lines, respectively. Coefficients for all fits are provided in Table 8.

The regression models for anisotropic lamellar (Fig. 29-a) and honeycomb (Fig. 29-c) structures account for ~63% and ~67% of variance in $\sigma_c^*/\sigma_f^0$, respectively (Table 8). These values are higher than the value of $R^2$ obtained for the regression model describing all anisotropic ceramics; *i.e.*, a reduction in data scatter is observed when lamellar and honeycomb structures are considered independently. Conversely, a decreased value of $R^2$=0.44 is obtained for describing dendritic structures (Fig. 29-b; Table 8). Nevertheless, as shown by the overlapping regression and open-cell model lines in Fig. 29(b), in comparison to other anisotropic structures, dendritic structures most closely follows the open-cell model; $\beta_1$ calculated via regression is 1.6±0.3. The values of $\beta_1$calculated by regression for lamellar and honeycomb structures are 2.5±0.3 and 2.3±0.2, respectively, which are substantially higher than the open-cell value of $\beta_1$=1.5. Even so, the open-cell model is a statistically better fit over the honeycomb model for each of these anisotropic structures ($R^2$=0.52, 0.44, and 0.59 for lamellar, dendritic, and honeycomb open-cell fits, respectively, vs. 0.40, 0.39, and 0.46 for honeycomb fits; Table 8).

Isotropic cellular structures (Fig. 29-d) are found to have a similar value of $\beta_1$ ($\beta_1$=2.0±0.4 $N$=147, $R^2$=0.55, $p$ <0.0001) as that obtained for anisotropic lamellar and honeycomb structures. When water is utilized as a fluid, the distinction between resulting lamellar, dendritic, and cellular structures is largely subjective. These distinctions are often difficult to make and likely influenced by the quality and magnification of the microstructure images provided. These factors may be responsible for some of the scatter in Fig. 29(d) and may also explain why cellular structures seem to show behavior similar to lamellar and honeycomb structures. Conversely, the behavior of isotropic equiaxed structures is closer to that of dendritic structures; equiaxed structures show good agreement with the open-cell model (Table 8). Similar to anisotropic structures, the open-cell model offers a better fit over the honeycomb model for both isotropic structures ($R^2$=0.51 and 0.45 for cellular and equiaxed open-cell fits, respectively, vs. 0.40, 0.41 for honeycomb fits).

#### *4.1.2.2. Fluid*

Fig. 30(a-c) shows normalized compressive strength for anisotropic freeze-cast ceramics plotted against their respective relative densities and grouped by employed fluid type, including: (a) *tert*-butyl alcohol (TBA), (b) water, and (c) camphene. The solid black regression lines in Fig. 30 were obtained by fitting data to Eq. 5; the corresponding open-cell and honeycomb (out-of-plane) fits are shown as long and short-dashed lines, respectively. Coefficients for all fits are provided in Table 9.

The regression fits for $\beta_1$ from Eq. 5 are shown to decrease from 3.1±0.2 for TBA to 2.2±0.2 for water, and finally, 1.8±0.3 for camphene (Table 9). The predominant pore structures for TBA, water, and camphene are honeycomb, lamellar, and dendritic, respectively. Although lamellar structures are highly anisotropic, materials derived from aqueous suspensions are shown to exhibit a range of pore structures, including lamellar, dendritic, cellular, and honeycomb. On average, materials derived from aqueous suspensions are likely more interconnected than those derived from TBA, whereas the dendritic structures obtained using camphene are likely to be the most interconnected of the three. As previously discussed, highly anisotropic materials with low relative densities may fail by internal wall buckling under compression. However, as relative density increases, the width and interconnectedness of cell walls increase, which reduces their tendency to fail by buckling [147]. Thus, increased values of $\beta_1$ are expected for highly anisotropic materials, which is consistent with data in Fig. 29.

Deville *et al.*[158] performed a similar analysis investigating the effect of fluid type on compressive strength of freeze-cast materials. These researchers observed more scatter for materials derived from aqueous suspensions in comparison to those obtained using camphene or TBA as the fluid. Increased scatter for aqueous processing was attributed to an increased risk of defect development during processing and the large variety of pore structures attainable. Similar scatter is observed here for water (Fig. 29-b). However, the scatter observed in Fig. 29(c), for camphene, is statistically higher than that observed for water ($R^2$=0.54 and 0.57 for camphene and water, respectively; $p$<0.0001, for both). The previous analysis was conducted on ceramic and metal materials together, whereas these are separated here. This, coupled with an increase in data currently available, may be responsible for the discrepancy. Here, the greatest predictability in compressive strength properties is found with TBA, where ~81% of variance in $\sigma_c^*/\sigma_f^0$ is predicted by $\rho^*/\rho^0$ ($R^2$=0.81, $p$<0.0001; Table 9).

*4.1.2.3. Test specimen geometry*

The geometry of the testing specimen, including test-specimen shape, aspect ratio, and volume are important considerations for evaluating experimental data scatter of ceramics. An increased propensity for macroscopic buckling failure (as opposed to internal wall failure) is observed for specimens exhibiting high aspect ratios [509]. In contrast, at low aspect ratios (<2), confinement due to friction at the top and bottom of the specimen can increase compressive strength [509]. Specimens of greater volume are known to fail at lower stresses due to an increased probability that the specimen will contain a defect that will act as a stress concentrator [510]. Data shown in Fig. 30 are filtered to obtain data where specimen aspect ratio, $A$ (the ratio the length of the long axis to the short axis) and specimen volume ($V$) are reported. Geometrical characteristics of these data are shown in Fig. 31, where (a) is $\rho^*/\rho^0$, (b) is $A$, (c) is $V$, and (d) is test-specimen shape. For each subplot, TBA, water, and camphene are shown in green, blue, and red, while the combination of these three are shown in grey. For the box plots shown in Fig. 31(a-c), the mean for each condition is shown as a black diamond, the median as a black bar, and outliers are shown as +.

The distributions of $\rho^*/\rho^0$ appear similar, however, a statistically significant difference between the mean values of $\rho^*/\rho^0$ are observed between water and TBA ($t$=2.9, multi-test corrected $p$<0.05); a statistically significant difference is not detected for water/camphene nor TBA/camphene. Aspect ratio distributions are shown in Fig. 32(b). For all fluids, the mean value of $A$ is 1.4±0.7 ($N$=649, median=1.3, min=0.4, max=7.5), and a statistically significant difference among the means between fluid types is detected (water/TBA: $t$=8.1, $p$<0.0001;

water/camphene: $t$=3.3, $p$<0.01; TBA/camphene: $t$=2.8, $p$<0.05; all $p$-values are corrected for multiple tests). The mean $A$ is highest for water (1.5±0.7; $N$=403) and lowest for TBA (1.0±0.5; $N$=116). With regard to specimen volume (Fig. 32-c), the mean value of $V$ for all fluid types is 1.7±1.6 $cm^3$ ($N$=649, median=0.9 $cm^3$, min=0.03 $cm^3$, max=12.1 $cm^3$); a statistically significant difference among means is not detected by one-way ANOVA. Lastly, Fig. 31(d) shows that the majority of all test-specimen shapes, for all fluid types, are cylinders.

A multivariable regression was performed to investigate whether variation in test specimen volume and/or aspect ratio was partially responsible for the scatter observed here. Multicollinearity was observed when including both $A$ and $V$ in the model. Subsequent testing showed that the inclusion of $A$ offered better model fits over $V$ for all fluid types. Thus, the employed model takes the following form:

$$\ln\left(\frac{\sigma_c^*}{\sigma_f^0}\right) = C_3 + \beta_1 \cdot \ln\left(\frac{\rho^*}{\rho^0}\right) + \beta_2 \cdot \ln(A), \qquad \text{Eq. 7}$$

Fig. 32(a-c) shows normalized compressive strength as a function of $\rho^*/\rho^0$ and $A$ for (a) TBA, (b) water, and (c) camphene. In all cases, condition numbers (used here to detect multicollinearity) for the models are <10 and values of $adj.\ R^2$ are within 98.8% of $R^2$. Adjusted $R^2$ for the model shown for TBA in Fig. 30(a) is 0.96, meaning that 96% of variance can be explained by Eq. 7 ($\sigma_c^*/\sigma_f^0 \propto (\rho^*/\rho^0)^{2.5} \cdot A^{-1.0}$; Table 10), whereas only ~88% can be explained by $\rho^*/\rho^0$ alone (Table 10). The Bayesian Information Criterion (BIC) decreases from 187 for $\sigma_c^*/\sigma_f^0 \propto \rho^*/\rho^0$ to 64 for Eq. 7, providing evidence that, for TBA, Eq. 7 is a better predictor of $\sigma_c^*/\sigma_f^0$ than $\rho^*/\rho^0$ alone. Similarly, the BIC decreases from 956 to 952 for water (Fig. 30-b; Table 10) and from 200 to 176 for camphene (Fig. 30-c; Table 10).

For freeze-cast ceramics derived from water and TBA suspensions, increased values of $A$ are predicted to yield decreased compressive strength. This effect is predicted to be stronger for TBA, where $\beta_2$=-1.0±0.1 vs. $\beta_2$=-0.4±0.2 for water (Table 10). This observation agrees with our hypothesis that microstructures templated by TBA are less interconnected, on average, than those templated by water, because increased test-specimen aspect ratios would have a greater, negative effect on more highly anisotropic microstructures by increasing the probability of wall buckling. A reverse trend is observed for camphene, where $\beta_2$=0.7±0.3; that is, test specimens with higher aspect ratios are predicted to fail at increased compressive strengths relative to their lower aspect ratio counterparts.

The relatively high negative value of $\beta_2$ for TBA and positive value for camphene may be influenced by the distribution of test-specimen shapes shown in Fig. 31(d). Cylinders make up the majority of all test-specimen shapes, however, the percentage of cylinders is highest for camphene (~80% of test-specimens) and lowest for TBA (~50%). Conversely, most cube specimens are represented by TBA whereas no data are available for cube specimens for camphene. Research on the influence of test-specimen shape on compressive properties is largely limited to that for concrete. Increased aspect ratios are reported to have a larger, negative effect on the compressive strength of bar-shaped specimens vs. cylinders [511]. Increased aspect ratio has also been reported to increase the compressive strength of cylinders and decrease the compressive strength of bars [512]. The over-representation of cylinder specimens for camphene relative to other fluids may be partially responsible for the reverse trend of *A* for camphene.

To test the influence of specimen shape on resulting compressive properties, a dummy variable for test-specimen shape ($S$) was added to Eq. 7, where bar and cube specimens were grouped (under the assumption that predictor variable *A* captures the difference between a bar and a cube) and given a value of *S*=1, whereas for cylinders, *S*=0 (Eq. 8).

$$\ln\left(\frac{\sigma_c^*}{\sigma_f^0}\right) = C_3 + \beta_1 \cdot \ln\left(\frac{\rho^*}{\rho^0}\right) + \beta_2 \cdot \ln(A) + \beta_3 \cdot \ln(S), \quad \text{Eq. 8}$$

The resulting coefficient of *S* for TBA, $\beta_3 = \exp(-0.6 \pm 0.2)$ predicts a reduction of ~60% in compressive strength for bar-shaped vs. cylinder specimens (holding all other variables constant). The reverse trend is found for water, where $\beta_3 = \exp(0.4 \pm 0.2)$; that is, bar-shaped specimens are predicted to yield ~40% higher compressive strengths relative to cylinder-shaped specimens (holding all other variables constant). Inclusion of *S* as a dummy predictor variable for camphene was statistically insignificant. As shown in Fig. 31(d), the distribution of cylinder and bar specimens is highly skewed, thus it is difficult to draw any conclusions here. Nevertheless, this highlights an important issue regarding the need for adherence to generalized testing procedures. The mean value of *A* for cylinder-shaped test specimens is rather low; 1.3±0.3 and 0.6±0.3 for aqueous and TBA systems, respectively. A combination of friction at the interfaces and a low aspect ratio leads to a anisotropic stress distribution within the samples during testing and may result in barreling of cylindrically-shaped test specimens [513], resulting in decreased predictability of strength properties.

In Fig. 32(b) for water, data point color denotes pore structure, where blue is lamellar, red is dendritic, and green is honeycomb. Scatter is observed for dendritic structures above and below the regression line in Fig. 32(b). However, given that dendritic branching increases interconnectedness of pore walls, a transition to a dendritic structure should result in increased compressive strengths, especially at low values of $\rho^*/\rho^0$. Data in Fig. 32(b) do not support this prediction, which is likely attributable to the difficulty in drawing qualitative distinctions between what constitutes lamellar vs. dendritic morphologies, as discussed earlier. A quantitative approach, such as that employed by Ghosh *et al.* [514] through the use of a "bridge density" factor, where the number of interconnections between lamellae are countered and related to area of the region, holds much utility provided its quantification is utilized in more studies. On the other hand, data corresponding to honeycomb pore structures (green) are mostly observed above the regression line.

#### *4.1.2.4. Comparison with previous modeling approaches*

These results contradict those reported in previous reviews [158, 515, 516], where the authors determined that the open-cell micromechanical model was inappropriate for describing the compressive response of freeze-cast metals and ceramics. Instead, the Zhang and Ashby [490] out-of-plane honeycomb model was utilized to describe freeze-cast ceramic and metal materials:

$$\sigma_c^* = 6E_0 \cdot \left(\frac{\rho^*}{\rho}\right)^3, \qquad \text{Eq. 9}$$

where $E_0$ is the elastic modulus of the bulk material. The above relation describes failure by elastic buckling and is an approximation based on a model for describing elastomeric honeycomb materials in compression, where $E_0$=0.9 GPa [490] Failure by elastic buckling has been reported for some anisotropic freeze-cast ceramics, including zeolite/bentonite composites [293] and titanium dioxide ($TiO_2$)[295], and we find some support for it here. However, there are at least two issues with extending Eq. 9 to ceramic materials. First, the elastic deformation region for elastomeric materials is comparatively longer than that of ceramics; even still, the elastomeric materials studied in [490] failed by buckling only when $\rho^*/\rho^0$<0.12, above which, they failed by fracture. Second, the value of $E_0$ for ceramic materials is significantly higher than that for typical elastomeric materials; given the prefactor of 6 in Eq. 1, predicted values of $\sigma_c^*$ are unreasonably high (*e.g.*, for an $Al_2O_3$ material with $\rho^*/\rho^0$=0.4 and $E_0$=376 GPa [517]; by Eq. 8, $\sigma_c^* \approx$ 144 GPa, which is ~50 times higher than the compressive strength of fully-dense $Al_2O_3$, 2.95 GPa [518]). Alternative models of elastic buckling failure, which predict more reasonable values of compressive strength for anisotropic ceramic materials, can be found in refs.[293, 295].

Upon analyzing freeze-cast metals and ceramics independently, we find that the open-cell model is appropriate for describing anisotropic freeze-cast metals, especially when the value of the intercept is obtained by fitting experimental data to the open-cell model. The open-cell model accounts for ~53% variance in values of normalized compressive strength for freeze-cast ceramics, which is lower than that explained for metals (~67%). However, we do not find a significant difference between values of the coefficient of determination when fitting anisotropic vs. isotropic freeze-cast ceramics to the open-cell model; albeit, 78% of these data represent anisotropic materials whereas only 22% of these data are obtained from isotropic structures. Upon categorizing by pore structure for ceramics, we find that the open-cell model is a statistically better fit for anisotropic dendritic and isotropic equiaxed structures in comparison to other pore structures; for lamellar, honeycomb, and cellular structures, we find that neither the open-cell nor the honeycomb model offers a good fit to these data. The scatter observed for ceramics may also be due to other factors that are inherent to ceramic processing; *e.g.*, microporosity within cell walls, variance in test specimen volume/aspect ratio, etc.

## 5. Concluding remarks

The newly established FreezeCasting.net open-data repository contains, as of August 2017, experimental data from over 800 papers published on the freeze-casting technique pertaining to data which relate processing conditions to microstructure and material properties. The aim of the database and the present paper is to facilitate broad dissemination of relevant data to freeze-casting researchers so as to promote better informed experimental design, to encourage industrial use of freeze-casting, and to accelerate modeling efforts that relate processing conditions to microstructure formation and material properties.

Microstructural characteristics of freeze-cast materials are summarized and key processing-structure-property relationships are reviewed and tested against literature data contained within the FreezeCasting.net database. The following conclusions are drawn:

**All materials**

1. The relationship between porosity and solid particle fraction in the suspension is minimally dependent on the type of fluid employed, but highly correlated with the material type (*i.e.,* metal, ceramic, or polymer). Sintering shrinkage is also an important factor, wherein, materials that exhibit the lowest rates of sintering shrinkage (*e.g.,* polymers) show greater dependence of porosity on initial solid fraction.
2. The range of pore width attainable during freeze-casting is not highly dependent on total porosity.

**Anisotropic materials derived from aqueous suspensions**

3. Structure wavelength - defined as the sum of the widths of one lamellar pore and its adjacent wall - correlates with solidification velocity; this relationship is statistically weaker when considering dependencies of only pore width or only wall width on solidification velocity and strongest when considering sublimated, but not sintered samples in comparison to sintered samples.
4. Cooling rate - usually defined as the rate at which the freezing substrate (on which the sample mold is placed at the start of solidification) is cooled to a set, minimal temperature – does not significantly correlate with microstructural parameters in freeze-cast materials.

**Metals**

5. For anisotropic freeze-cast, sintered metals, the relationship between compressive strength and relative density is best described by the Gibson-Ashby open-cell model when the relative density of the material is less than 0.5. At higher relative densities, the out-of-plane honeycomb model is a better fit.

**Ceramics**

6. For both anisotropic and isotropic freeze-cast ceramics, the Gibson and Ashby open-cell model offers a better fit to experimental data than the honeycomb model; these results contradict previous reviews on this topic.
7. For anisotropic freeze-cast ceramics derived from particle suspensions using water, camphene, and *tert*-butyl alcohol as the fluid, improved regression models predicting compressive strength are obtained when considering test specimen aspect ratio along with relative density. However, a decrease in compressive strength is predicted for increased test-specimen aspect ratios for materials derived from water and *tert*-butyl alcohol suspensions, whereas the reverse is found for camphene. Further testing is needed to clarify this issue.

**Acknowledgements**

This work was supported by grants from NASA's Physical Sciences Research Program, MaterialsLab Open Science Campaign (NNH15ZTT002N) and NASA Office of Education and the Science Mission Directorate (NNH15ZDA010C). KS was supported by a scholarship through the Jack Kent Cooke Foundation. The authors acknowledge Dr. Sylvain Deville (CNRS, France) for numerous discussions on the FreezeCasting.net website, database, and preparation of this manuscript; the following individuals are also acknowledged for their helpful feedback and suggestions on the FreezeCasting.net website design: Mr. Jon Chambers (School of the Art Institute of Chicago, USA), Mr. Maxime Garnier (ETH, Switzerland), Dr. Adam Stevenson (CNRS, France). The authors also acknowledge Mr. Aaron Shelhamer (Northern Illinois University, USA) for his assistance with the website interactive plotting application.

**Tables**

Table 1. Fitting parameters for regression models describing dependency of volume fraction on total porosity, where *a* and *b* are the slope and intercept, respectively, from Eq. 1 ($\phi_p = a \cdot \phi_s + b$), where $\phi_p$, $\phi_s$, and *b* are expressed as percent, *N* is the number of samples, *r* and $R^2$ are the coefficients of correlation and determination, respectively. Confidence intervals are shown in brackets and *p* values for regression models and coefficient values are expressed as one, two, three and four stars for $p < 0.05$, 0.01, 0.001, and 0.0001; *ns* indicates the value was not significant.

| | Coefficients | | | | | |
|---|---|---|---|---|---|---|
| | a | b | N | r | $R^2$ | References |
| ALL | -1.0[-1.3,-0.7]**** | 86[80,92]**** | 2855 | -0.12 | 0.02**** | |
| Frozen | ns | ns | 23 | ns | ns | [262, 410] |
| Green | -0.9[-1.0, 0.8]**** | 92[90,93]**** | 551 | -0.62 | 0.39**** | [18, 25, 54, 180, 198, 223, 230, 245, 281, 307, 314, 321, 323, 341, 350, 352, 390, 413, 416, 426, 427, 436, 440, 455, 519-562] |
| Sintered | -0.9[-1.3,-0.5]**** | 82[72,92]**** | 2281 | -0.09 | 0.01**** | [6, 17, 24, 38, 49, 107, 109-112, 114, 117, 120, 121, 124, 141, 148, 149, 157, 181, 184, 185, 187, 189, 195-197, 199, 217-220, 224, 226, 228, 232, 239, 240, 261, 262, 271, 274, 282-286, 295, 296, 298, 299, 302, 304, 309-312, 316, 320, 324, 325, 328, 329, 336, 339, 341, 343-345, 347, 351, 353, 379-388, 391-394, 396, 403, 405, 406, 408, 410, 412, 414, 415, 425, 429, 432-435, 446, 448, 449, 453, 458, 460, 462, 489, 492, 493, 496-501, 504, 505, 514, 516, 531, 532, 543, 563-739][740] |
| FLUIDS | | | | | | |
| Water | -1.1[-1.2,-1.0]**** | 87[86,88]**** | 2044 | -0.66 | 0.43**** | |
| Green | -0.9[-1.0,-0.8]**** | 91[90,93]**** | 458 | -0.66 | 0.43**** | [54, 198, 223, 245, 281, 323, 341, 390, 413, 426, 427, 436, 519-524, 528-531, 533, 535-537, 539-544, 547-549, 551, 552, 555, 557-562, 710, 741] |
| Sintered | -1.0[-1.1,-1.0]**** | 84[82,85]**** | 1586 | -0.59 | 0.35**** | [6, 17, 24, 49, 107, 109, 111, 112, 114, 117, 120, 124, 141, 148, 149, 157, 181, 184, 185, 187, 195-197, 199, 217-220, 224, 228, 232, 239, 240, 261, 282-286, 295, 296, 298, 299, 302, 304, 309-312, 316, 320, 324, 325, 328, 329, 336, 339, 341, 343-345, 353, 379, 384, 386, 387, 392, 393, 396, 403, 405, 406, 408, 414, 415, 425, 429, 432-434, 446, 448, 449, 458, 460, 462, 489, 493, 496, 497, 505, 506, 514, 516, 531, 532, 543, 563-567, 569-572, 577-579, 581-590, 594, 595, 598-603, 609-611, 613-617, 619-639, 642-646, 648-652, 655-662, 665, 667, 669, 670, 679, 681-684, 688, 689, 696, 701, 705, 707, 709-714, 717-719, 721, 722, 727-732, 735, 742, 743] |
| Camphene | -1.2[-1.3,-1.0]**** | 81[77,85]**** | 290 | -0.61 | 0.37**** | |
| Frozen | ns | ns | 23 | ns | ns | [262, 410] |
| Sintered | -1.2[-1.4,-1.0]**** | 82[78,86]**** | 267 | -0.63 | 0.39**** | [24, 38, 226, 228, 262, 271, 347, 380-383, 385, 388, 391, 410, 453, 498-501, 568, 573-576, 604-608, 618, 640, 653, 654, 668, 671-673, 675, 676, 685-687, 691, 692, 694, 695, 697, 702, 719, 720, 723-726, 740] |
| TBA | ns | ns | 408 | ns | ns | [110, 121, 189, 224, 274, 325, 351, 394, 435, 492, 504, 591-593, 596, 597, 647, 663, 664, 666, 674, 677, 678, 680, 690, 693, 698-700, 703, 704, 706, 708, 715, 716, 733, 734, 737-739] |
| CERAMICS | -0.9[-1.3,-0.4]**** | 82[72,92]**** | 2183 | -0.08 | 0.01**** | |
| Water | -1.0[-1.1,-0.9]**** | 84[83,86]**** | 1506 | -0.60 | 0.36**** | [6, 49, 107, 109, 111, 112, 117, 120, 124, 148, 149, 184, 185, 187, 195-197, 199, 217-220, 224, 228, 239, 240, 245, 261, 282-286, 295, 296, 298, 299, 302, 304, 309-312, 316, 320, 324, 325, 328, 329, 336, 339, 341, 343-345, 353, 379, 384, 386, 387, 390, 392, 393, 396, 403, 405, 408, 414, 415, 425, 429, 432, 433, 446, 448, 449, 458, 460, 505, 506, 514, 516, 520, 521, 523, 524, 528, 530-532, 536, 537, 539-541, 543, 544, 549, 558-563, 566, 567, 569-572, 577-579, 581-590, 594, 595, 598-603, 609, 610, 615-617, 619-639, 642-646, 648, 649, 651, 652, 655-662, 665, 667, 670, 679, 681-684, 688, 689, 696, 701, 705, 707, 709-714, 717-719, 721, 722, 727-731, 735, 742, 744] |
| Camphene | -1.1[-1.3,-0.9]**** | 81[77,85]**** | 259 | -0.61 | 0.37**** | [38, 226, 228, 262, 271, 347, 380, 381, 383, 385, 388, 391, 410, 453, 568, 573-576, 606-608, 618, 640, 653, 654, 668, 671-673, 675, 676, 685, 691, 692, 694, 695, 697, 702, 719, 720, 723-726, 740] |
| TBA | ns | ns | 400 | ns | ns | [110, 121, 189, 224, 274, 325, 351, 394, 435, 492, 504, 591-593, 596, 597, 647, 663, 664, 666, 674, 677, 678, 680, 690, 693, 698-700, 703, 704, 706, 708, 715, 716, 733, 734, 737-739] |
| METALS | -1.6[-1.9,-1.3]**** | 88[82,94]**** | 131 | -0.72 | 0.51**** | |
| Water | -1.6[-1.9,-1.3]**** | 88[82,94]**** | 104 | -0.73 | 0.53**** | [17, 141, 157, 181, 232, 406, 434, 462, 489, 493, 496, 497, 564, 565, 732] |
| Camphene | -1.6[-2.3,-0.9]**** | 88[71,99]**** | 27 | -0.67 | 0.44**** | [382, 498-501, 604, 605, 686, 687] |
| POLYMER | -0.8[-0.9,-0.7]**** | 93[92,95]**** | 248 | -0.62 | 0.38**** | [180, 198, 230, 350, 352, 525-527, 545] |
| Water | -0.8[-0.9,-0.7]**** | 95[94,96]**** | 204 | -0.77 | 0.59**** | [54, 198, 223, 281, 323, 413, 426, 427, 436, 519, 522, 529, 533, 535, 541, 542, 547, 548, 551, 552, 555, 557, 741] |

Table 2. Mean values of sintering shrinkage reported in freeze-casting literature expressed as a change in diameter, height, and/or volume for metals, ceramics, and polymer materials.

| | All | | % Diameter | | %Height | | %Volume | | |
|---|---|---|---|---|---|---|---|---|---|
| | N | Mean | N | Mean | N | Mean | N | Mean | References |
| Metals | 63 | 52±18 | 2 | 83 | 35 | 42±12 | 26 | 64±17 | |
| Ceramics | 644 | 21±13 | 136 | 20±9 | 418 | 21±13 | 90 | 25±13 | |
| Polymers | 23 | 12±13 | 14 | 6±7 | 1 | 5 | 8 | 23±14 | |
| Anisotropic | | | | | | | | | |
| Metals | 58 | 52±19 | 2 | 83 | 35 | 42±12 | 21 | 67±18 | [55, 141, 181, 434, 462, 497, 565] |
| Ceramics | 554 | 21±11 | 123 | 20±8 | 349 | 21±12 | 82 | 25±13 | [6, 110, 148, 156, 185, 228, 239, 240, 262, 286, 295, 296, 316, 341, 383, 384, 390, 408, 414, 415, 566, 573, 574, 586, 590, 608, 619, 622, 623, 637, 638, 647, 652, 655, 656, 660, 665, 666, 674, 683, 688, 690, 691, 700, 712, 714, 720, 731, 737, 738, 745-751] |
| Polymers | 21 | 12±22 | 14 | 6±7 | 1 | 5 | 6 | 26±14 | [352, 519, 535, 547, 752] |
| Isotropic | | | | | | | | | |
| Metals | 5 | 51±7 | - | - | - | - | 5 | 51±7 | |
| Ceramics | 90 | 21±19 | 13 | 12±16 | 69 | 23±20 | 8 | 19±15 | [108, 199, 588, 591, 592, 596, 698, 750, 753-755] |
| Polymer | 2 | 12±8 | - | - | - | - | 2 | 12±8 | [519] |

Table 3. Fitting parameters for least square regression analysis describing dependency microstructural parameters on solidification velocity, where $N$ is the number of samples, $C_1$ and $k$ are from the log transformation of Eq. 2 ($\ln(\lambda) = C_1 + k \cdot \ln(v)$), and $R^2$ is the coefficient of determination. Confidence intervals are shown in brackets and $p$ values are indicted by one, two, three and four stars for $p$< 0.05, 0.01, 0.001, and 0.0001; *ns* indicates the value was not significant.

| | Coefficients | | | | |
|---|---|---|---|---|---|
| | ln ($C_1$) | k | N | $R^2$ | References |
| STRUCTURE WAVELENGTH | | | | | |
| ALL | 5.2[4.9,5.5]**** | -0.5[-0.6,-0.4]**** | 286 | 0.29**** | |
| Ceramics | 5.2[4.9,5.5]**** | -0.6[-0.7,-0.4]**** | 260 | 0.28**** | |
| Solidification | 3.6[3.4,3.7]**** | -0.0[-0.1,-0.0]* | 12 | 0.28* | [243, 398] |
| Green | 6.3[5.9,6.7]**** | -0.9[-1.1,-0.7]**** | 63 | 0.59**** | [236-238] |
| Sintered | 4.4[3.9,4.9]**** | -0.3[-0.5,-0.1]**** | 185 | 0.07*** | [120, 147, 150, 156, 228, 229, 239, 296, 379, 396, 411, 418, 421, 448, 514, 689, 756, 757] |
| Polymers | | | | | |
| Green | 7.0[5.7,8.3]**** | -1.0[-1.4,-0.7]**** | 7 | 0.93*** | [223, 419, 758-760] |
| Composites | | | | | |
| Ceramic-polymer | 4.5[4.0,4.9]**** | -0.4[-0.6,-0.2]**** | 18 | 0.59*** | [307] |
| Carbon-polymer | ns | ns | 1 | ns | [440] |
| PORE WIDTH | | | | | |
| ALL | 4.3[3.9,4.7]**** | -0.4[-0.6,-0.3]**** | 157 | 0.24**** | [401, 440, 441, 445] |
| Ceramics | 3.6[3.2,4.1]**** | -0.3[-0.4,-0.1]**** | 124 | 0.13**** | |
| Solidification | 3.6[3.6,3.7]**** | -0.2[-0.2,-0.1]**** | 6 | 0.97*** | [243, 398] |
| Green | ns | ns | 1 | ns | [417] |
| Sintered | 3.6[3.1,4.1]**** | -0.3[-0.4,-0.1]** | 117 | 0.07** | [120, 156, 228, 229, 239, 296, 379, 396, 406, 411, 418, 420, 440, 445, 448, 514, 649, 689, 749, 756, 761-763] |
| Metals | | | | | |
| Sintered | 18[8.6,27]* | -4.6[-7.9,-1.4]* | 3 | 0.99* | [406] |
| Polymers | | | | | |
| Green | 6.4[5.8,6.9]**** | -0.9[-1.0,-0.8]**** | 19 | 0.89**** | [223, 323, 417, 420, 758, 759, 763, 764] |
| Composites | | | | | |
| Carbon-polymer | ns | ns | 4 | ns | [440, 445] |
| Ceramic-polymer | ns | ns | 4 | ns | [323] |
| WALL WIDTH | | | | | |
| ALL | 2.8[2.2,3.4]**** | -0.2[-0.4,0.0]* | 140 | 0.04* | [445] |
| Ceramics | ns | ns | 132 | ns | [765] |
| Solidification | ns | ns | 5 | 0.91* | [398] |
| Frozen | ns | ns | 2 | ns | [765] |
| Sintered | 3.1[2.4,3.8]**** | -0.3[-0.5,0.1]** | 125 | 0.06** | [120, 156, 228, 229, 239, 296, 379, 396, 397, 411, 418, 448, 514, 689, 713, 756, 766] |
| Polymers | 5.4[2.7,8.1]** | -0.9[-1.5,0.2]* | 6 | 0.77* | |
| Green | 5.4[2.7,8.1]** | -0.9[-1.5,0.2]* | 6 | 0.77* | [223, 758, 759] |
| Composites | | | | | |
| Carbon-polymer | ns | ns | 4 | ns | [445] |

Table 4. Fitting parameters for least square regression analysis describing dependency microstructural parameters on cooling rate, where *N* is the number of samples, $C_1$ and *k* are from the log transformation of Eq. 2 ($\ln(\lambda) = C_1 + k \cdot \ln(v)$), where *v* is the applied cooling rate, and $R^2$ is the coefficient of determination. Confidence intervals are shown in brackets and *p* values are indicted by one, two, three and four stars for $p < 0.05$, 0.01, 0.001, and 0.0001; *ns* indicates the value was not significant.

| | Coefficients | | | | |
|---|---|---|---|---|---|
| | $C_1$ | k | N | $R^2$ | References |
| STRUCTURE WAVELENGTH | | | | | |
| ALL | 3.9[3.7,4.0]$^{****}$ | -0.2[-0.3,-0.1]$^{****}$ | 212 | 0.11$^{****}$ | |
| Ceramics | 3.9[3.8,4.1]$^{****}$ | -0.2[-0.3,-0.1]$^{****}$ | 185 | 0.12$^{****}$ | |
| Solidification | ns | ns | 5 | ns | [398, 765] |
| Sintered | 4.0[3.8,4.2]$^{****}$ | -0.2[-0.3,-0.2]$^{****}$ | 180 | 0.13$^{****}$ | [109, 149, 290-293, 295, 339, 396, 400, 403, 408, 411, 418, 421, 516, 569, 626, 743, 767-769] |
| Polymers | ns | ns | 3 | ns | |
| Green | ns | ns | 2 | ns | [419, 770] |
| Solidification | ns | ns | 1 | ns | [771] |
| Composites | | | | | |
| Ceramic/polymer | 3.5[3.4,3.6]$^{****}$ | -0.1[-0.2, 0.0]$^{*}$ | 24 | 0.22$^{*}$ | [307] |
| PORE WIDTH | | | | | |
| ALL | 3.8[3.6,4.0]$^{****}$ | -0.4[-0.5,-0.3]$^{****}$ | 334 | 0.13$^{****}$ | |
| Ceramics | 3.7[3.5,3.9]$^{****}$ | -0.3[-0.4,-0.2]$^{****}$ | 280 | 0.10$^{****}$ | |
| Solidification | ns | ns | 5 | ns | [398] |
| Green | ns | ns | 11 | ns | [390, 561, 772, 773] |
| Sintered | 3.7[3.5,3.9]$^{****}$ | -0.3[-0.4,-0.2]$^{****}$ | 264 | 0.10$^{****}$ | [107, 109, 149, 233, 282, 283, 286, 290-293, 295, 311, 339, 396, 400, 403, 408, 411, 414, 418, 456, 516, 569, 570, 582, 626, 660, 667, 728, 743, 767-769, 774-777] |
| Polymers | 4.4[4.1,4.7]$^{****}$ | -0.3[-0.5,-0.1]$^{**}$ | 31 | 0.28$^{**}$ | |
| Green | 4.4[4.1,4.7]$^{****}$ | -0.3[-0.5,-0.1]$^{**}$ | 30 | 0.28$^{**}$ | [54, 281, 297, 323, 420, 423, 542, 770, 778-780] |
| Solidified | ns | ns | 1 | ns | [771] |
| Composites | | | | | |
| Ceramic/polymer | 3.9[3.3,4.5]$^{****}$ | -0.4[-0.8,-0.1]$^{*}$ | 17 | 0.28$^{*}$ | [297, 307, 323, 728, 781] |
| WALL WIDTH | | | | | |
| ALL | 2.8[2.6,2.0]$^{****}$ | -0.3[-0.5,-0.2]$^{****}$ | 190 | 0.18$^{****}$ | |
| Ceramics | 2.9[2.7,3.1]$^{****}$ | -0.4[-0.5,-0.3]$^{****}$ | 180 | 0.28$^{****}$ | |
| Solidification | ns | ns | 5 | ns | [398, 765] |
| Sintered | 3.1[2.9,3.2]$^{****}$ | -0.4[-0.5,-0.4]$^{****}$ | 173 | 0.38$^{****}$ | [109, 149, 290-293, 295, 302, 339, 396, 400, 403, 408, 411, 418, 516, 569, 626, 696, 767-769] |
| Metals | | | | | |
| Sintered | ns | ns | 2 | ns | [494] |
| Polymers | ns | ns | 2 | ns | |
| Green | ns | ns | 1 | ns | [770, 771] |
| Solidified | ns | ns | 1 | ns | [771] |
| Composites | | | | | |
| Ceramic/polymer | ns | ns | 6 | ns | [307] |

Table 5. Summary of demonstrated porosity for anisotropic (lamellar, dendritic, and honeycomb) and isotropic (cellular and equiaxed) freeze-cast microstructures. $N$ is the number of samples; mean ($\bar{x}$) values are expressed as $\bar{x} \pm$ standard deviation (SD). Summary data representing data obtained during *in-situ* solidification is termed "solidification"; data summarizing frozen, sublimated-not-sintered-, and sintered samples are termed, "green", "frozen", and "sintered", respectively.

| | | N | Mean $\bar{x} \pm \mathbf{SD}$ | Median ($\tilde{x}$) | References |
|---|---|---|---|---|---|
| Anisotropic structures | | | | | |
| Lamellar | Green | 206 | 80±18 | 91 | [25, 54, 223, 281, 307, 314, 321, 390, 416, 426, 427, 440, 519, 522, 523, 528, 529, 534, 536, 538, 539, 541-544, 546, 551-554, 557, 558, 560-562, 741, 779, 782-787] |
| | Sintered | 1116 | 65±14 | 65 | [6, 17, 24, 49, 107, 109, 111, 112, 117, 120, 124, 141, 147, 149, 150, 181, 184, 185, 195, 217-220, 224, 228, 232, 239, 240, 261, 282-286, 295, 296, 298, 299, 302, 309-312, 316, 325, 336, 339, 345, 353, 379, 384, 386, 390, 392, 396, 397, 403, 405, 406, 408, 412, 414, 415, 425, 446, 448, 458, 460, 462, 488, 489, 493, 494, 496, 497, 505, 506, 514, 516, 543, 563, 564, 566, 567, 569, 570, 572, 579, 581, 582, 586, 601-603, 611, 613, 614, 616, 624-628, 630-633, 635, 637-639, 642, 644, 645, 649, 651, 655, 659, 660, 662, 667, 670, 680, 682, 688, 689, 693, 696, 705, 711-714, 718, 721, 722, 724, 725, 728, 732-735, 742, 747, 767, 768, 788-797] |
| Dendritic | Frozen | 23 | 66±18 | 50 | [262, 410] |
| | Green | 74 | 80±15 | 82 | [180, 230, 281, 436, 455, 519, 521, 525, 528, 533, 540, 546, 555] |
| | Sintered | 398 | 61±17 | 60 | [6, 148, 187, 196, 224, 228, 262, 271, 296, 302, 304, 309, 311, 320, 324, 328, 329, 347, 348, 353, 357, 379-383, 385, 388, 393, 394, 396, 410, 449, 453, 461, 498-501, 567, 568, 570, 573-576, 584, 586, 589, 604-608, 613, 614, 617-620, 622-624, 632, 640, 641, 652, 653, 656, 657, 665, 666, 668, 672, 673, 675, 676, 681, 683-687, 691, 692, 694, 695, 697, 702, 707, 719, 723, 726, 731, 736, 797] |
| Honeycomb | Green | 101 | 87±17 | 94 | [281, 323, 341, 350, 352, 525, 530, 534, 535, 798, 799] |
| | Sintered | 334 | 82±18 | 66 | [110, 121, 189, 224, 274, 299, 339, 341, 342, 351, 387, 394, 429-435, 492, 569, 593-595, 647, 650, 663, 664, 678, 690, 700, 703, 704, 715, 716, 737-739, 750, 800] [504, 750] |
| Isotropic structures | | | | | |
| Cellular | Green | 199 | 82±18 | 89 | [198, 223, 230, 245, 314, 413, 426, 427, 519, 520, 522, 524, 526, 527, 529, 531, 536, 537, 540, 545, 547-551, 553, 554, 559, 560, 562, 710, 741, 801-805] |
| | Sintered | 541 | 49±24 | 52 | [38, 39, 107, 114, 157, 199, 220, 226, 271, 274, 282, 284, 285, 304, 310, 325, 347, 379, 391, 453, 492, 499, 531, 532, 565, 571, 577, 578, 585, 590-593, 596-598, 612, 615, 617, 619, 621, 624, 625, 629, 634, 636, 641, 642, 644, 648, 654, 660, 661, 668, 669, 671, 674, 677, 684, 696, 706, 708-710, 712-714, 717, 719-721, 727, 729-731, 735, 745, 754, 790, 793, 795, 806-809] |
| Equiaxed | Green | 1 | 95 | - | [784] |
| | Sintered | 58 | 76±10 | 75 | [197, 343, 580, 583, 588, 609, 610, 643, 658, 679, 698, 699, 755] |

Table 6. Summary of demonstrated pore and wall width for anisotropic (lamellar, dendritic, and honeycomb) and isotropic (cellular and equiaxed) freeze-cast microstructures. $N$ is the number of samples; mean ($\bar{x}$) values are expressed as $\bar{x} \pm$ standard deviation (SD). Summary data representing data obtained during *in-situ* solidification is termed "solidification"; data summarizing frozen, sublimated-not-sintered-, and sintered samples are termed, "green", "frozen", and "sintered", respectively.

| | | N | Mean $\tilde{\boldsymbol{x}} \pm \mathbf{SD}$ | Median ($\bar{\boldsymbol{x}}$) | References |
|---|---|---|---|---|---|
| **PORE WIDTH** (μm) | | | | | |
| Lamellar | Solidification | 8 | 29±13 | 31 | [243, 398, 771, 810] |
| | Green | 221 | 73±90 | 42 | [54, 155, 191, 223, 281, 287-289, 294, 297, 307, 314, 321, 390, 401, 417, 420, 423, 439-441, 445, 459, 519, 528, 529, 536, 538, 539, 542, 544, 551, 553, 557, 558, 561, 562, 741, 752, 758, 759, 763, 764, 770, 772, 773, 778-781, 783, 785, 787, 811-825] |
| | Sintered | 653 | 44±85 | 21 | [6, 17, 24, 49, 55, 106, 107, 109, 111, 117, 120, 149, 156, 181, 184, 191, 217-220, 228, 229, 233, 239, 282-286, 290-293, 295, 296, 298, 299, 309, 311, 312, 339, 345, 353, 379, 396, 400, 403, 405, 406, 408, 411, 412, 414, 418, 425, 448, 456, 458, 460, 462, 488, 489, 493, 496, 514, 516, 543, 563, 564, 567, 569, 570, 582, 601, 602, 616, 626, 628, 632, 633, 637, 645, 649, 659, 660, 667, 670, 680, 689, 722, 724, 728, 732-734, 743, 747-749, 751, 756, 761, 762, 767-769, 774-777, 789, 795, 797, 826-835] |
| Dendritic | Solidification | 39 | 45±5 | 45 | [2] |
| | Frozen | 22 | 86±45 | 75 | [262, 410] |
| | Green | 31 | 17±20 | 7 | [44, 52, 180, 281, 389, 519, 525, 540, 836-838] |
| | Sintered | 267 | 58±87 | 20 | [6, 148, 187, 196, 228, 262, 296, 303, 309, 311, 320, 329, 348, 353, 357, 381, 382, 388, 393, 394, 396, 410, 418, 449, 453, 461, 498-500, 567, 568, 570, 575, 584, 604-607, 618-620, 632, 653, 656, 673, 675, 676, 685, 686, 691, 692, 694, 719, 726, 797, 831, 839-852] |
| Honeycomb | Green | 201 | 129±148 | 80 | [1, 281, 313, 323, 349, 350, 352, 424, 428, 438, 459, 530, 798, 799, 815, 823, 853-862] [225, 813] |
| | Sintered | 101 | 44±60 | 21 | [189, 274, 276, 299, 339, 351, 428-430, 432-435, 569, 593, 594, 647, 650, 663, 678, 690, 716, 749, 800] [504, 847] |
| Cellular | Green | 189 | 130±120 | 100 | [67, 198, 223, 245, 314, 413, 417, 420, 437, 524, 526, 527, 529, 536, 537, 540, 545, 547-551, 553, 741, 781, 802-804, 813, 836, 853, 858, 863-875] |
| | Sintered | 136 | 42 ±87 | 15 | [114, 199, 220, 226, 274, 284, 285, 499, 577, 593, 597, 612, 619, 621, 648, 661, 674, 677, 684, 706, 708, 710, 719, 720, 730, 745, 753, 795, 806, 876-884] |
| Equiaxed | Green | 6 | 43±39 | 29 | [197, 344, 583, 609, 643, 679, 698, 699, 755] |
| | Sintered | 28 | 348±533 | 220 | [316, 885] |
| **WALL WIDTH** (μm) | | | | | |
| Lamellar | Solidification | 6 | 3±2 | 2 | [398] |
| | Frozen | 2 | 23±11 | 23 | [765] |
| | Green | 69 | 24±47 | 8 | [223, 307, 314, 439, 445, 447, 459, 539, 553, 562, 741, 758, 759, 770, 771, 783, 787, 821, 886] |
| | Sintered | 404 | 17±24 | 10 | [49, 109, 117, 120, 147, 149, 156, 228, 229, 239, 284, 290-293, 295, 296, 299, 339, 379, 396, 397, 400, 403, 405, 408, 411, 418, 448, 460, 462, 493, 494, 496, 497, 505, 514, 516, 567, 569, 626, 639, 642, 659, 689, 696, 713, 724, 732, 743, 747, 756, 766-769, 795, 826] |
| Dendritic | Solidification | 39 | 7±3 | 7 | [2] |
| | Sintered | 30 | 31±67 | 8 | [296, 357, 396, 418, 568, 653, 691, 692] |
| Honeycomb | Green | 13 | 3±3 | 3 | [350, 352, 424, 459, 887, 888] |
| | Sintered | 39 | 18±24 | 10 | [351, 429, 430, 433, 569, 593, 647] |
| Cellular | Green | 15 | 16±29 | 8 | [314, 553, 802, 804, 863] |
| | Sintered | 19 | 50±63 | 25 | [226, 492, 696, 710, 720, 795] |
| Equiaxed | Green | - | - | - | |
| | Sintered | 8 | 95±11 | 98 | [698, 699] |

Table 7. Micromechanical model and regression parameters describing relationships between relative density and compressive strength of freeze-cast metals. $C_2$ and $\beta_1$ are from the log transformation of Eq. 6 ($\ln(\sigma_c^*/\sigma_c^0) = C_2 + \beta_1 \cdot \ln(\rho^*/\rho^0)$); confidence intervals are shown in brackets and *p* values are indicted by one, two, three and four stars for $p<0.05$, 0.01, 0.001, and 0.0001; *ns* indicates the value was not significant. *N* is the number of samples, and $R^2$ is the coefficient of determination.

| Model | Coefficients | | N | $R^2$ | References |
|---|---|---|---|---|---|
| | $C_2$ | $\boldsymbol{\beta_1}$ | | | |
| ALL | | | | | |
| Regression | -0.34[-0.62,-0.05]* | 1.6[1.3,1.9]**** | 60 | 0.67**** | |
| Open-cell | -0.45[-0.54,-0.36] | 1.5 | 60 | 0.67**** | |
| Honeycomb | -0.90[-1.00,-0.79] | 1.0 | 60 | 0.57**** | |
| Relative density, $\boldsymbol{\rho^*/\rho^0}$ < 0.5 | | | | | |
| Regression | ns | 1.7[1.3,2.2]**** | 46 | 0.60**** | [17, 141, 406, 462, 489, 493, 496-501, 604, 605, 686, 687] |
| Open-cell | -0.46[-0.57,-0.35] | 1.5 | 46 | 0.60**** | |
| Honeycomb | -0.96[-1.08,-0.84] | 1.0 | 46 | 0.49**** | |
| Relative density, $\boldsymbol{\rho^*/\rho^0}$ > 0.5 | | | | | |
| Regression | -0.59[-1.04,-0.15]* | 1.1[0.5,1.8]** | 32 | 0.30** | [141, 157, 406, 489, 493, 494, 497, 499, 501, 604, 605, 686, 687] |
| Open-cell | -0.35[-0.47,-0.23] | 1.5 | 32 | 0.27** | |
| Honeycomb | -0.69[-0.80,-0.57] | 1.0 | 32 | 0.30** | |

**Bulk material compressive strength values** (all units in MPa). CES Edupack [889] was utilized for the following materials: Ti-6Al-4V (950), Fe (210 ), Ti-5wt.% W (910), and W (610). Remaining values were obtained from literature: Ti (900 and 483 for 4.0 [17] and 0.4 [890] wt.% $O_2$) and Cu (176 [833]).

Table 8. Micromechanical model and regression parameters describing relationships between relative density and compressive strength of anisotropic and isotropic freeze-cast ceramics. $C_2$ and $\beta_1$ are from the log transformation of Eq. 6 ($\ln(\sigma_c^*/\sigma_c^0) = C_2 + \beta_1 \cdot \ln(\rho^*/\rho^0)$); confidence intervals are shown in brackets and $p$ values are indicted by one, two, three and four stars for $p$<0.05, 0.01, 0.001, and 0.0001; *ns* indicates the value was not significant. *N* is the number of samples, and $R^2$ is the coefficient of determination.

| Model | Coefficients | | N | $R^2$ | References |
|---|---|---|---|---|---|
| | $C_2$ | $\beta_1$ | | | |
| ALL | | | | | |
| Regression | -0.37[-0.51,-0.23]$^{****}$ | 2.2[2.1,2.3]$^{****}$ | 1032 | 0.59$^{****}$ | |
| Open-cell | -1.14[-1.20,-1.08] | 1.5 | 1032 | 0.53$^{****}$ | |
| Honeycomb | -1.71[-1.78,-1.65] | 1.0 | 1032 | 0.42$^{****}$ | |
| | | | | | |
| ANISOTROPIC | | | | | |
| Regression | -0.24[-0.39,-0.01]$^{**}$ | 2.3[2.1,2.4]$^{****}$ | 844 | 0.61$^{****}$ | |
| Open-cell | -1.12[-1.18,-1.06] | 1.5 | 844 | 0.54$^{****}$ | |
| Honeycomb | -1.70[-1.78,-1.63] | 1.0 | 844 | 0.42$^{****}$ | |
| Lamellar | | | | | [112, 117, 120, 149, 195, 219, 224, 239, 240, 282, 283, 285, 286, 295, 296, 302, 312, 336, 339, 379, 384, 386, 390, 405, 412, 414, 415, 446, 448, 460, 488, 505, 506, 514, 516, 563, 566, 569, 572, 579, 581, 624, 626-628, 633, 635, 637, 651, 655, 662, 667, 670, 682, 688, 689, 705, 711, 712, 721, 742, 747, 792, 795] |
| Regression | ns | 2.5[2.4,2.7]$^{****}$ | 481 | 0.63$^{****}$ | |
| Open-cell | -1.05[-1.13,-0.97] | 1.5 | 481 | 0.52$^{****}$ | |
| Honeycomb | -1.62[-1.71,-1.53] | 1.0 | 481 | 0.40$^{****}$ | |
| Dendritic | | | | | [196, 224, 262, 296, 302, 320, 324, 379-381, 385, 388, 410, 461, 575, 576, 584, 606-608, 617-620, 623, 624, 640, 652, 665, 666, 668, 672, 675, 681, 683, 691, 692, 697, 723, 726] |
| Regression | -1.17[-1.52,-0.83]$^{****}$ | 1.6[1.3,1.9]$^{****}$ | 150 | 0.44$^{****}$ | |
| Open-cell | -1.27[-1.39,-1.14] | 1.5 | 150 | 0.44$^{****}$ | |
| Honeycomb | -1.82[-1.95,-1.69] | 1.0 | 150 | 0.39$^{****}$ | |
| Honeycomb | | | | | |
| Regression | ns | 2.3[2.0,2.5]$^{****}$ | 207 | 0.67$^{****}$ | [110, 224, 339, 341, 342, 351, 429-431, 492, 504, 569, 595, 647, 678, 690, 700, 703, 704, 737, 738, 750] [504, 750] |
| Open-cell | -1.19[-1.33,-1.06] | 1.5 | 207 | 0.59$^{****}$ | |
| Honeycomb | -1.82[-1.98,-1.66] | 1.0 | 207 | 0.46$^{****}$ | |
| | | | | | |
| ISOTROPIC | | | | | |
| Regression | -0.74[-1.06,-0.42]$^{****}$ | 2.0[1.7,2.2]$^{****}$ | 187 | 0.55$^{****}$ | |
| Open-cell | -1.25[-1.41,-1.08] | 1.5 | 187 | 0.52$^{****}$ | |
| Honeycomb | -1.78[-1.96,-1.60] | 1.5 | 187 | 0.41$^{****}$ | |
| Cellular | | | | | |
| Regression | -0.67[-1.00,-0.30]$^{****}$ | 2.0[1.7,2.4]$^{****}$ | 147 | 0.55$^{****}$ | |
| Open-cell | -1.23[-1.43,-1.03] | 1.5 | 147 | 0.51$^{****}$ | [38, 39, 226, 285, 379, 391, 590, 615, 617, 619, 629, 634, 668, 671, 674, 677, 708, 712, 720, 721, 745, 754, 795, 809] |
| Honeycomb | -1.72[-1.94,-1.50] | 1.5 | 147 | 0.40$^{****}$ | |
| Equiaxed | | | | | |
| Regression | -1.40[-2.16,-0.65]$^{****}$ | 1.4[0.9,1.9]$^{****}$ | 40 | 0.45$^{****}$ | [344, 583, 588, 609, 610, 658, 679, 698, 699] |
| Open-cell | -1.30[-1.52,-1.07] | 1.5 | 40 | 0.45$^{****}$ | |
| Honeycomb | -1.99[-2.23,-1.76] | 1.5 | 40 | 0.41$^{****}$ | |

**Bulk material flexural strength values** (all units in MPa). CES Edupack [889] was utilized for the following materials: $Al_2O_3$ (400), biphasic calcium phosphate (BCP; 45), bioglass (110), calcium phosphate (55), hydroxyapatite (HAP; 120), yttria-stabilized zirconia (YSZ, 500), $ZrO_2$ (460), $Si_3N_4$ (758), SiC (250), tricalcium phosphate (TCP; 45), $TiO_2$ (180), SiAlON (800), $CaSiO_3$ (100), whereas literature values were taken for: $Y_2SiO_5$ (116 [891]), $ZrB_2$ (300 [892]), lead zirconate titanate (PZT; 90 [893]), $Yb_2SiO_5$ (150 [894]), $Ca_3ZrSiO_9$ (98 [895]), and mullite (240 [507]). Rule of mixtures, using the aforementioned values, was used to approximate values for mixed particle systems.

Table 9. Micromechanical model and regression parameters describing relationships between relative density and compressive strength of anisotropic freeze-cast ceramics based on fluid type. $C_2$ and $\beta_1$ are from the log transformation of Eq. 6 ($\ln(\sigma_c^*/\sigma_c^0) = C_2 + \beta_1 \cdot \ln(\rho^*/\rho^0)$); confidence intervals are shown in brackets and $p$ values are indicted by one, two, three and four stars for $p$<0.05, 0.01, 0.001, and 0.0001; *ns* indicates the value was not significant. *N* is the number of samples, and $R^2$ is the coefficient of determination.

| Model | Coefficients | | N | $R^2$ | References |
|---|---|---|---|---|---|
| | $C_2$ | β1 | | | |
| **Water** | | | | | |
| Regression | -0.22[-0.42,-0.02]** | 2.2[2.1,2.4]**** | 542 | 0.57**** | [112, 117, 120, 149, 196, 219, 224, 239, 240, 282, 283, 285, 286, 295, 296, 302, 312, 320, 324, 336, 339, 341, 342, 379, 384, 386, 390, 405, 414, 415, 429-431, 448, 460, 488, 505, 506, 514, 516, 563, 566, 569, 572, 579, 581, 584, 595, 617, 619, 620, 623, 624, 626-628, 633, 635, 637, 651, 652, 655, 662, 665, 670, 681-683, 688, 689, 705, 711, 712, 721, 742, 747, 792, 795] |
| Open-cell | -1.05[-1.12,-0.98] | 1.5 | 542 | 0.51**** | |
| Honeycomb | -1.63[-1.71,-1.55] | 1.0 | 542 | 0.40**** | |
| **Camphene** | | | | | |
| Regression | -0.85[-1.26,-0.43]** | 1.8[1.4,2.1]**** | 89 | 0.54**** | [262, 380, 381, 388, 410, 461, 575, 576, 606-608, 618, 640, 668, 672, 675, 691, 692, 697, 723, 726] |
| Open-cell | -1.15[-1.31,-1.00] | 1.5 | 89 | 0.53**** | |
| Honeycomb | -1.70[-1.87,-1.54] | 1.0 | 89 | 0.44**** | |
| **TBA** | | | | | |
| Regression | 0.52[0.25, 0.80]**** | 3.1[2.9,3.3]**** | 168 | 0.81**** | [110, 224, 351, 492, 504, 647, 666, 678, 690, 700, 703, 704, 737, 738, 750] |
| Open-cell | -1.23[-1.38,-1.09] | 1.5 | 168 | 0.59**** | |
| Honeycomb | -1.78[-1.95,-1.62] | 1.0 | 168 | 0.44**** | |

**Bulk material flexural strength values** (all units in MPa). CES Edupack [889] was utilized for the following materials: $Al_2O_3$ (400), biphasic calcium phosphate (BCP; 45), bioglass (110), calcium phosphate (55), hydroxyapatite (HAP; 120), yttria-stabilized zirconia (YSZ, 500), $ZrO_2$ (460), $Si_3N_4$ (758), SiC (250), tricalcium phosphate (TCP; 45), $TiO_2$ (180), SiAlON (800), $CaSiO_3$ (100), whereas literature values were taken for: $Y_2SiO_5$ (116 [891]), $ZrB_2$ (300 [892]), lead zirconate titanate (PZT; 90 [893]), $Yb_2SiO_5$ (150 [894]), $Ca_3ZrSiO_9$ (98 [895]), and mullite (240 [507]). Rule of mixtures, using the aforementioned values, was used to approximate values for mixed particle systems.

Table 10. Regression models comparing model fits between Eq. 6 ($\sigma_c^*/\sigma_f^0 = e^{C_3} \cdot (\rho^*/\rho^0)^{\beta_1}$ ) and Eq. 7 ($\sigma_c^*/\sigma_f^0 = e^{C_3} \cdot (\rho^*/\rho^0)^{\beta_1} \cdot (A)^{\beta_2}$) for anisotropic ceramic materials derived from particle suspensions using tert-butyl alcohol (TBA), water, and camphene as the fluid. Eq. 6 describes normalized compressive strength ($\sigma_c^*/\sigma_f^0$) of freeze-cast ceramics in terms of relative density ($\rho^*/\rho^0$) and Eq. 7 also considers the effect of test specimen aspect ratio (A). Confidence intervals for coefficients are shown in brackets and p values are indicted by one, two, three and four stars for p<0.05, 0.01, 0.001, and 0.0001; ns indicates the value was not significant; N is the number of samples, BIC is the Bayesian Information Criterion, and *adj.* $R^2$ is the adjusted coefficient of determination.

| | Coefficients | | | | | | |
|---|---|---|---|---|---|---|---|
| Model | $C_3$ | $\beta_1$ | $\beta_2$ | N | BIC | adj.$R^2$ | Refs. |
| TBA | | | | | | | |
| Eq. 8 | 0.7[0.4,0.9]**** | 3.1[2.9,3.4]**** | - | 116 | 187 | 0.88**** | [110, 224, 351, 492, 504, 647, 678, 690, 700, 703, 704] |
| Eq. 10 | ns | 2.5[2.4,2.7]**** | -1.0[-1.1,-0.9]**** | | 64 | 0.96**** | |
| Water | | | | | | | |
| Eq. 8 | ns | 2.4[2.2,2.6] **** | - | 403 | 956 | 0.61**** | [112, 120, 149, 196, 219, 224, 239, 240, 282, 283, 285, 286, 295, 296, 302, 312, 320, 336, 339, 341, 390, 405, 414, 415, 429, 430, 448, 460, 514, 516, 563, 569, 579, 581, 595, 617, 619, 620, 623, 626-628, 633, 635, 651, 652, 670, 689, 705, 711, 747] |
| Eq. 10 | *ns* | 2.4[2.2,2.6] **** | -0.4[-0.6,-0.1]** | | 952 | 0.62**** | |
| Camphene | | | | | | | |
| Eq. 8 | -0.8[-1.3,-0.4]**** | 1.8[1.4,2.1] **** | - | 89 | 200 | 0.54**** | [262, 380, 381, 388, 410, 461, 575, 576, 606-608, 618, 640, 668, 672, 675, 691, 692, 697, 723, 726] |
| Eq. 10 | -0.5[-0.9,-0.2]** | 2.1[1.8,2.5]**** | 0.7[0.5, 1.0]*** | | 176 | 0.66**** | |

**Bulk material flexural strength values** (all units in MPa). CES Edupack [889] was utilized for the following materials: $Al_2O_3$ (400), biphasic calcium phosphate (BCP; 45), bioglass (110), calcium phosphate (55), hydroxyapatite (HAP; 120), yttria-stabilized zirconia (YSZ, 500), $ZrO_2$ (460), $Si_3N_4$ (758), SiC (250), tricalcium phosphate (TCP; 45), $TiO_2$ (180), SiAlON (800), $CaSiO_3$ (100), whereas literature values were taken for: $Y_2SiO_5$ (116 [891]), $ZrB_2$ (300 [892]), lead zirconate titanate (PZT; 90 [893]), $Yb_2SiO_5$ (150 [894]), $Ca_3ZrSiO_9$ (98 [895]), and mullite (240 [507]). Rule of mixtures, using the aforementioned values, was used to approximate values for mixed particle systems.

**Figures**

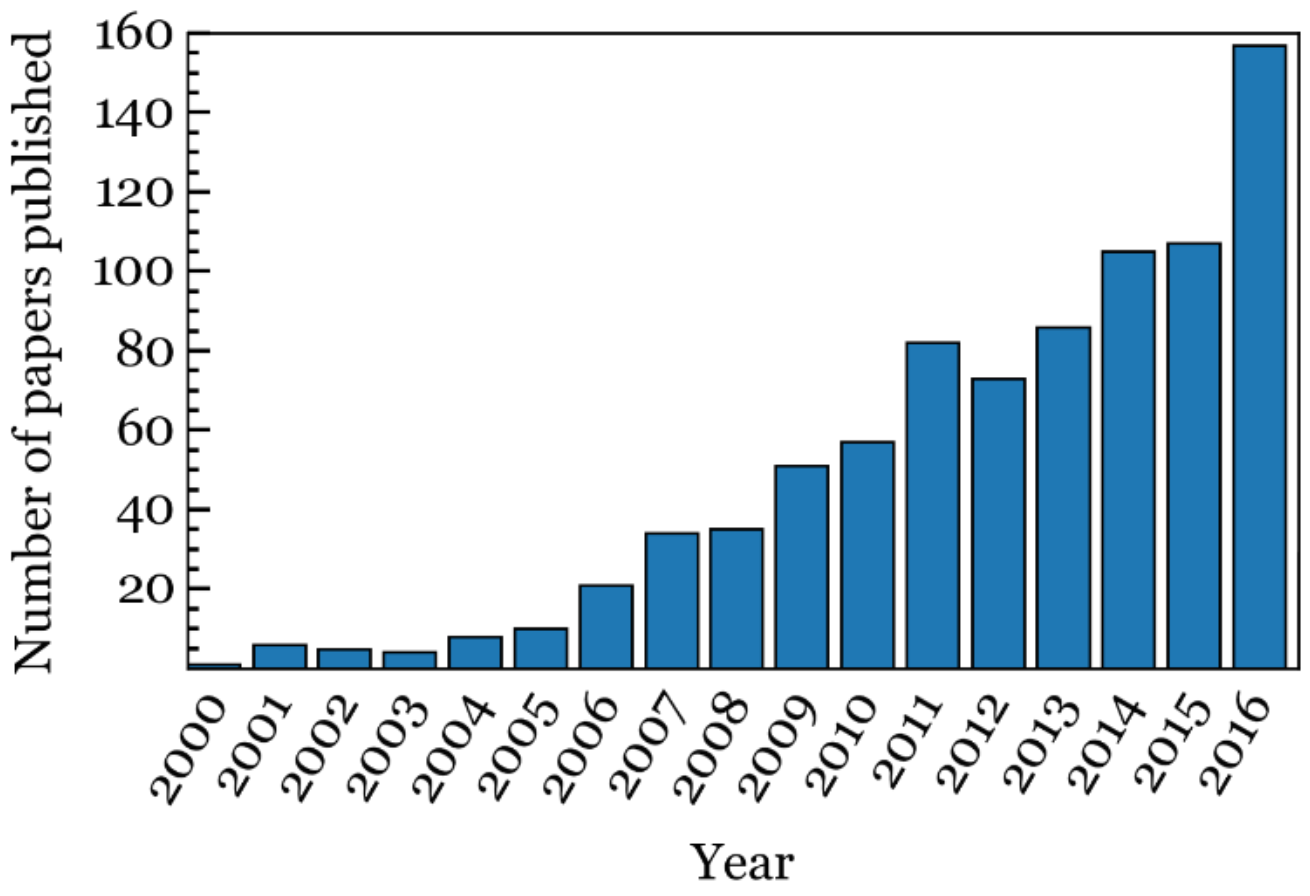

**Fig. 1.** Number of freeze-casting papers published in peer-reviewed journals per year, since 2000.

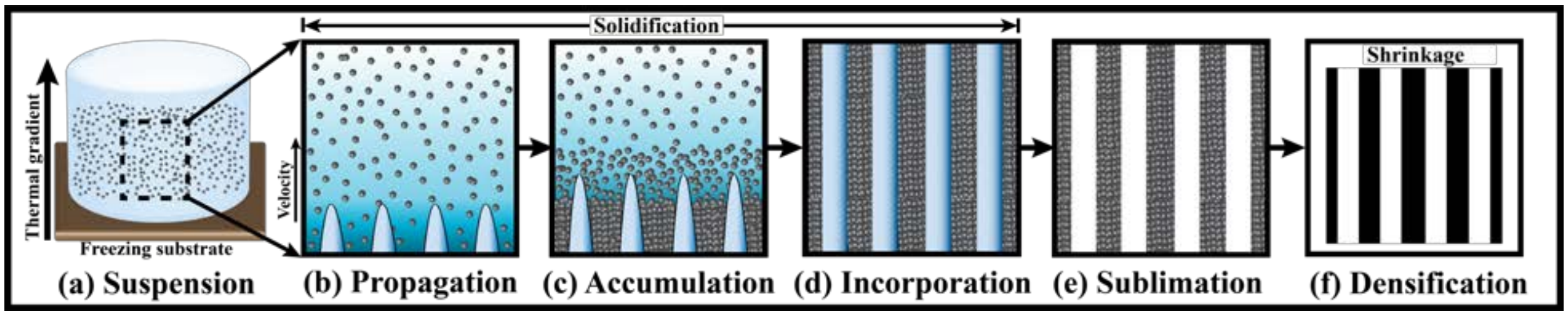

Fig. 2. Unidirectional freeze-casting process. A suspension consisting of particles (grey) dispersed in a fluid (dark blue) is (a) placed onto a freezing substrate; after nucleation, (b) dendrites (light blue) propagate in the direction of the thermal gradient while pushing particles away from the moving front. An (c) accumulation region of rejected particles develops ahead of the solid/liquid interface, inducing particle packing within interdendritic space. After (d) complete solidification, the solidified fluid is (e) removed via sublimation leaving elongated pores (white) which template the dendrites. In the case of ceramic and metallic materials, the remaining particle scaffold is (f) sintered to densify particle-packed walls (black).

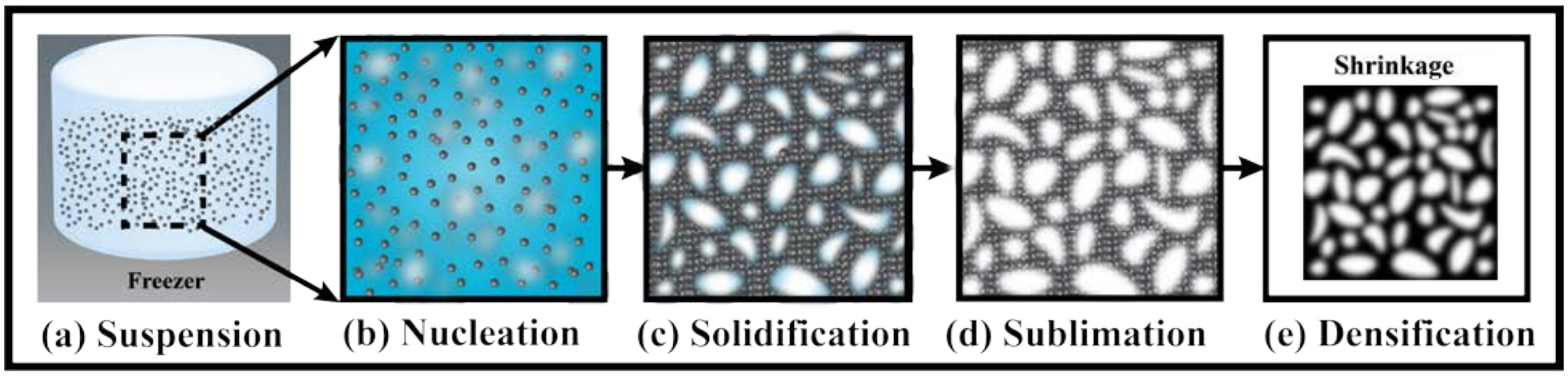

Fig. 3. Isotropic freeze-casting process. A suspension contained within a thermally insulating mold is (a) placed in a cooled environment and (b) solid nucleation occurs randomly throughout the suspension. After nucleation, (c) solid crystal growth occurs at random orientations throughout the volume and no preferential growth direction is observed. The resulting pore structures are nearly isotropic following (d) sublimation and (e) sintering.

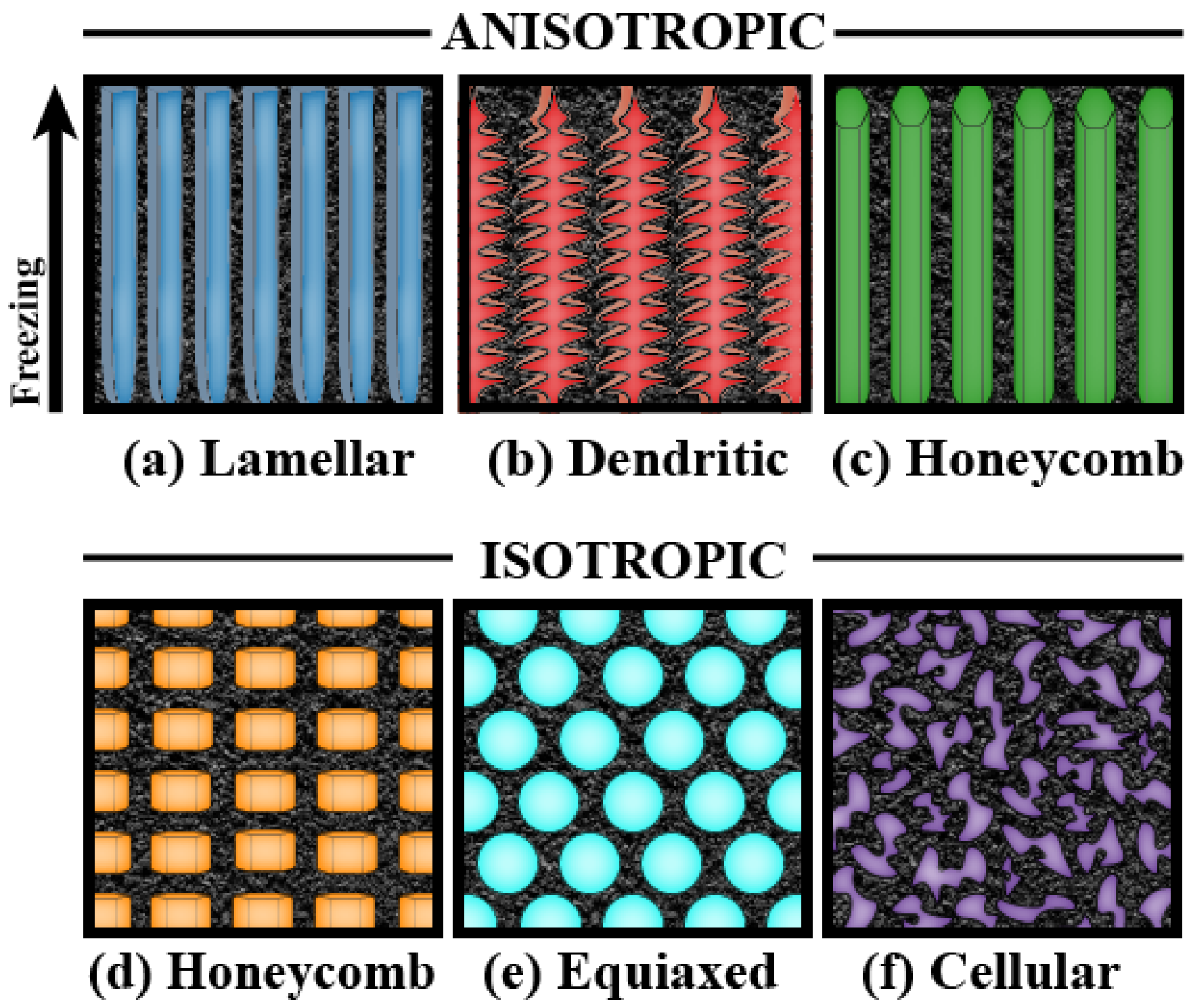

Fig. 4. Schematic representation of the most commonly observed pore structures in freeze-cast materials, including anisotropic (a) lamellar, (b) dendritic, and (c) elongated honeycomb, and (d) cellular-honeycomb, (e) equiaxed, and (f) cellular pore structures. Here, the solid, colored regions represent the morphology of the solidified fluid which, after removal, largely defines the pore structure in the final material.  The dark regions represent the particle-rich regions, which form are templated during the solidification process. These particle-rich regions can be further sintered into dense walls.

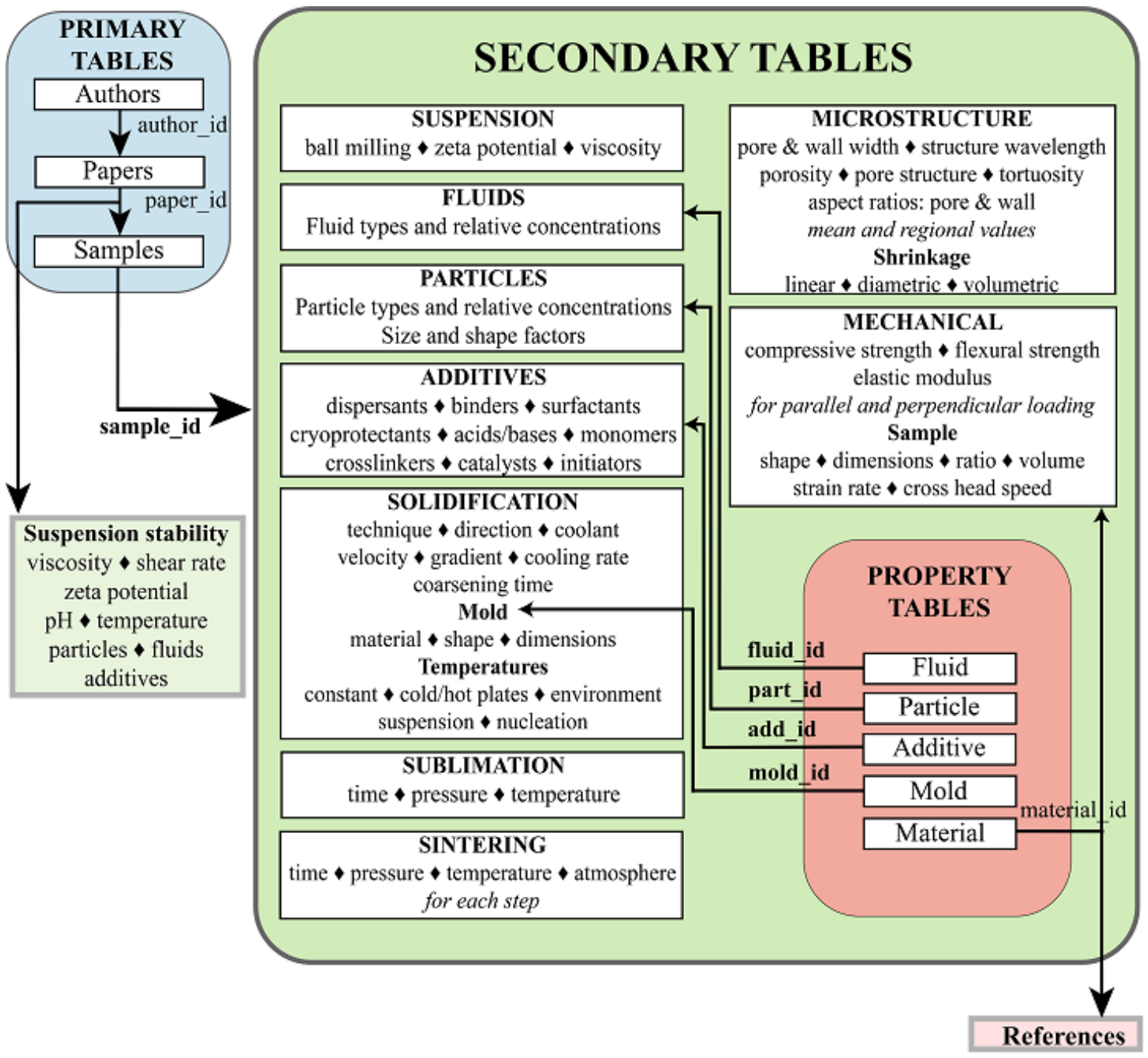

Fig. 5. FreezeCasting.net database schema. Primary tables (blue) contain citation data and assign sample identification numbers for each processing condition explored by each individual paper. Secondary tables (green) contain experimental data and the property tables (red) contain property values for fluids, particles, additives, molds, and bulk materials. Lastly, the suspension stability table (green) is a supplementary table that contains experimental data extracted from suspension stabilization studies.

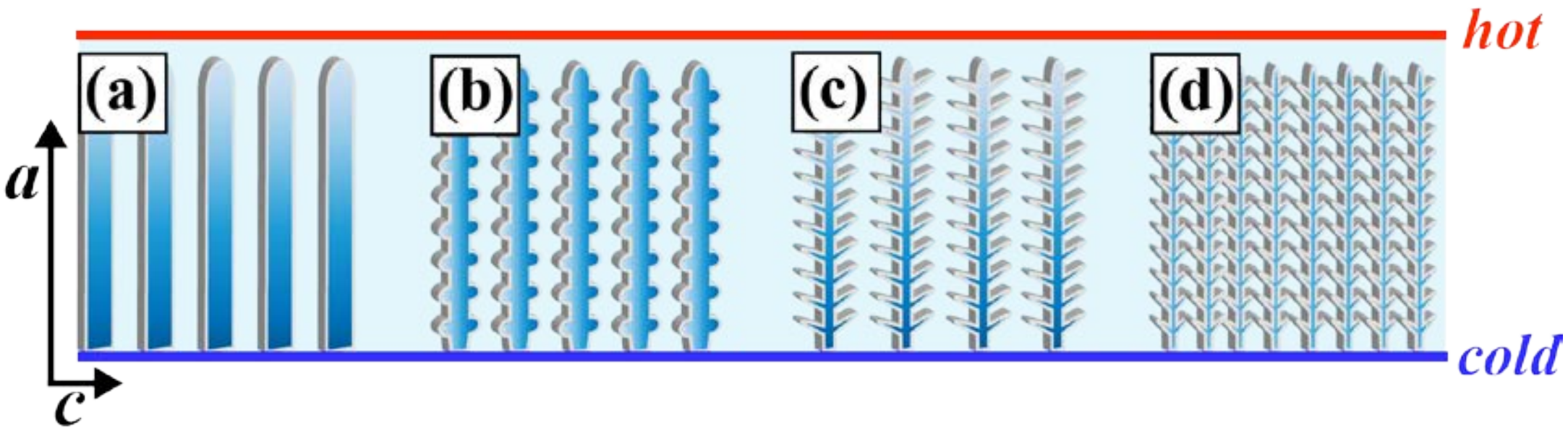

Fig. 6. Morphological transition of ice from lamellar to dendritic structures, where (a) a lamellar morphology is depicted. The onset of dendritic side branching is shown in (b); these “secondary arms” increase in both length and density (number of arms) (c), where the ice morphology has completely transitioned from lamellar to dendritic. As side branching increases, the diameter of the primary branch decreases. Although the morphological structure of ice is still dendritic in (d), the resulting porous structure is likely to be cellular as a result of the highly interconnected ice network that has formed.

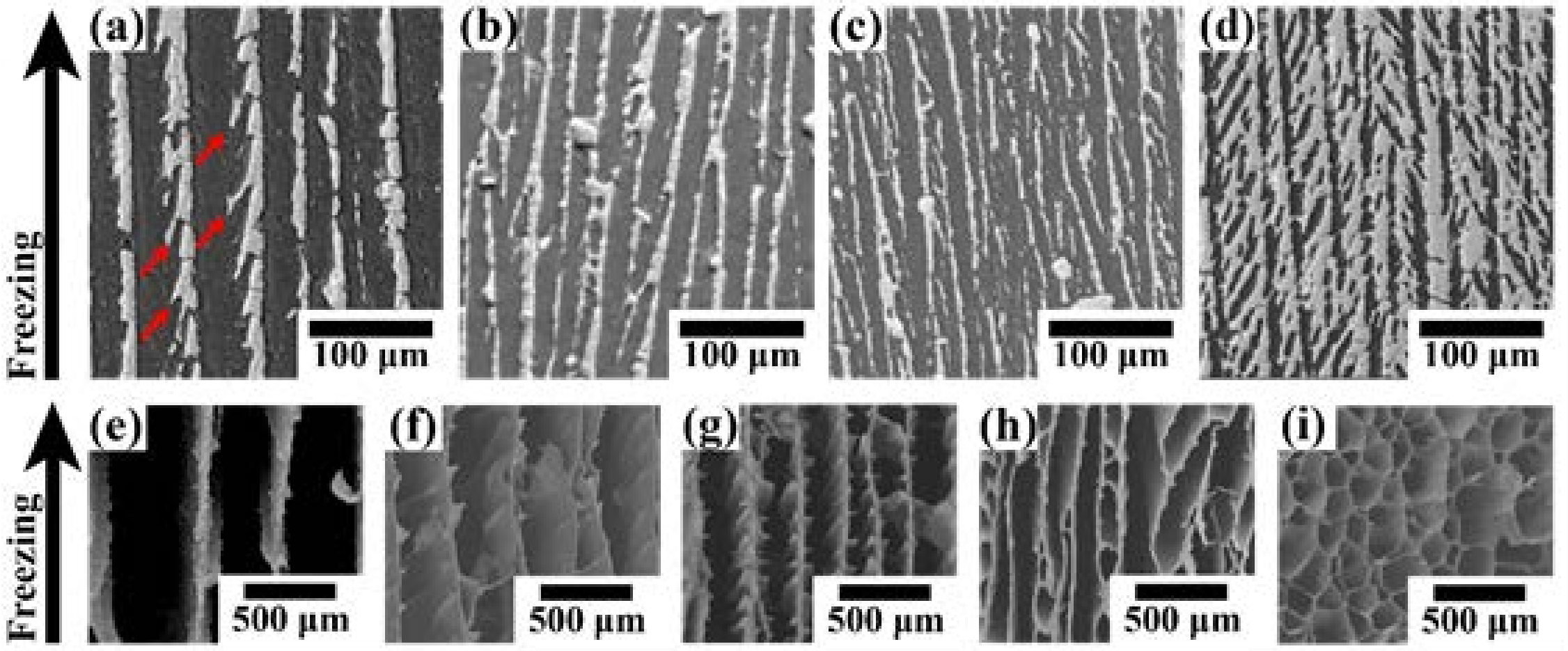

Fig. 7. Effect of binder on anisotropic freeze-cast structures derived from aqueous suspensions. Sintered alumina ($Al_2O_3$) with increasing amounts of binder, polyvinyl alcohol (PVA) [309] is shown in (a-d). Cross-sections are taken parallel to the solidification direction; dark regions represent pores and the light regions are sintered, $Al_2O_3$ walls; (a) shows the microstructure obtained when no binder is added; the pore morphology is lamellar and one-sided secondary dendritic features (red arrows) are present. Microstructures obtained with the inclusion of 5, 15, and 20 wt.% PVA (with respect to $Al_2O_3$) are shown in (b) through (d), respectively. Non-sintered tungsten disulfide ($WS_2$) is shown in (e-i) [314]. The lamellar microstructure shown in (e) is obtained with the inclusion of 1 wt.% gelatin (with respect to $WS_2$); the binder content increases to 2, 3, 4, and 5 wt.% in (f) through (h).

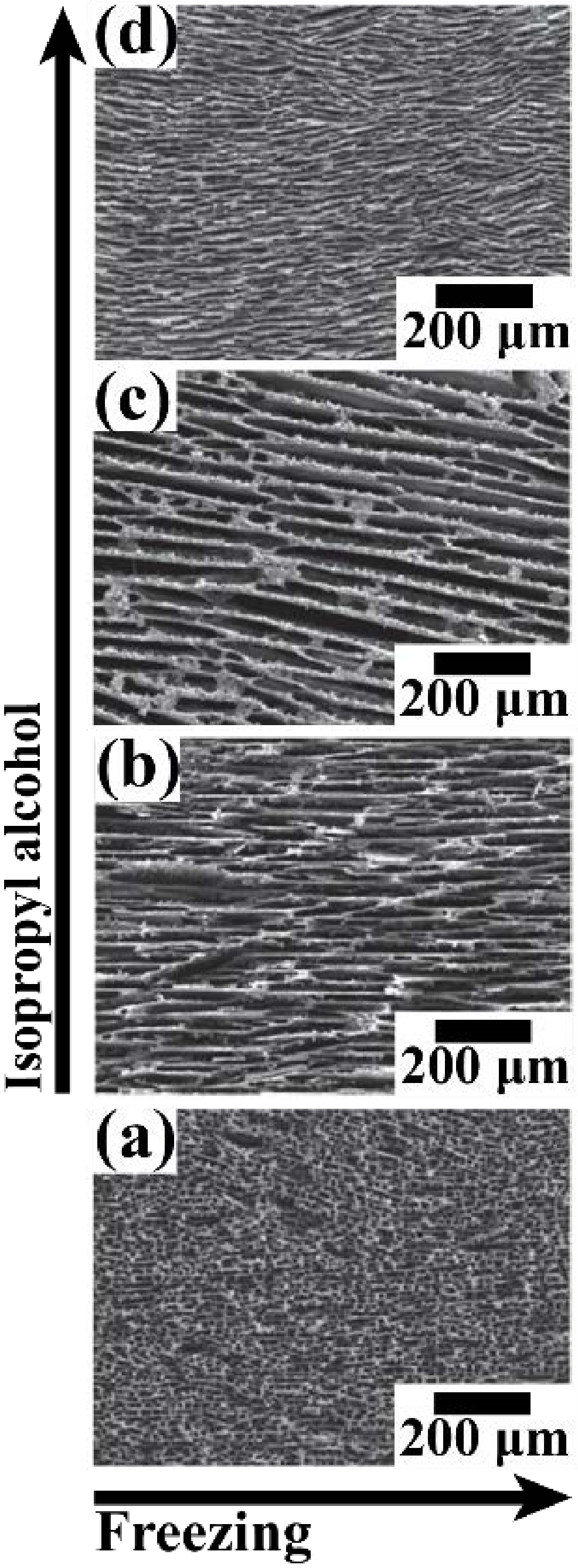

Fig. 8. Effect of isopropyl alcohol (IPA) additives on anisotropic freeze-cast microstructures derived from aqueous suspensions. These microstructural cross-sections are taken parallel to the solidification direction; (a) shows sintered 10 vol.% $TiO_2$ [295] obtained without IPA, and (b) - (d) with the addition of 1, 5, and 30 vol.% IPA (with respect to fluid volume).

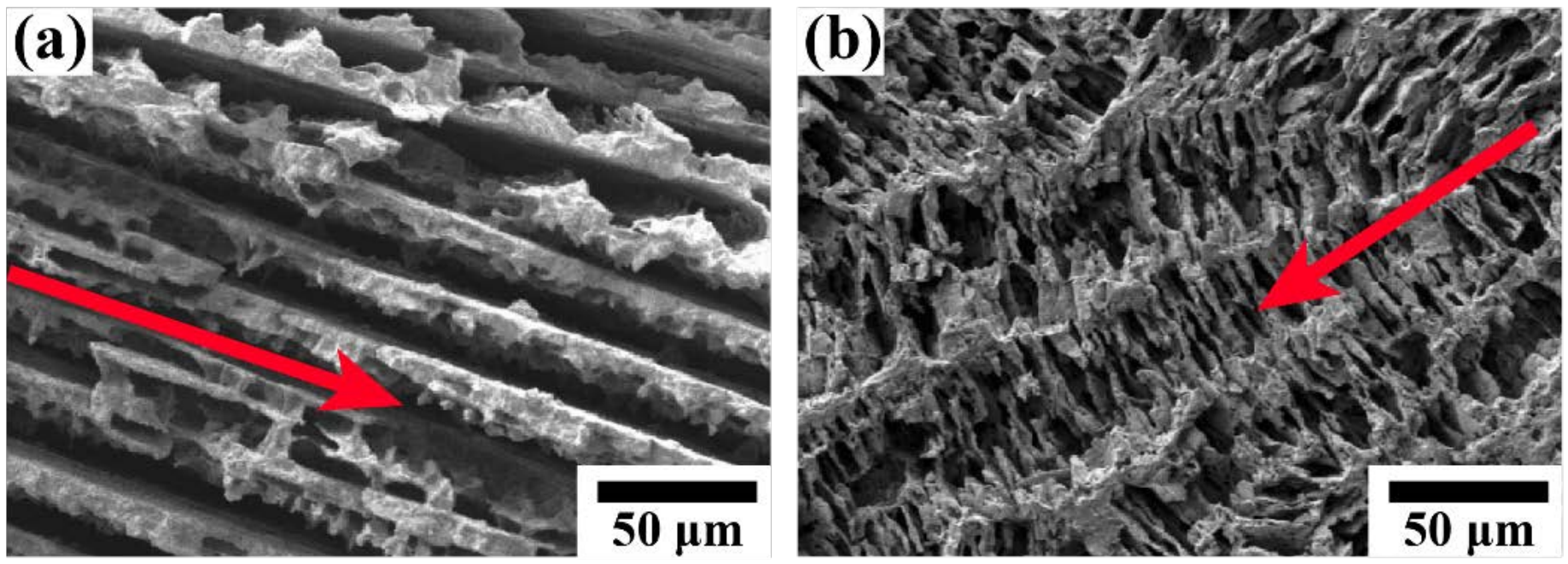

Fig. 9. Effect of glycerol on anisotropic freeze-cast microstructures derived from aqueous suspensions. The lamellar structure in (a) was obtained after sintering microstructures templated using aqueous suspensions of 10 vol.% HAP [285]. In (b), 20 wt.% glycerol (with respect to water) was added to the initial suspension; the resulting microstructure consists of lamellar walls joined by bridges that form nearly perpendicular to, and join, the walls. Red arrows indicate the direction of freezing.

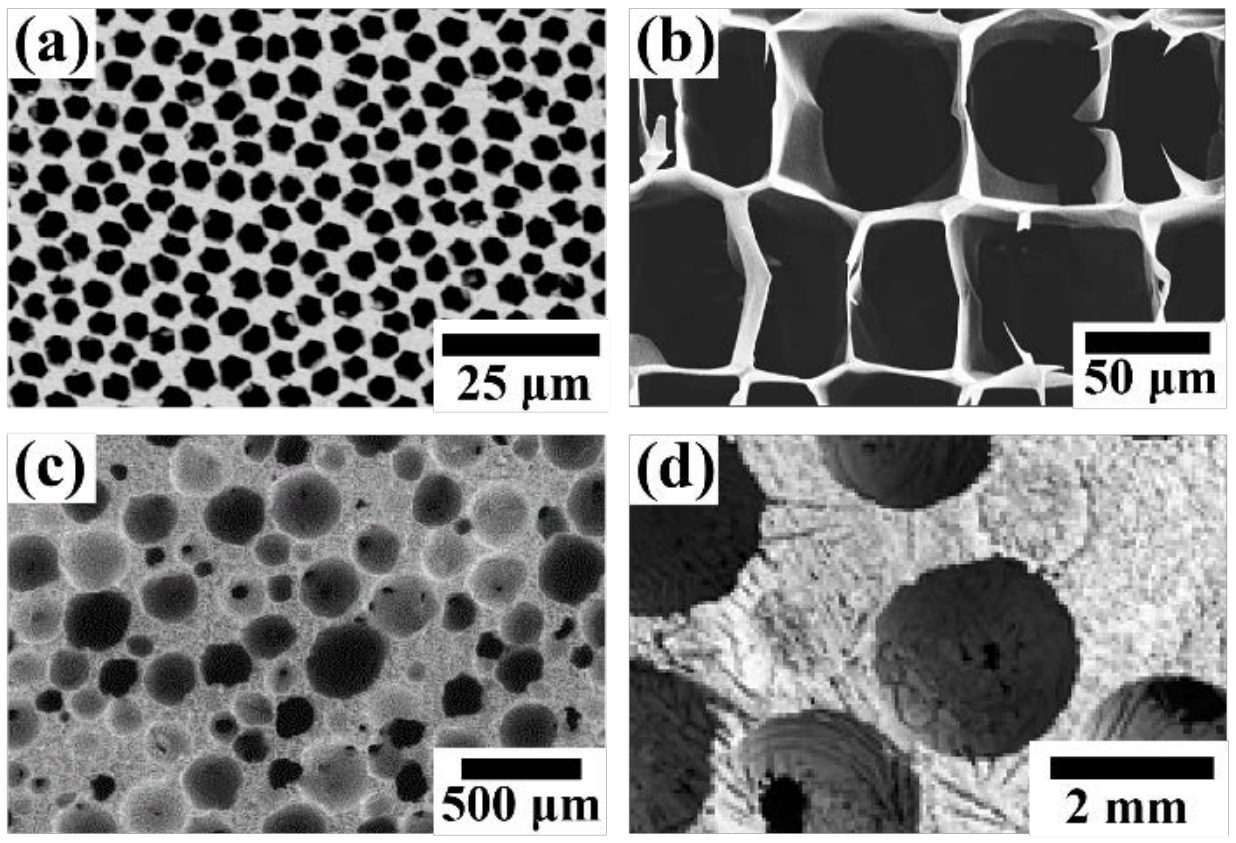

Fig. 10. Non-standard pore structures obtained using aqueous suspensions. The sintered honeycomb structures shown in (a) are obtained with the addition of a zirconium acetate complex (ZRA) for zirconia [276], whereas (b) was obtained for 5 wt.% gelatin [281]; anisotropic freezing was used for both radial cross-sections in (a) and (b). The equiaxed structures shown in (c) and (d) were both obtained by employing isotropic freezing techniques. In (c), air entrapped during the tumbling step was responsible for templating the equiaxed macropores in the sintered silica [197] material, whereas polystyrene space-holders were used for the sintered mullite foam [343] shown in (d).

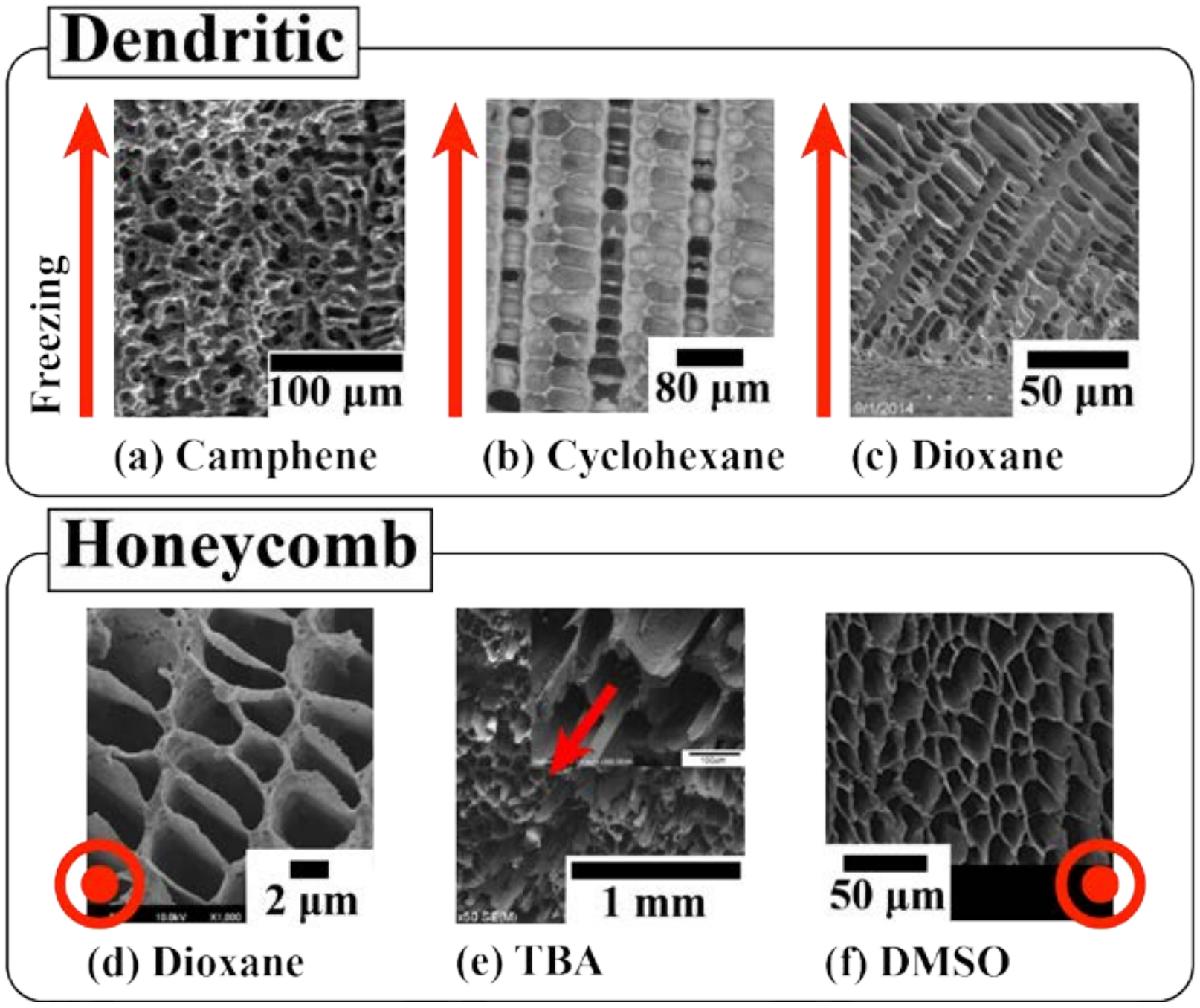

Fig. 11. Overview of microstructures obtained using non-aqueous fluids under directional freezing conditions. Red arrows indicate the direction of freezing for cross-sections taken parallel to the solidification direction; cross-sections taken perpendicular to the solidification direction are marked with a circle. Dendritic structures are shown in (a-c), where (a) is sintered, 50 vol.% $Al_2O_3$ templated with camphene [347], (b) is sintered, 20 wt.% silicon oxycarbide (SiOC) obtained using cyclohexane, and (c) is polyurethane templated with dioxane [180]. Honeycomb structures are shown in (d-f), where (d) is 30 wt.% polystyrene obtained with a fluid mixture of polyethylene glycol and dioxane [349], (e) is sintered 10 vol.% $Al_2O_3$-$ZrO_2$ obtained using tert-butyl alcohol [351], and (f) is poly (L-lactic acid) obtained using dimethyl sulfoxide (DMSO) [352].

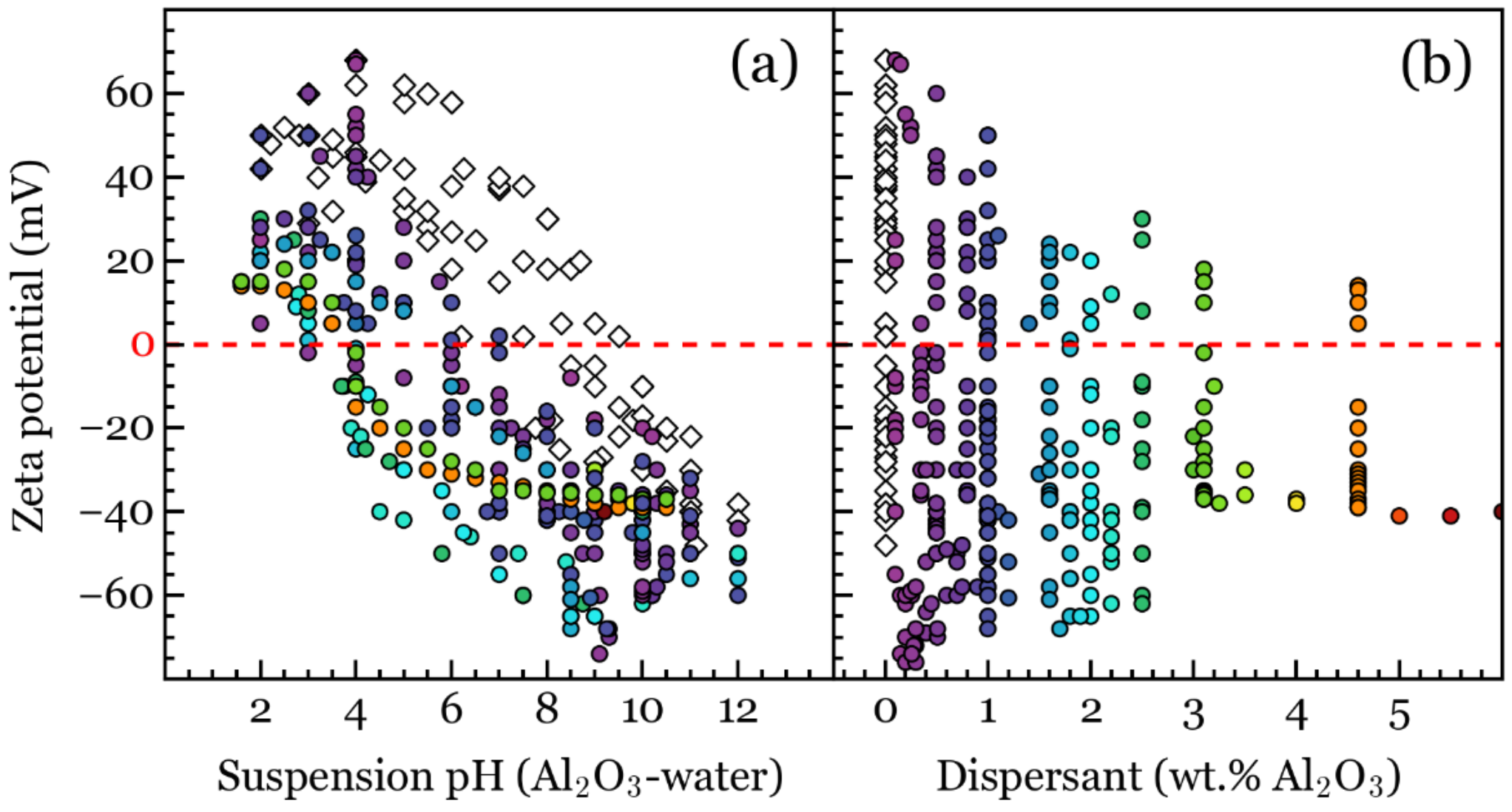

Fig. 12. Dependency of zeta potential on (a) suspension pH and (b) amount of dispersant (wt.%, relative to particle dry mass) for aqueous suspensions of alumina [304, 353-367, 896]. Data point colors are mapped by dispersant weight percent, ranging from < 1 wt.% (violet) to 6 wt.% (red); white data points indicate no dispersant was used. Plot created via the FreezeCasting.net open-data repository.

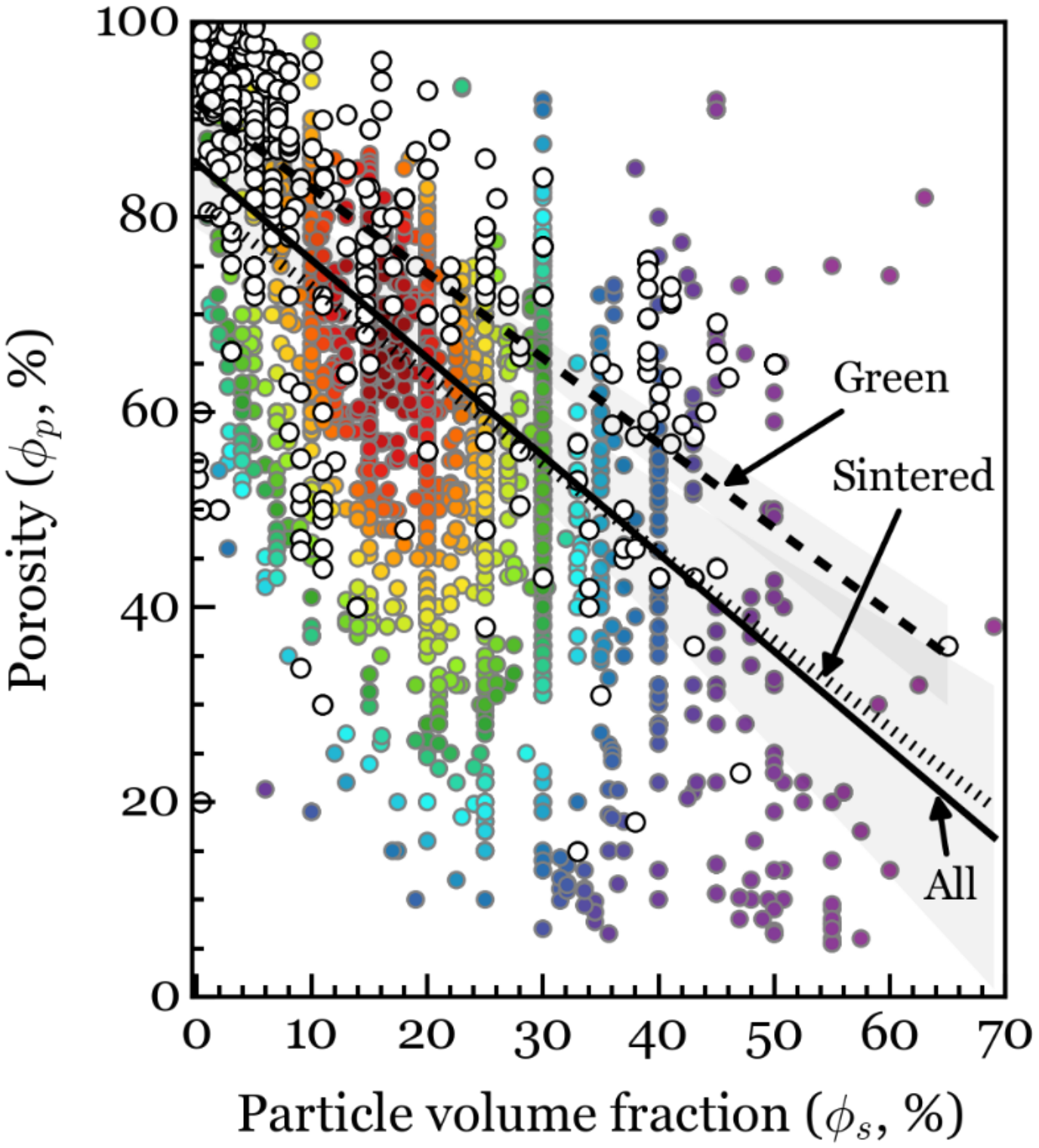

Fig. 13. Plot showing dependence of total porosity on particle volume fraction within the suspension (all volume fraction~porosity data from the FreezeCasting.net database included). Data points that describe sintered samples are colored based on relative point-density (red circles indicate high density regions and purple circles represent low density regions of the plot, whereas data points that describe green samples (sublimated, not sintered) are shown with white circles [6, 17, 18, 24, 25, 38, 49, 54, 107, 109-112, 114, 117, 120, 121, 124, 141, 148, 149, 157, 180, 181, 184, 185, 187, 189, 195-199, 217-220, 223, 224, 226, 228, 230, 232, 239, 240, 245, 261, 262, 271, 274, 281-286, 295, 296, 298, 302, 304, 307, 309-312, 314, 316, 320, 321, 323-325, 328, 329, 336, 339, 341, 343-345, 347, 350-353, 379-388, 390-394, 396, 403, 405, 406, 408, 410, 412-416, 425-427, 429, 432-436, 440, 446, 448, 449, 453, 455, 458, 460, 462, 489, 492, 493, 496-501, 504-506, 514, 516, 519-555, 557-570, 572-579, 581-736, 738, 739, 741, 742, 825]. The solid line is obtained by fitting all data to Eq. 1 ($\%\phi_p$= -1.0±0.3 · %; $\phi_s$ + 86%±6%; $N$ = 2,855, $r$ = -0.12, $R^2$ = 0.01, $p$ < 0.0001), whereas the long- and short-dashed lines are obtained by fitting green ($N$=551) and sintered ($N$=2,281) data, respectively. The grey, shaded regions represent the 95% confidence interval about the regression lines. When only data obtained from green (sublimated, not sintered) samples is considered, the fit to Eq. 1 improves and ~40% variance in porosity is predicted by Eq. 1 (Table 1). Plot created via the FreezeCasting.net open-data repository.

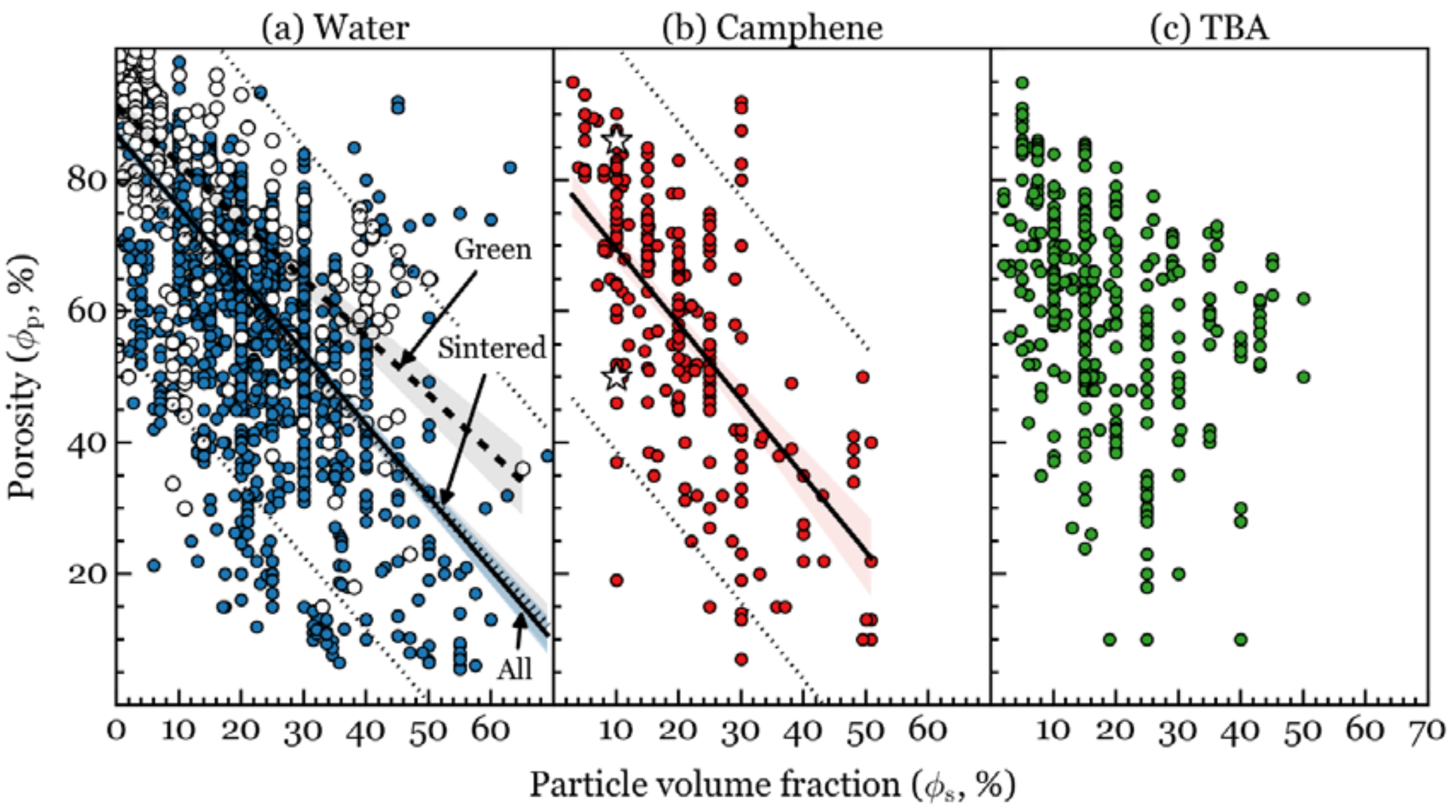

Fig. 14. Plots of total porosity vs. particle volume fraction categorized by fluid type (irrespective of material type), where (a) is water (*N*=2,040), (b) is camphene (*N*=290), and (c) is tert-butyl alcohol (*N*=411). In (a), data obtained from non-sintered samples are shown with white circles [54, 198, 223, 245, 281, 323, 341, 390, 413, 426, 427, 436, 519-524, 528-531, 533, 535-537, 539-544, 547-549, 551, 552, 555, 557-562, 710, 741], and blue circles represent sintered samples [6, 17, 24, 49, 107, 109, 111, 112, 114, 117, 120, 124, 141, 148, 149, 157, 181, 184, 185, 187, 195-197, 199, 217-220, 224, 228, 232, 239, 240, 261, 282-286, 295, 296, 298, 299, 302, 304, 309-312, 316, 320, 324, 325, 328, 329, 336, 339, 341, 343-345, 353, 379, 384, 386, 387, 392, 393, 396, 403, 405, 406, 408, 414, 415, 425, 429, 432-434, 446, 448, 449, 458, 460, 462, 489, 493, 496, 497, 505, 506, 514, 516, 531, 532, 543, 563-567, 569-572, 577-579, 581-590, 594, 595, 598-603, 609-611, 613-617, 619-639, 642-646, 648-652, 655-662, 665, 667, 669, 670, 679, 681-684, 688, 689, 696, 701, 705, 707, 709-714, 717-719, 721, 722, 727-732, 735, 742, 743]. Porosity data for camphene suspensions shown in (b) as white stars [262, 410] represent data obtained from frozen samples; sintered data are red circles [24, 38, 226, 228, 262, 271, 347, 380-383, 385, 388, 391, 410, 453, 498-501, 568, 573-576, 604-608, 618, 640, 653, 654, 668, 671-673, 675, 676, 685-687, 691, 692, 694, 695, 697, 702, 719, 720, 723-726, 740]. All data in (c) were obtained from sintered samples [110, 121, 189, 224, 274, 325, 351, 394, 435, 492, 504, 591-593, 596, 597, 647, 663, 664, 666, 674, 677, 678, 680, 690, 693, 698-700, 703, 704, 706, 708, 715, 716, 733, 734, 737-739]. Solid lines are obtained using least squares regression (coefficients provided in Table 1); the shaded region and dotted lines represent the 95% confidence interval and upper and lower boundaries of the 95% prediction region, respectively. Plot created via the FreezeCasting.net open-data repository.

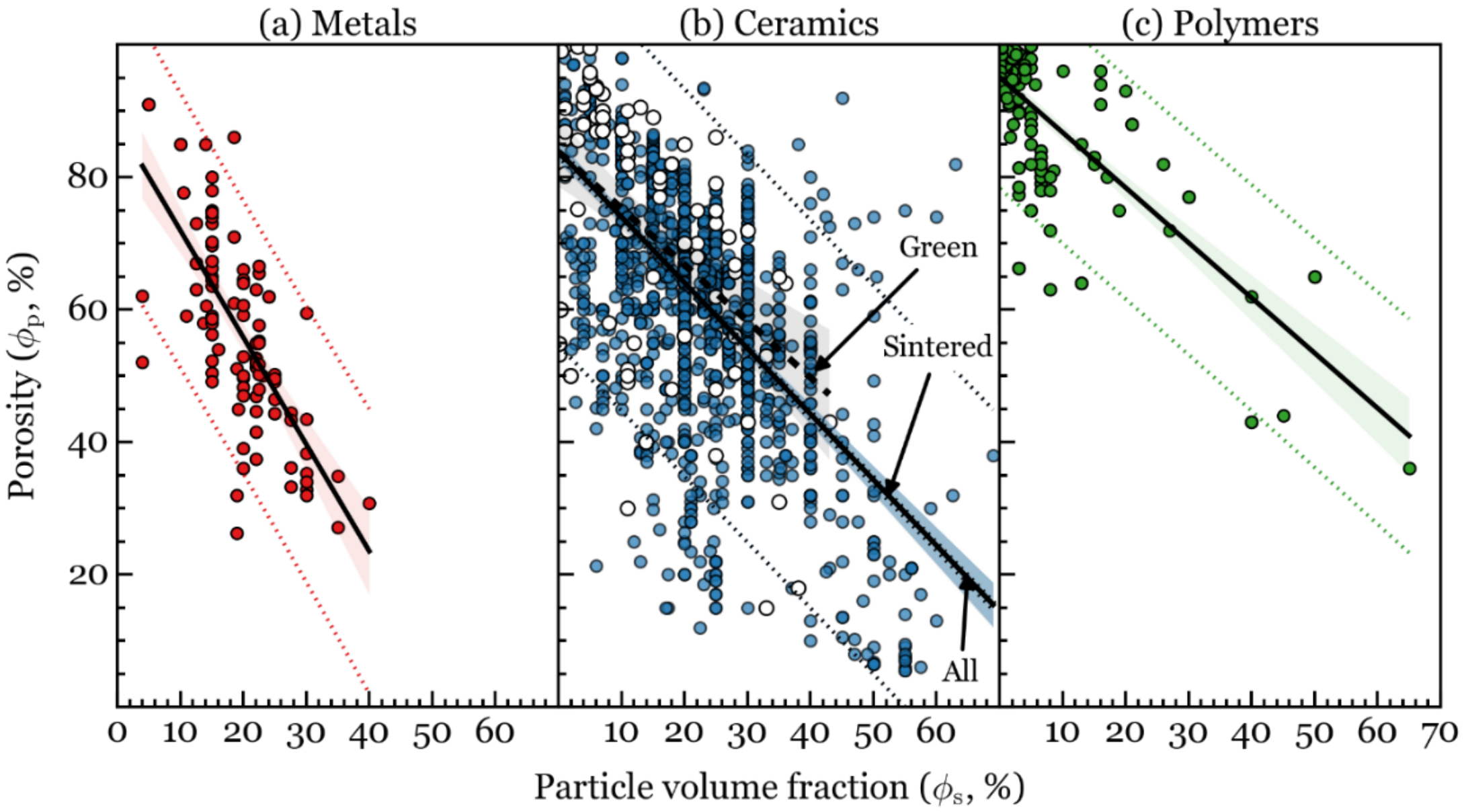

Fig. 15. Plot of total porosity vs. particle volume fraction for aqueous suspensions categorized by material type. Sintered samples are shown in red for (a) metals ($N$=131, [17, 141, 157, 181, 232, 406, 434, 462, 489, 493, 496, 497, 564, 565, 732]), blue for (b) ceramics ($N$=2,185, [6, 49, 107, 109, 111, 112, 117, 120, 124, 148, 149, 184, 185, 187, 195-197, 199, 217-220, 224, 228, 239, 240, 245, 261, 282-286, 295, 296, 298, 299, 302, 304, 309-312, 316, 320, 324, 325, 328, 329, 336, 339, 341, 343-345, 353, 379, 384, 386, 387, 390, 392, 393, 396, 403, 405, 408, 414, 415, 425, 429, 432, 433, 446, 448, 449, 458, 460, 505, 506, 514, 516, 520, 521, 523, 524, 528, 530-532, 536, 537, 539-541, 543, 544, 549, 558-563, 566, 567, 569-572, 577-579, 581-590, 594, 595, 598-603, 609, 610, 615-617, 619-639, 642-646, 648, 649, 651, 652, 655-662, 665, 667, 670, 679, 681-684, 688, 689, 696, 701, 705, 707, 709-714, 717-719, 721, 722, 727-731, 735, 742, 744]), and green for (c) polymers ($N$=248, [54, 198, 223, 281, 323, 413, 426, 427, 436, 519, 522, 529, 533, 535, 541, 542, 547, 548, 551, 552, 555, 557, 741]). Non-sintered samples are shown as white circles in (b); all data points in (c) describe non-sintered materials. Solid lines are obtained using least squares regression (coefficients provided in Table 1); the shaded region and dotted lines represent the 95% confidence interval and upper and lower boundaries of the 95% prediction region, respectively. Plot created via the FreezeCasting.net open-data repository.

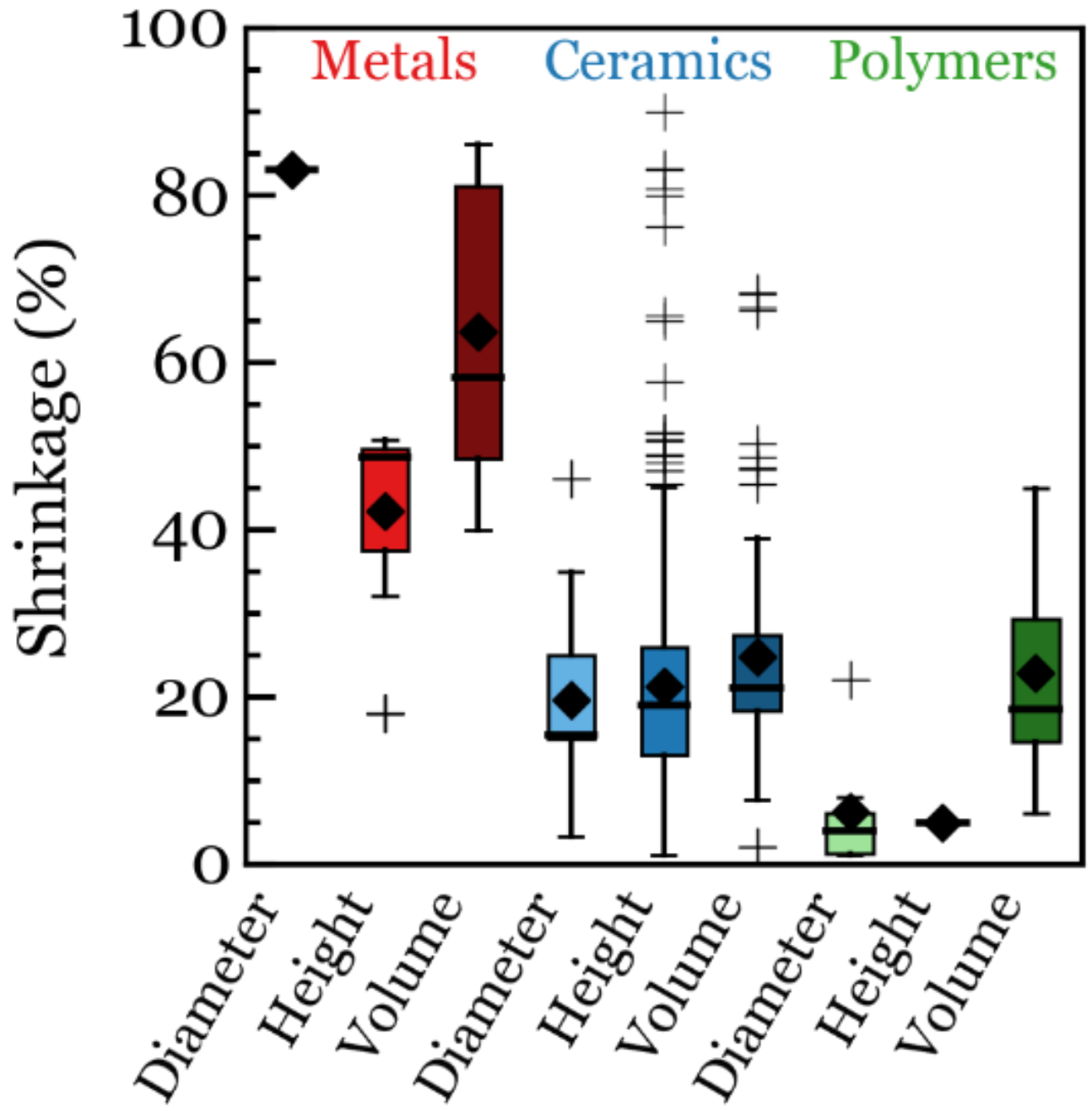

Fig. 16. Box plot summarizing sintering shrinkage data measured as change in diameter, height or volume for metals (red, *N*=63; [55, 141, 181, 434, 462, 497, 565]), ceramics (blue, *N*=644; [6, 110, 148, 156, 185, 228, 239, 240, 262, 286, 295, 296, 316, 341, 383, 384, 390, 408, 414, 415, 566, 573, 574, 586, 590, 608, 619, 622, 623, 637, 638, 647, 652, 655, 656, 660, 665, 666, 674, 683, 688, 690, 691, 700, 712, 714, 720, 731, 737, 738, 745-751]), and polymers (green, *N*=23; [352, 519, 535, 547, 752]). The mean for each condition is shown as a black diamond, the median as a black bar; outliers are shown as +, and error bars are derived using standard deviation; boxes extend from lower to upper quartile values. Plot created via the FreezeCasting.net open-data repository.

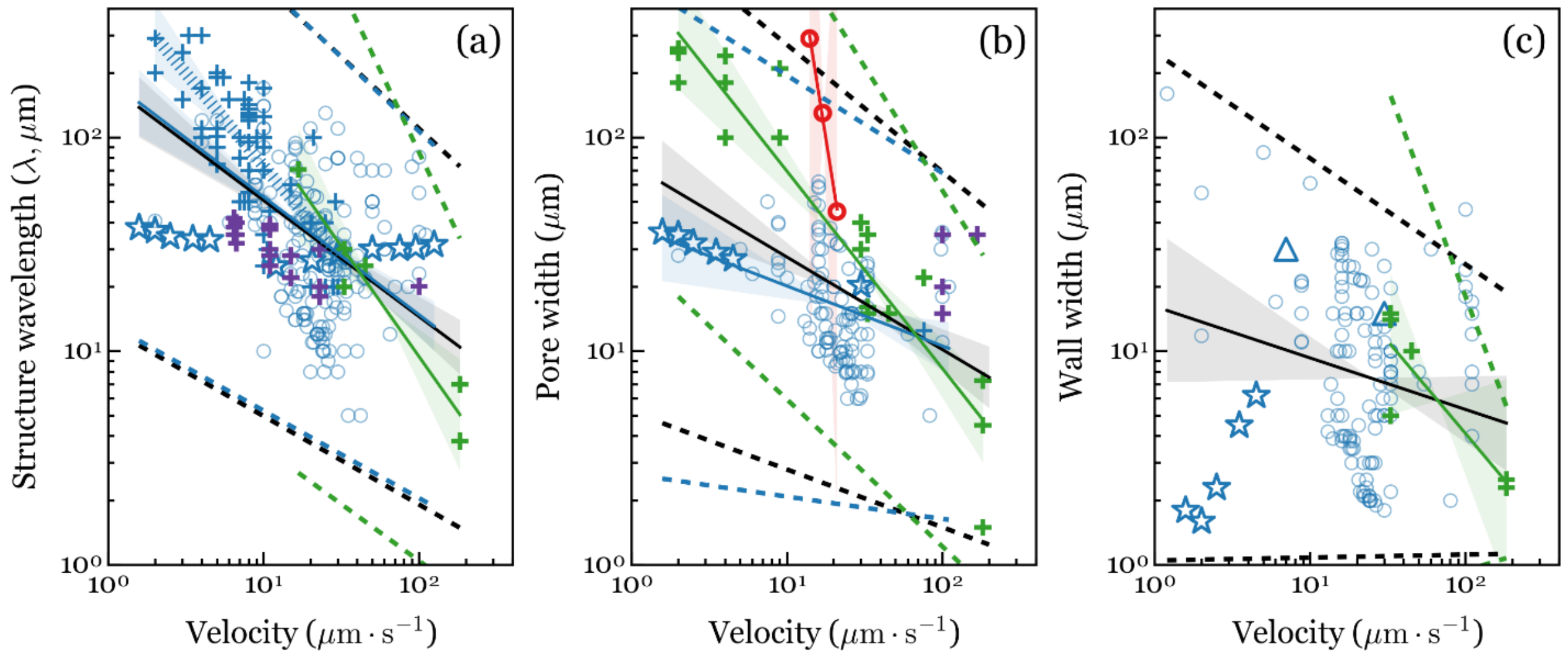

Fig. 17. Dependency of microstructural parameters on solidification velocity for materials exhibiting lamellar pore structures, including: (a) structure wavelength (ceramics, sintered; *N*=185, [120, 147, 150, 156, 228, 229, 239, 296, 379, 396, 411, 418, 421, 448, 514, 689, 756, 757], green; *N*=63, [236-238], and *in-situ*; *N*=12, [243, 398]; polymers; *N*=7, [223, 419, 758-760]; ceramic/polymer composites; *N*=18 [307]; carbon/polymer composites; *N*=1, [440]), (b) pore width (ceramics, sintered; *N*=117, [120, 156, 228, 229, 239, 296, 379, 396, 406, 411, 418, 420, 440, 445, 448, 514, 649, 689, 749, 756, 761-763], green; *N*=1, [417], and *in-situ*; *N*=6, [243, 398, 810]; polymers; *N*=19, [268, 302, 356, 359, 700, 701, 709, 710; metals; *N*=3, [406]; ceramic/polymer; *N*=4, [323]; carbon/polymer; *N*=4, [440, 445]), and (c) wall width (ceramics, sintered; *N*=125, [120, 156, 228, 229, 239, 296, 379, 396, 397, 411, 418, 448, 514, 689, 713, 756, 766], frozen; *N*=2, [765], *in-situ*; *N*=5, [398]; polymers; *N*=6, [223, 758, 759]; carbon/polymer; *N*=4, [445]). Ceramics, polymers, metals, and ceramic/polymer composites are shown in blue, green, red, and purple, respectively. Data obtained from sintered samples are shown in as ○, non-sintered as +, and data obtained from in-situ investigations as ☆, and frozen as △. Solid lines are obtained using least squares regression (coefficients provided in Table 3); the shaded region and dotted lines represent the 95% confidence interval and upper and lower boundaries of the 95% prediction region, respectively. Plot created via the FreezeCasting.net open-data repository.

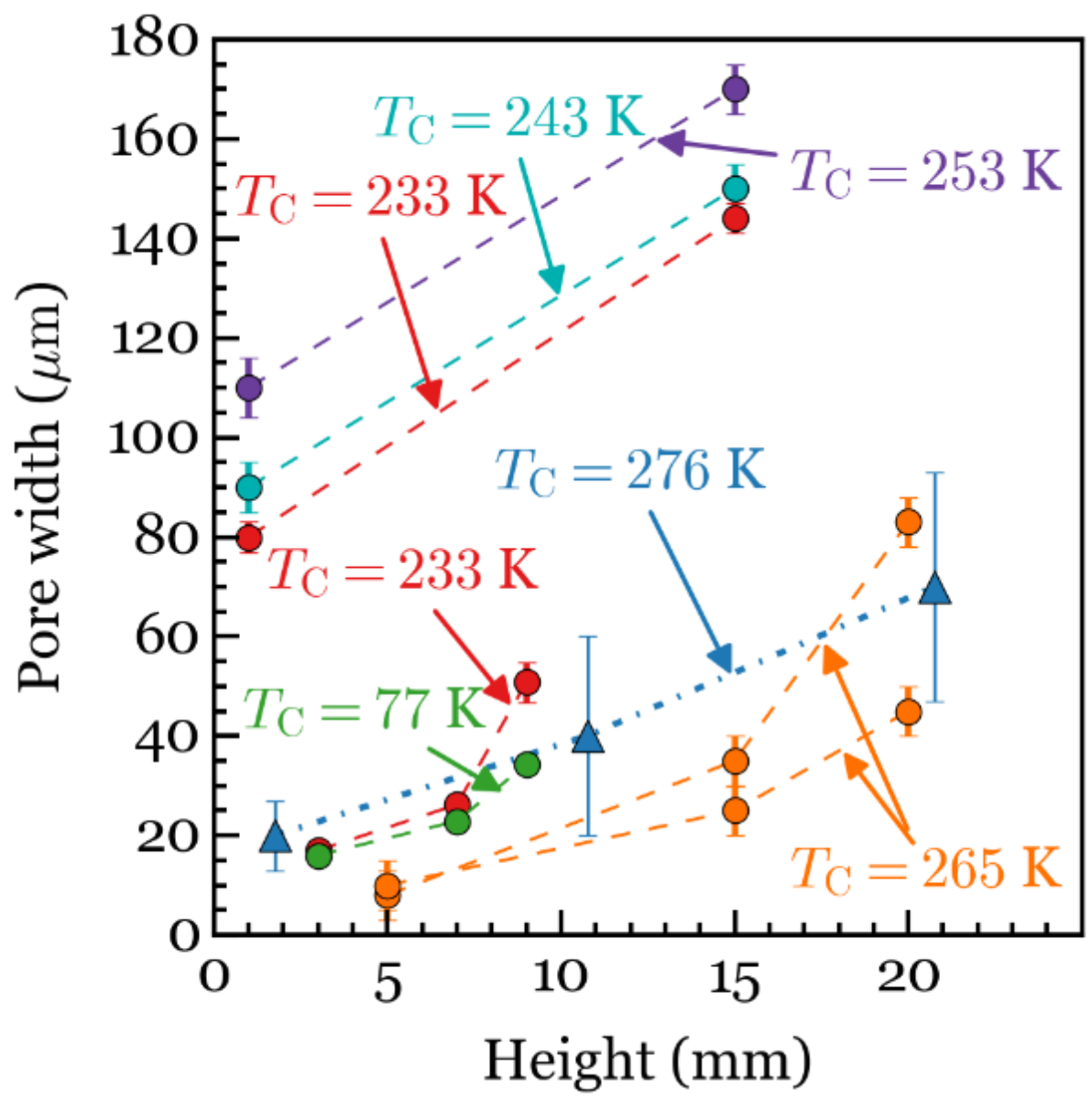

Fig. 18. Plot showing pore width at corresponding heights for samples directionally solidified using a constant substrate temperature [191, 229, 403-406]. Data obtained from samples of aqueous suspensions are shown as circles; camphene suspensions are shown as triangles. Plot created via the FreezeCasting.net open-data repository.

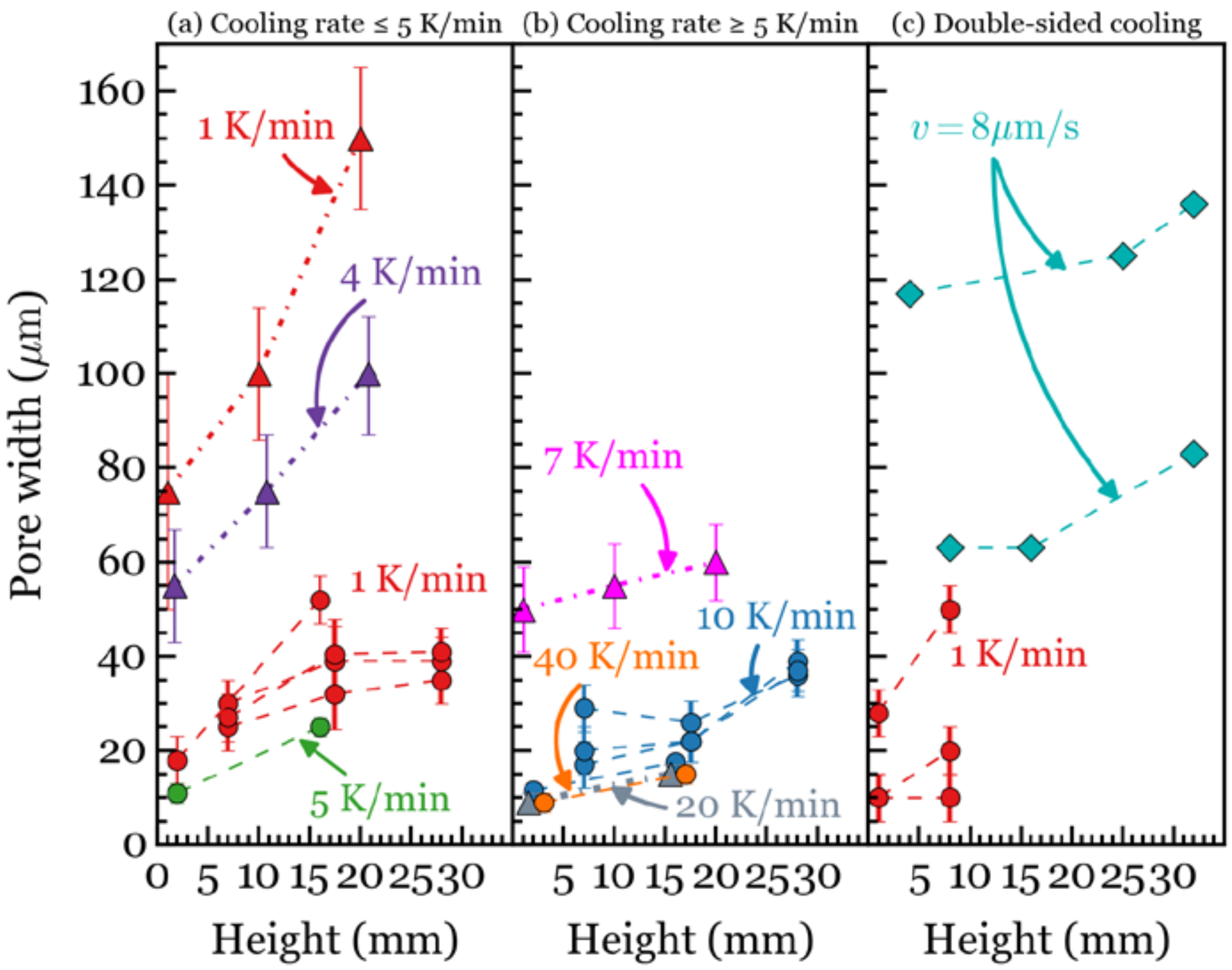

Fig. 19. Plots showing pore width at corresponding heights for samples directionally solidified using (a) linear cooling rates ≤ 5 K/min [307, 403, 410], (b) linear cooling rates ≥ 5 K/min [307, 403, 410], and (c) double-sided cooling with a linear cooling rate [237, 411]. Data obtained from samples of aqueous suspensions are shown as circles; camphene suspensions are shown as triangles. In (c), teal, diamond-shaped data points describe data obtained from green samples derived from aqueous suspensions [237]. All other data points represent sintered samples. Plot created via the FreezeCasting.net open-data repository.

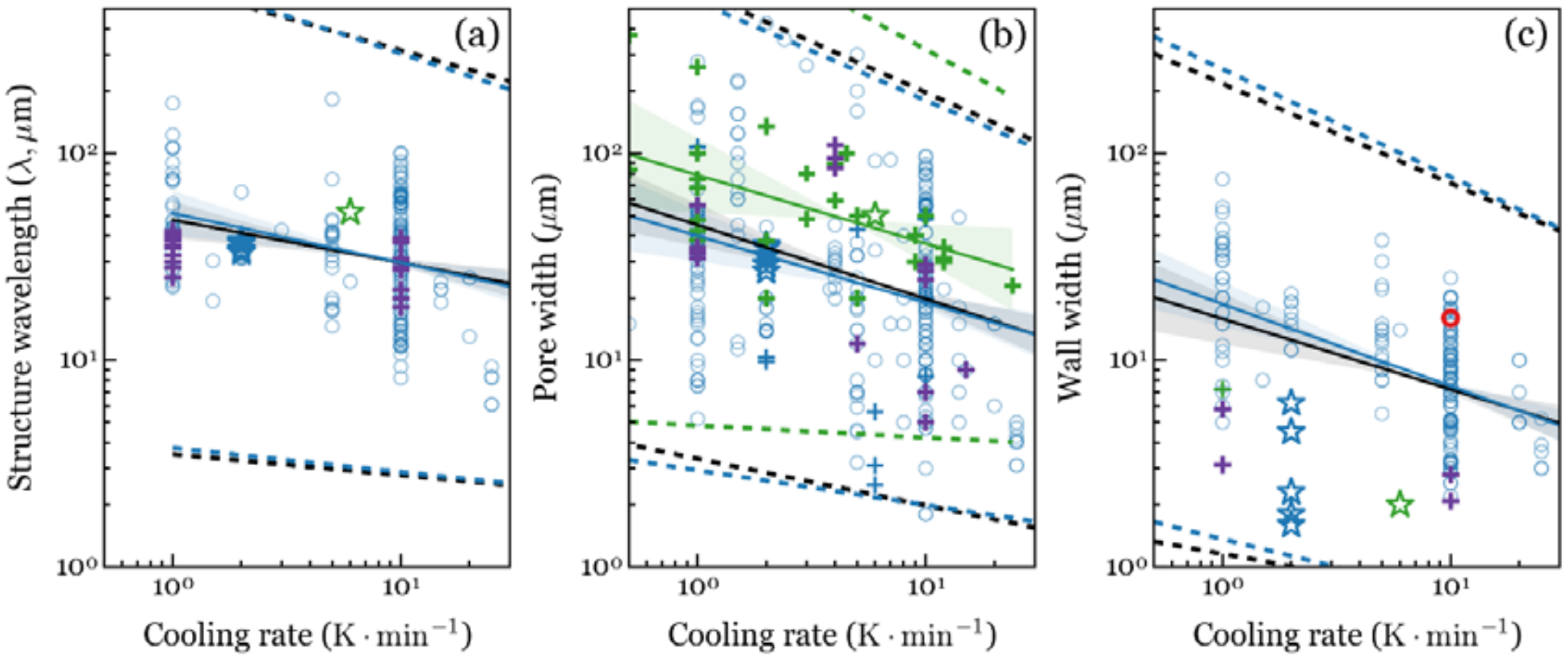

Fig. 20. Dependency of microstructural parameters on cooling rate for materials exhibiting lamellar pore structures, including: (a) structure wavelength (ceramics, sintered; *N*=180, [109, 149, 290-293, 295, 339, 396, 400, 403, 408, 411, 418, 421, 516, 569, 626, 743, 767-769], and *in-situ*; *N*=5, [398, 765]; polymers, green; *N*=2, [419, 770] and *in-situ*; *N*=1, [771]; ceramic/polymer composites; *N*=24, [307]), (b) pore width (ceramics, sintered; *N*=258, [107, 109, 149, 233, 282, 283, 286, 290-293, 295, 311, 339, 396, 400, 403, 408, 411, 414, 418, 456, 516, 569, 570, 582, 626, 660, 667, 728, 743, 767-769, 774-777], green; *N*=11, [390, 561, 772, 773], and *in-situ*; *N*=5, [398]; polymers, green; *N*=30, [54, 281, 297, 323, 420, 423, 542, 770, 778-780] and *in-situ*; *N*=1, [771]; ceramic/polymer composites; *N*=17, [297, 307, 323, 728, 781]), and (c) wall width (ceramics, sintered; *N*=173, [109, 149, 290-293, 295, 302, 339, 396, 400, 403, 408, 411, 418, 516, 569, 626, 696, 767-769] and *in-situ* [398, 765]; polymers, green [770, 771] and *in-situ*; *N*=5, [771]; metals; *N*=2, [494]; ceramic/polymer; *N*=6, [307]). Ceramics, polymers, metals, and ceramic/polymer composites are shown in blue, green, red, and purple, respectively. Data obtained from sintered samples are shown in as ○, non-sintered as +, and data obtained from in-situ investigations as ☆, and frozen as △. Solid lines are obtained using least squares regression (coefficients provided in Table 4); the shaded region and dotted lines represent the 95% confidence interval and upper and lower boundaries of the prediction region, respectively. Plot created via the FreezeCasting.net open-data repository.

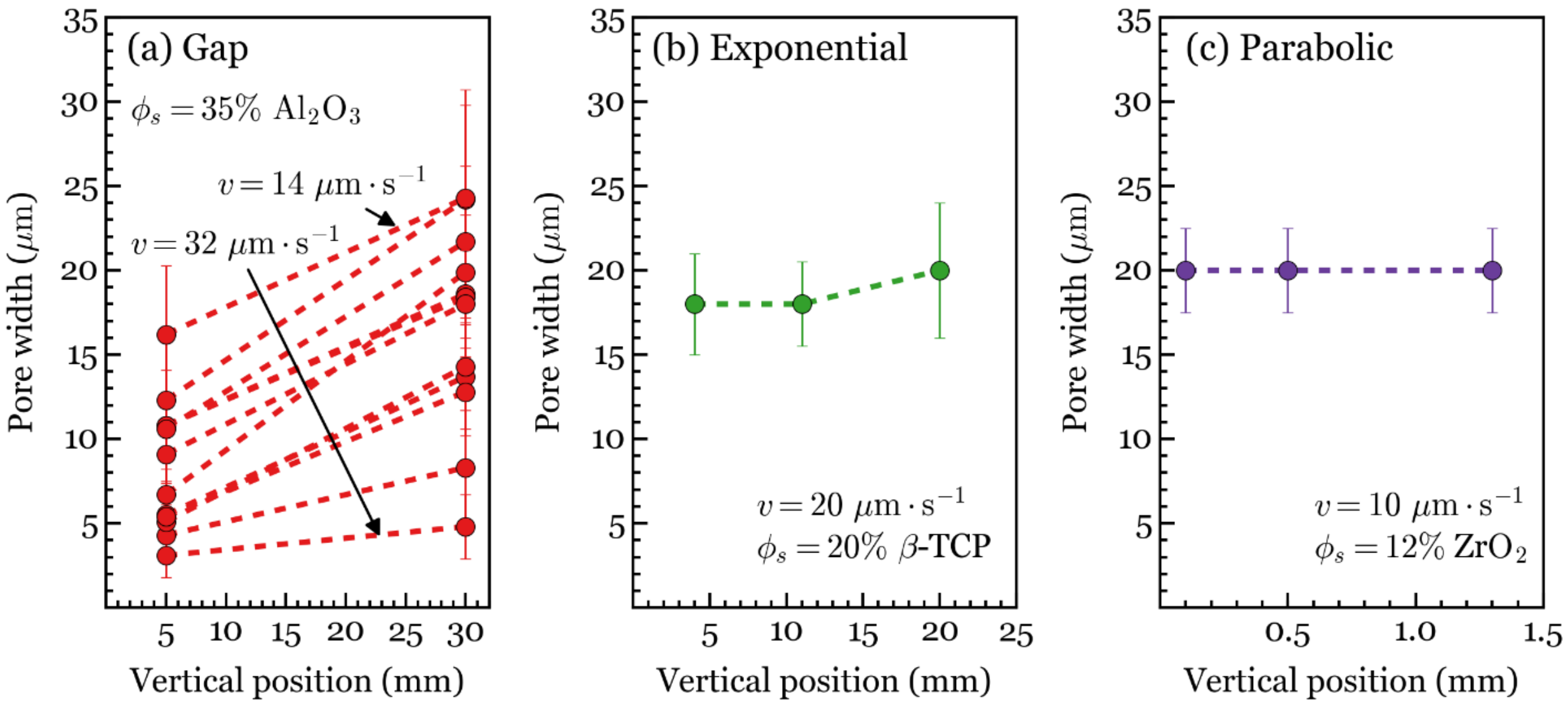

Fig. 21. Plots showing pore width at corresponding heights for aqueous suspensions directionally solidified using the (a) gap method [296], (b) exponential (tricalcium phosphate; TCP) [239], and (c) parabolic [449] cooling functions for achieving constant velocities. Plot created via the FreezeCasting.net open-data repository.

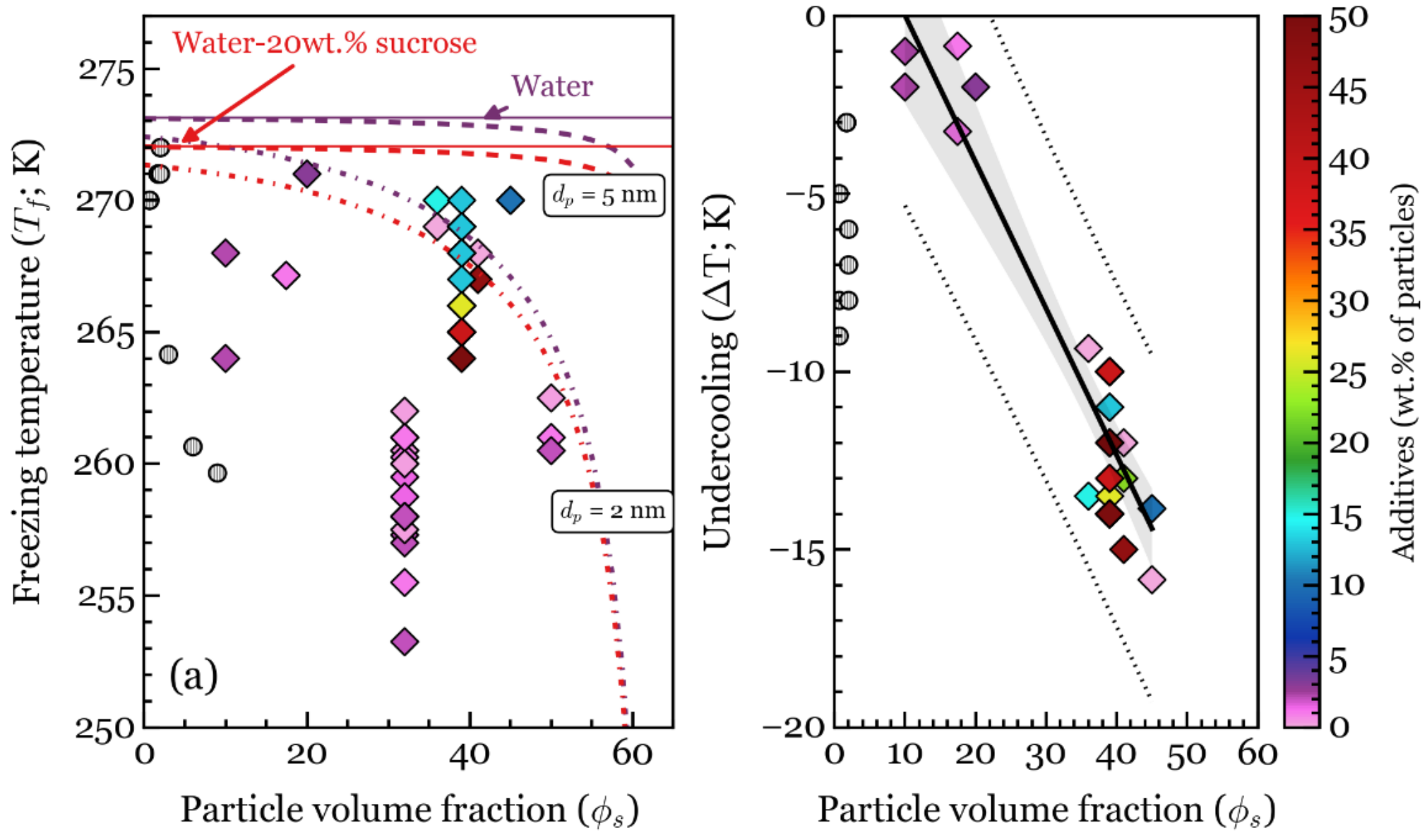

Fig. 22. (a) Plot of freezing temperature as a function of particle volume fraction. Data points for studies using dissolved polymer [404, 408, 413, 420] and suspended particles [25, 220, 224, 311, 454-456] are shown as circles and diamonds, respectively. Dotted lines are calculated for predicted freezing temperature for 2 and 5 nm particles (as marked) using the expression for thermodynamic freezing temperature (Eq. 3). (b) Plot of undercooling as a function of particle volume fraction. The solid line is obtained using least squares regression ($\Delta T = -41K \cdot \phi_s + 4.1K$), the shaded region and dotted lines represent the 95% confidence interval and upper and lower boundaries of the 95% prediction region, respectively. Plot created via the FreezeCasting.net open-data repository.

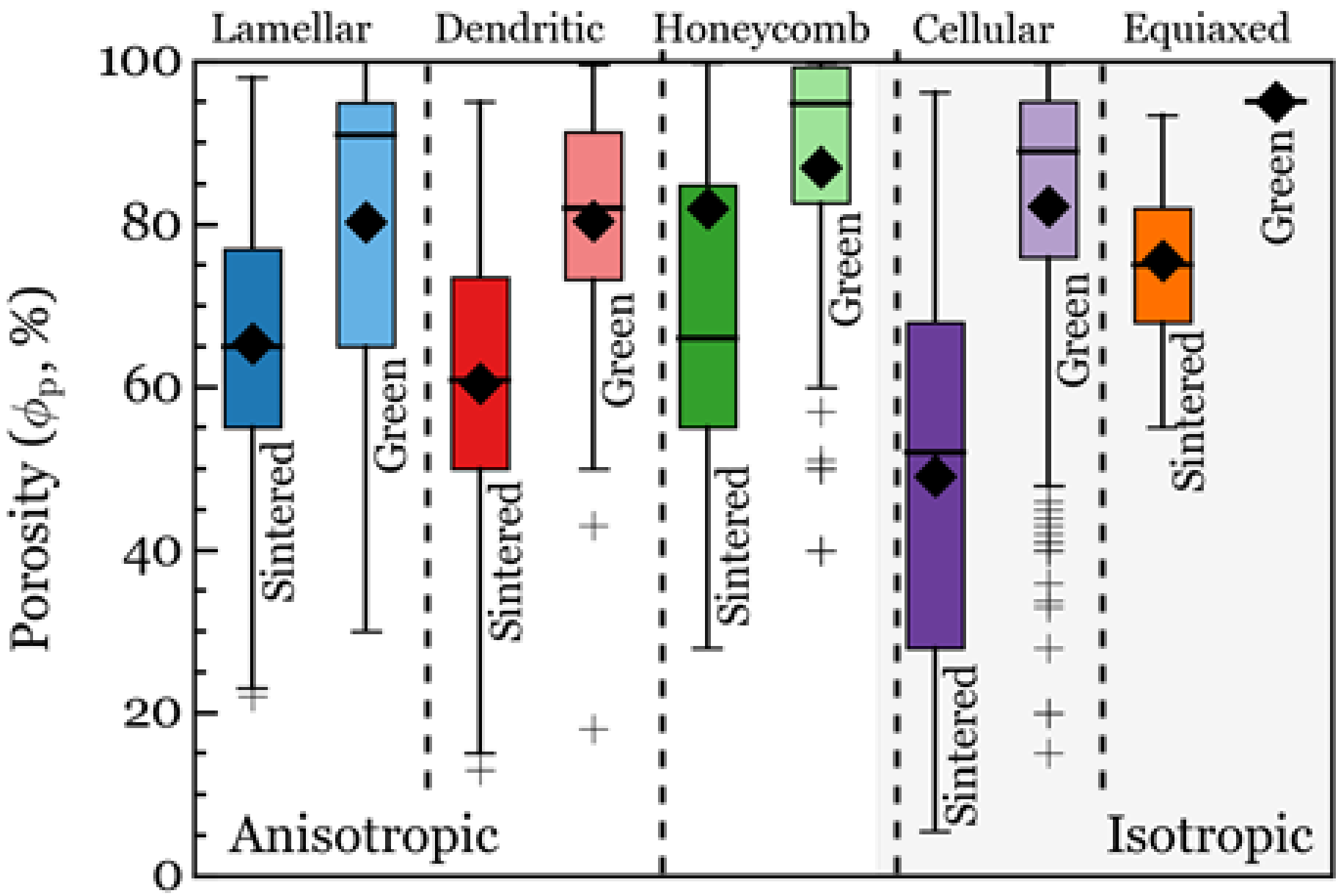

Fig. 23. Box plots showing range of demonstrated porosity for (a) anisotropic and (b) isotropic freeze-cast materials, including: (a) lamellar (sintered, *N* =1,116 [6, 17, 24, 49, 107, 109, 111, 112, 117, 120, 124, 141, 147, 149, 150, 181, 184, 185, 195, 217-220, 224, 228, 232, 239, 240, 261, 282-286, 295, 296, 298, 299, 302, 309-312, 316, 325, 336, 339, 345, 353, 379, 384, 386, 390, 392, 396, 397, 403, 405, 406, 408, 412, 414, 415, 425, 446, 448, 458, 460, 462, 488, 489, 493, 494, 496, 497, 505, 506, 514, 516, 543, 563, 564, 566, 567, 569, 570, 572, 579, 581, 582, 586, 601-603, 611, 613, 614, 616, 624-628, 630-633, 635, 637-639, 642, 644, 645, 649, 651, 655, 659, 660, 662, 667, 670, 680, 682, 688, 689, 693, 696, 705, 711-714, 718, 721, 722, 724, 725, 728, 732-735, 742, 747, 767, 768, 788-797]; green, *N* =206 [25, 54, 223, 281, 307, 314, 321, 390, 416, 426, 427, 440, 519, 522, 523, 528, 529, 534, 536, 538, 539, 541-544, 546, 551-554, 557, 558, 560-562, 741, 779, 782-787]), dendritic (sintered, *N* =398 [6, 148, 187, 196, 224, 228, 262, 271, 296, 302, 304, 309, 311, 320, 324, 328, 329, 347, 348, 353, 357, 379-383, 385, 388, 393, 394, 396, 410, 449, 453, 461, 498-501, 567, 568, 570, 573-576, 584, 586, 589, 604-608, 613, 614, 617-620, 622-624, 632, 640, 641, 652, 653, 656, 657, 665, 666, 668, 672, 673, 675, 676, 681, 683-687, 691, 692, 694, 695, 697, 702, 707, 719, 723, 726, 731, 736, 797]; green, *N*=74 [180, 230, 281, 436, 455, 519, 521, 525, 528, 533, 540, 546, 555]), and honeycomb (sintered, *N* =334 [110, 121, 189, 224, 274, 299, 339, 341, 342, 351, 387, 394, 429-435, 492, 569, 593-595, 647, 650, 663, 664, 678, 690, 700, 703, 704, 715, 716, 737-739, 750, 800] [504, 847]); green, *N*=01 [281, 323, 341, 350, 352, 525, 530, 534, 535, 798, 799]) and (b) cellular (sintered, *N*=541 [114, 199, 220, 226, 274, 284, 285, 499, 577, 593, 597, 612, 619, 621, 648, 661, 674, 677, 684, 706, 708, 710, 719, 720, 730, 745, 753, 795, 806, 876-884]; green, *N*=199 [67, 198, 223, 245, 314, 413, 417, 420, 437, 524, 526, 527, 529, 536, 537, 540, 545, 547-551, 553, 741, 781, 802-804, 836, 853, 858, 863-875]), and equiaxed (sintered *N*=58 [197, 344, 583, 609, 643, 679, 698, 699, 755]; green, *N*=1 [784]) pore structures. The mean for each condition is shown as a black diamond, the median as a black bar; outliers are shown as +, and error bars are derived using standard deviation; boxes extend from lower to upper quartile values. Plot created via the FreezeCasting.net open-data repository.

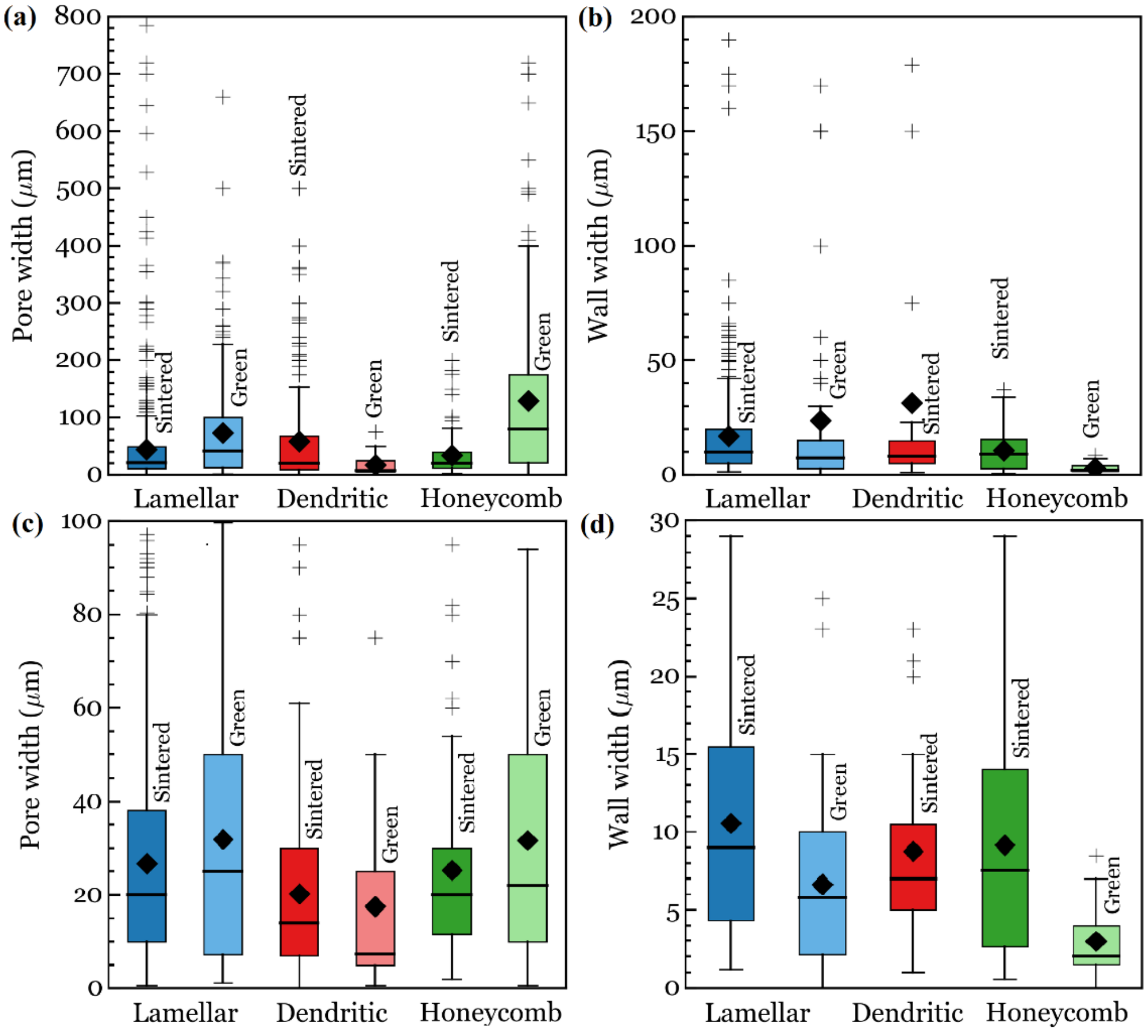

Fig. 24. Box plots showing range of measured (a) pore and (b) wall widths for anisotropic freeze-cast materials; full distributions are shown in (a) and (b), whereas (c) shows distribution for pore width < 100 um and (d) shows distribution for wall width < 30 um; these include: (a) pore width: lamellar (sintered, *N* =653 [6, 17, 24, 49, 55, 106, 107, 109, 111, 117, 120, 149, 156, 181, 184, 191, 217-220, 228, 229, 233, 239, 282-286, 290-293, 295, 296, 298, 299, 309, 311, 312, 339, 345, 353, 379, 396, 400, 403, 405, 406, 408, 411, 412, 414, 418, 425, 448, 456, 458, 460, 462, 488, 489, 493, 496, 514, 516, 543, 563, 564, 567, 569, 570, 582, 601, 602, 616, 626, 628, 632, 633, 637, 645, 649, 659, 660, 667, 670, 680, 689, 722, 724, 728, 732-734, 743, 747-749, 751, 756, 761, 762, 767-769, 774-777, 789, 795, 797, 826-835]; green, *N* =221 [54, 155, 191, 223, 281, 287-289, 294, 297, 307, 314, 321, 390, 401, 417, 420, 423, 439-441, 445, 459, 519, 528, 529, 536, 538, 539, 542, 544, 551, 553, 557, 558, 561, 562, 741, 752, 758, 759, 763, 764, 770, 772, 773, 778-781, 783, 785, 787, 811-825]), dendritic (sintered, *N* =267 [6, 148, 187, 196, 228, 262, 296, 303, 309, 311, 320, 329, 348, 353, 357, 381, 382, 388, 393, 394, 396, 410, 418, 449, 453, 461, 498-500, 567, 568, 570, 575, 584, 604-607, 618-620, 632, 653, 656, 673, 675, 676, 685, 686, 691, 692, 694, 719, 726, 797, 831, 839-852]; green, *N* =31 [44, 52, 180, 281, 389, 519, 525, 540, 836-838]), and honeycomb (sintered, *N* =101 [189, 274, 276, 299, 339, 351, 428-430, 432-435, 569, 593, 594, 647, 650, 663, 678, 690, 716, 749, 800]; green, *N* =201 [1, 281, 313, 323, 349, 350, 352, 424, 428, 438, 459, 530, 798, 799, 815, 823, 853-862]), (b) wall width, including: lamellar (sintered, *N* =404 [49, 109, 117, 120, 147, 149, 156, 228, 229, 239, 284, 290-293, 295, 296, 299, 339, 379, 396, 397, 400, 403, 405, 408, 411, 418, 448, 460, 462, 493, 494,

496, 497, 505, 514, 516, 567, 569, 626, 639, 642, 659, 689, 696, 713, 724, 732, 743, 747, 756, 766-769, 795, 826]; green, *N* =69 [223, 307, 314, 439, 445, 447, 459, 539, 553, 562, 741, 758, 759, 770, 771, 783, 787, 821, 886]), dendritic (sintered, *N* =30 [296, 357, 396, 418, 568, 653, 691, 692]), and honeycomb (sintered, *N* =39 [351, 429, 430, 433, 569, 593, 647]; green, *N* =13 [350, 352, 424, 459, 887, 888]). The mean for each condition is shown as a black diamond, the median as a black bar; outliers are shown as +, and error bars are derived using standard deviation; boxes extend from lower to upper quartile values. Plot created via the FreezeCasting.net open-data repository.

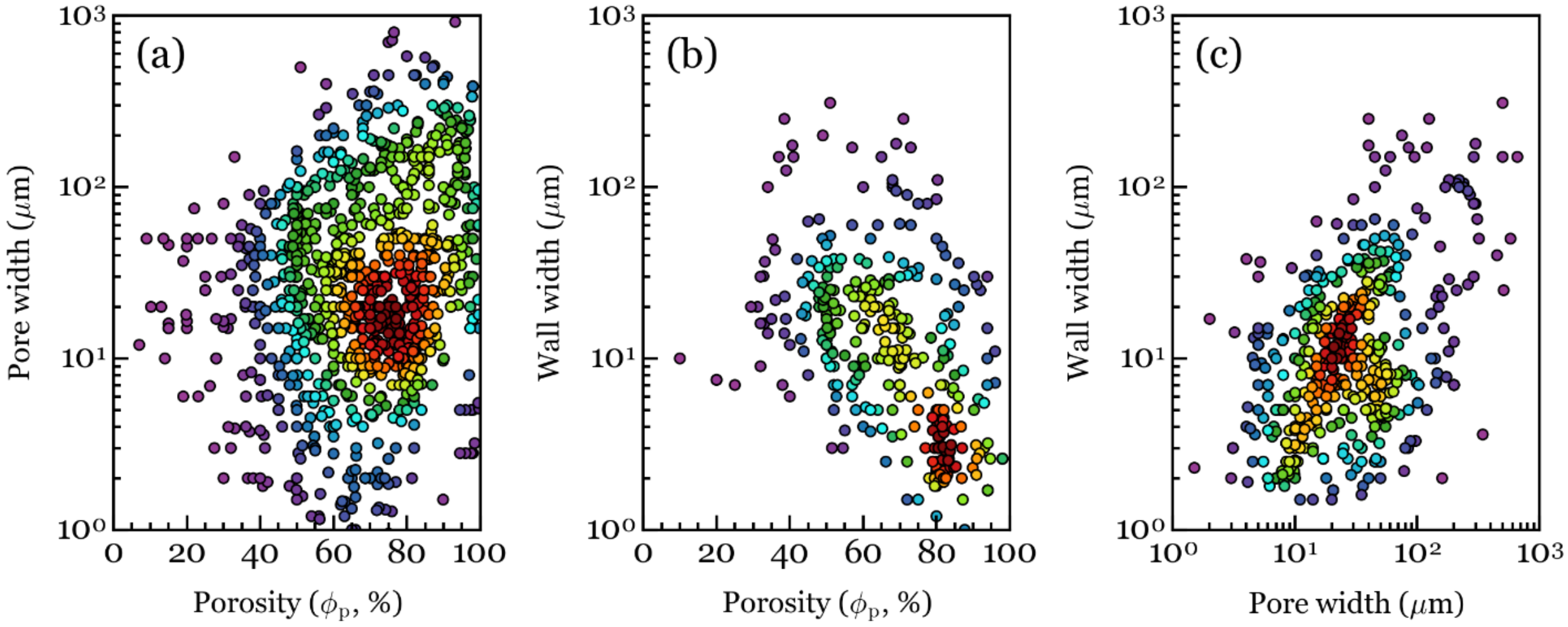

Fig. 25. Point density plots showing microstructural parameters in freeze-cast materials: (a) pore width vs. porosity [6, 17, 24, 49, 54, 107, 109, 111, 114, 117, 120, 148, 149, 180, 181, 184, 187, 189, 196-199, 217-220, 222, 223, 226, 228, 239, 245, 262, 274, 281-286, 295, 296, 299, 307, 309, 311, 312, 314, 320, 321, 329, 339, 344, 345, 348, 350-353, 357, 379, 381, 382, 388, 390, 393, 394, 396, 405, 406, 408, 410, 412-414, 425, 429, 430, 432-435, 440, 448, 449, 458, 460-462, 488, 489, 493, 496, 498-500, 514, 516, 519, 524-530, 536-540, 542-545, 547-551, 553, 556-558, 561-564, 567-570, 575, 582-584, 593, 594, 597, 601, 602, 604-607, 609, 612, 616, 618-621, 626, 628, 632, 633, 637, 643, 645, 647-650, 653, 656, 659-661, 663, 667, 670, 673-680, 684-686, 689-692, 694, 698, 699, 706, 708, 710, 716, 719, 720, 722, 724, 726, 728, 730, 732-734, 745, 747, 767, 768, 785, 787, 789, 795, 797-799, 802-804, 806], (b) wall widths vs. porosity [49, 109, 117, 120, 149, 223, 226, 228, 239, 284, 295, 296, 299, 307, 314, 350-352, 357, 379, 396, 405, 408, 429, 430, 433, 448, 460, 462, 492-494, 496, 497, 505, 514, 516, 539, 556, 567-569, 593, 604, 626, 639, 642, 647, 653, 659, 689, 691, 692, 696, 698, 699, 710, 713, 720, 724, 732, 747, 767, 768, 787, 795, 802, 804] and (c) wall widths vs. pore widths [2, 49, 106, 109, 117, 120, 149, 156, 223, 226, 228, 229, 239, 284, 290-293, 295, 296, 299, 307, 311, 314, 339, 350-352, 357, 379, 396, 398, 400, 405, 408, 411, 418, 424, 429, 430, 433, 439, 448, 459, 460, 462, 493, 496, 504, 514, 516, 539, 553, 556, 567-569, 593, 604, 626, 647, 659, 689, 691, 692, 698, 699, 710, 720, 724, 732, 743, 747, 756, 758, 759, 767-771, 783, 787, 795, 802, 804, 811, 821, 826, 847, 863]. Points are colored based on relative point-density; red points indicate high density regions and purple points represent low density regions. Plot created via the FreezeCasting.net open-data repository.

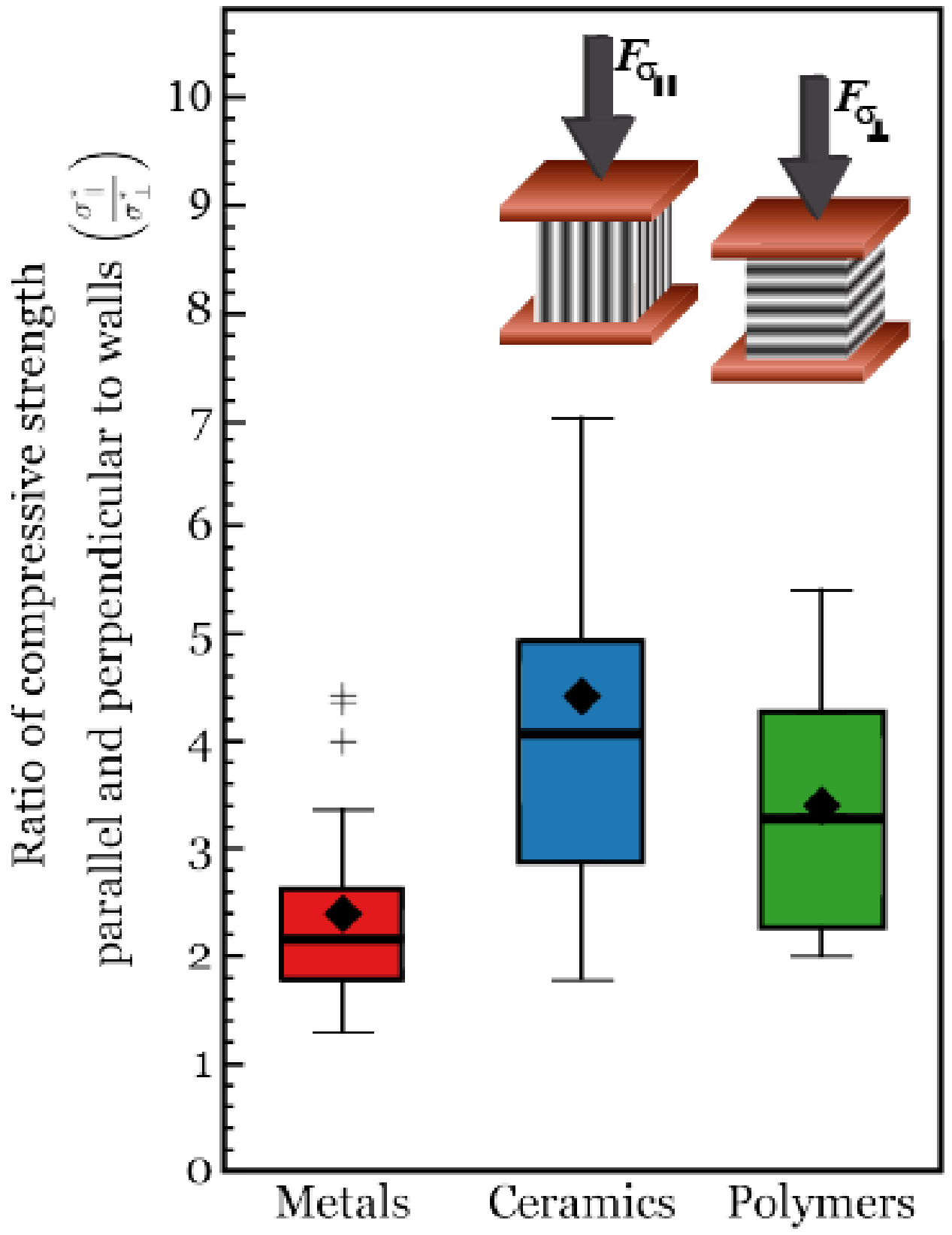

Fig. 26. Anisotropic compressive properties observed for freeze-cast materials (irrespective of pore structure). The insert (after ref. [462]) shows the loading direction for compressive strength taken parallel and perpendicular to the wall direction. The box plot on the right shows the distribution of reported ratios of compressive strength taken parallel and perpendicular to the wall direction ($\sigma^*_{||}/\sigma^*_{\perp}$) including (red) metals [17, 406, 462, 489, 604], (blue) ceramics [147, 149, 262, 285, 302, 336, 403, 461, 492, 575, 633], and (green) poymers [352, 519, 522, 552]. The mean for each condition is shown as a black diamond, the median as a black bar; outliers are shown as +, and error bars are derived using standard deviation; boxes extend from lower to upper quartile values. Plot created via the FreezeCasting.net open-data repository.

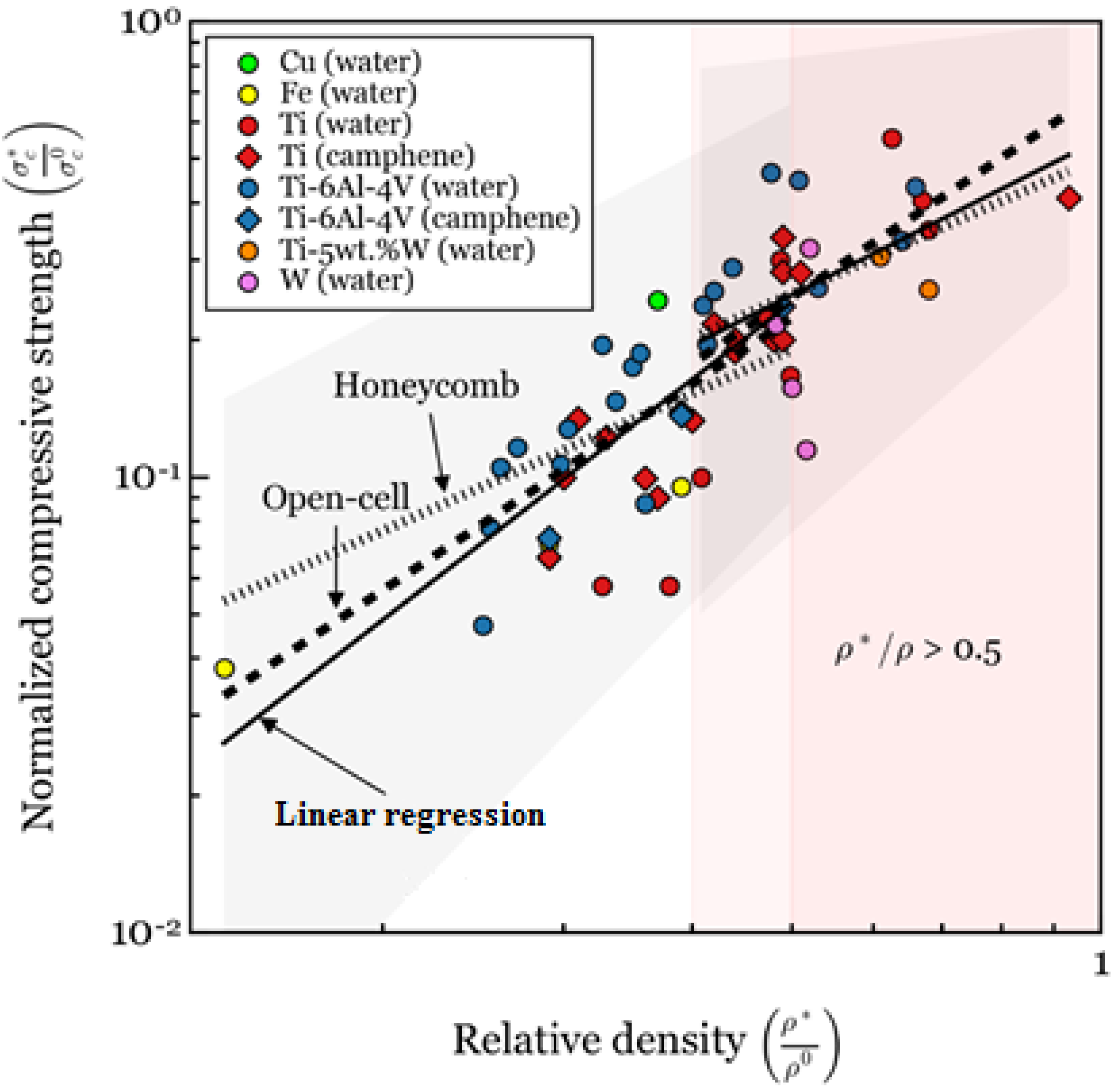

Fig. 27. Normalized longitudinal compressive strength of freeze-cast, sintered metals as a function of relative density (1 - $\phi_p$). All materials resulting from aqueous particle suspensions (circles) show lamellar pore structures; these include: Ti [17, 157, 489, 493, 494], Ti-6Al-4V [406], Ti-5wt.%W [157, 494], Cu [462], Fe [496], and W [497]. All materials employing camphene (diamonds) as the fluid exhibited dendritic microstructures; these are: Ti [498-500] and Ti-6Al-4V [501]. The linear regression line ($N$ = 60, $R^2$ = 0.67, $p$ < 0.0001) is shown as a solid black line; the shaded region about the regression line represents the 95% confidence interval about the regression line. Curve fittings for open-cell and honeycomb models (Eq. 5) are shown as long and short-dashed lines, respectively. Equation coefficients and bulk material strength ($\sigma_c^0$ ) are provided in Table 7. Plot created via the FreezeCasting.net open-data repository.

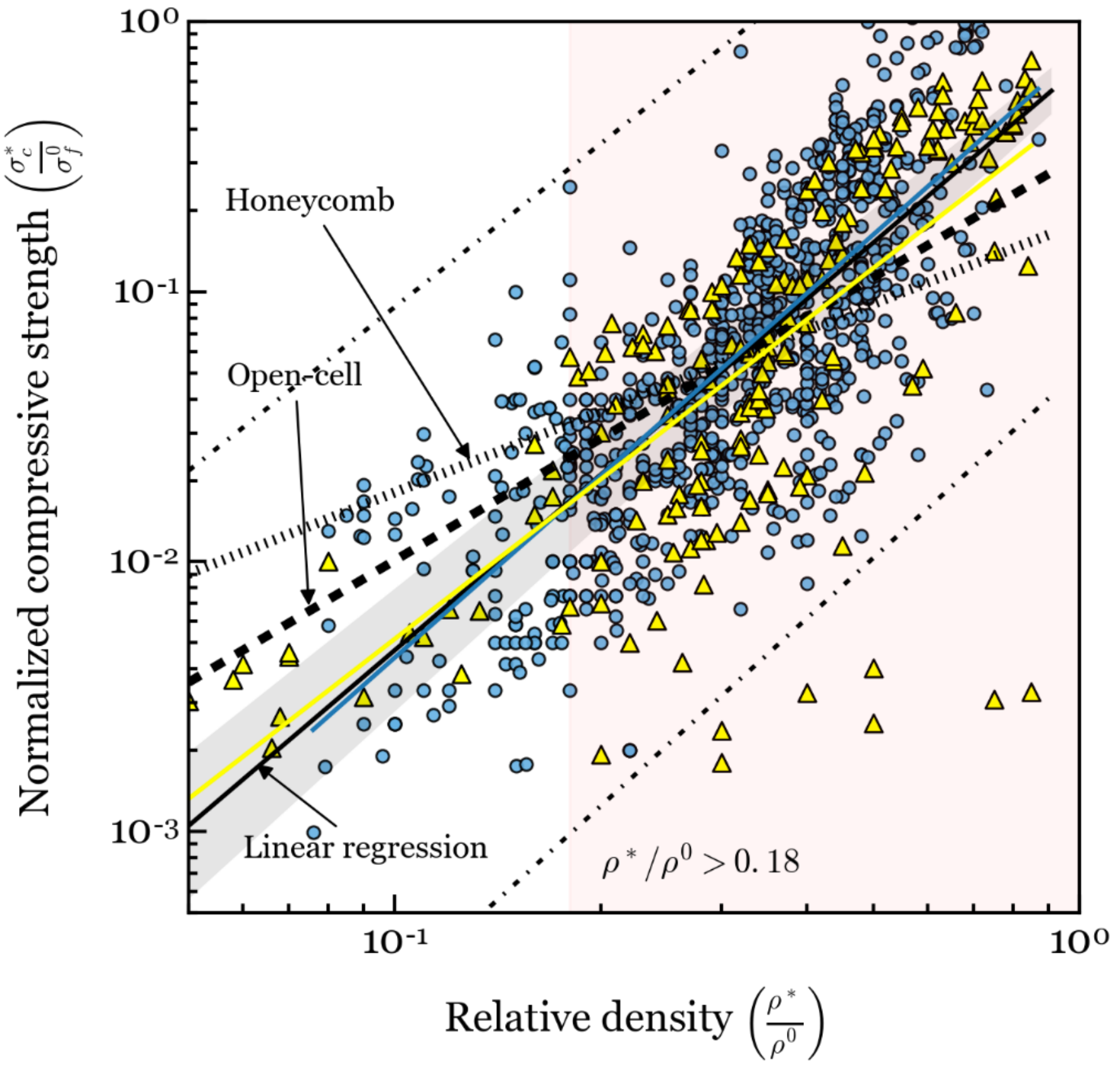

Fig. 28. Normalized compressive strength of freeze-cast ceramics as a function of relative density (1 - $\phi_p$); anisotropic (parallel loading) and isotropic structures are shown as blue circles and yellow triangles, respectively [38, 39, 110, 112, 117, 120, 149, 189, 195, 196, 219, 224, 226, 239, 240, 262, 282, 283, 285, 286, 295, 296, 302, 312, 320, 324, 336, 339, 341, 342, 344, 351, 379-381, 384-386, 388, 390, 391, 405, 410, 412, 414, 415, 429-431, 446, 448, 461, 488, 492, 504-506, 514, 516, 563, 566, 569, 572, 575, 576, 579, 581, 583, 584, 588, 590, 595, 606-610, 615, 617-620, 623, 624, 626-629, 633-635, 637, 640, 647, 651, 652, 655, 658, 662, 665-668, 670-672, 674, 675, 677-679, 681-683, 688-692, 697-700, 703-705, 708, 711, 712, 716, 720, 721, 723, 726, 742, 745, 747, 750, 754, 792, 795, 809] [483]. The linear regression line for all materials is shown as a solid black line ($N$=1,032, $R^2$=0.59, $p$<0.0001); blue and yellow lines represent anisotropic and isotropic materials, respectively. The shaded region about the regression line represents the 95% confidence interval about the regression line, whereas the dashed line of similar color represents the 95% prediction region. Curve fittings for open-cell and honeycomb models (Eq. 5) are shown as horizontal and vertical dashed lines, respectively. Coefficients and bulk material values of $\sigma_f^0$ are provided in Table 8. Plot created via the FreezeCasting.net open-data repository.

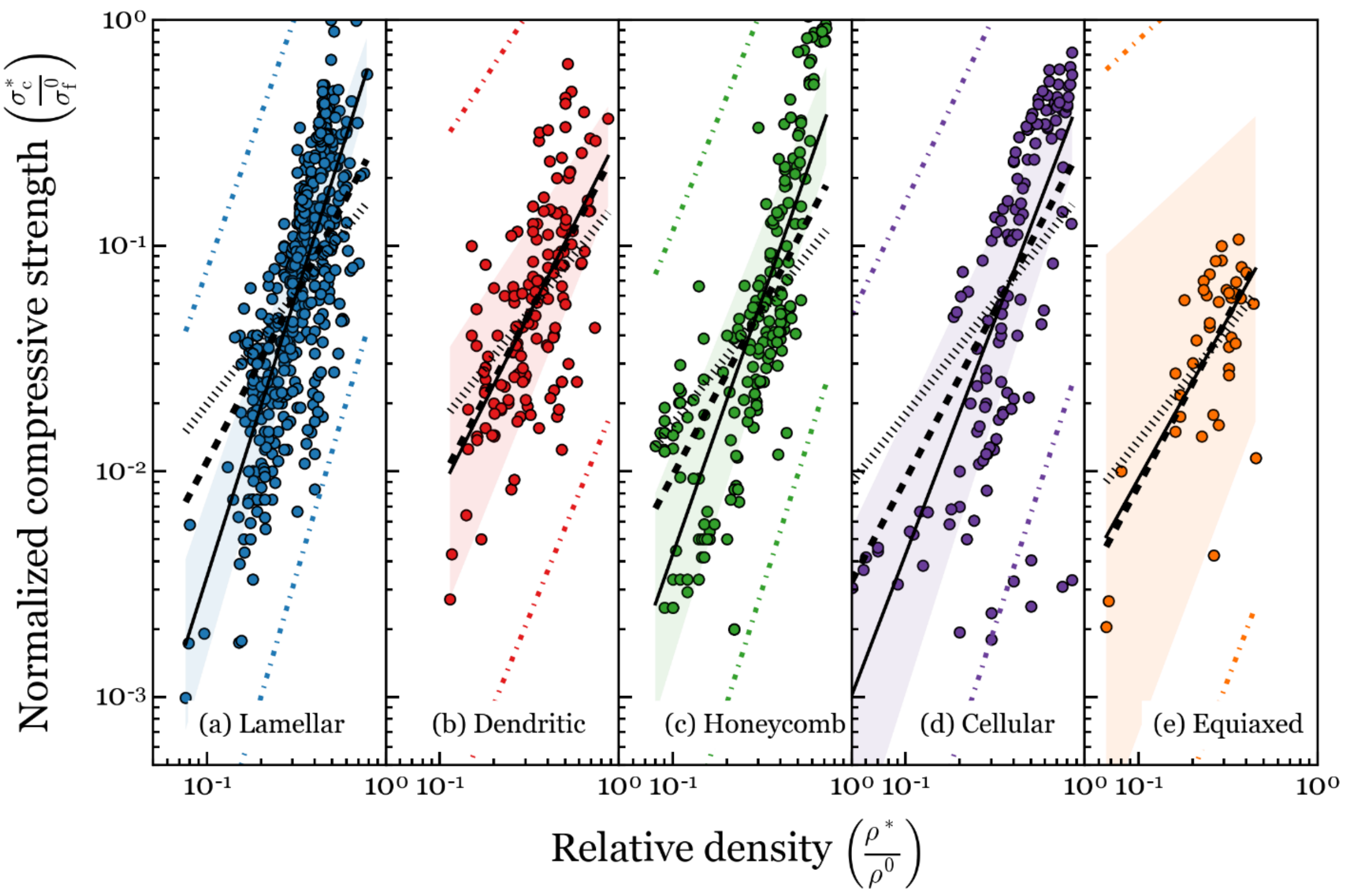

Fig. 29. Plots of normalized compressive strength of freeze-cast ceramics as a function of relative density (1 - $\phi_p$), categorized by pore structure, including: (a) lamellar (*N*=492 [112, 117, 120, 149, 195, 219, 224, 239, 240, 282, 283, 285, 286, 295, 296, 302, 312, 336, 339, 379, 384, 386, 390, 405, 412, 414, 415, 446, 448, 460, 488, 505, 506, 514, 516, 563, 566, 569, 572, 579, 581, 624, 626-628, 633, 635, 637, 651, 655, 662, 667, 670, 682, 688, 689, 705, 711, 712, 721, 742, 747, 792, 795]), (b) dendritic (*N*=143 [196, 224, 262, 296, 302, 320, 324, 379-381, 385, 388, 410, 461, 575, 576, 584, 606-608, 617-620, 623, 624, 640, 652, 665, 666, 668, 672, 675, 681, 683, 691, 692, 697, 723, 726]), (c) honeycomb (*N*=207 [110, 224, 339, 341, 342, 351, 429-431, 492, 504, 569, 595, 647, 678, 690, 700, 703, 704, 737, 738, 750]), (d) cellular (*N*=145 [38, 39, 226, 285, 379, 391, 590, 615, 617, 619, 629, 634, 668, 671, 674, 677, 708, 712, 720, 721, 745, 754, 795, 809]), and (e) equiaxed (*N*=40 [344, 583, 588, 609, 610, 658, 679, 698, 699]). All data in (a)-(c) correspond to anisotropic materials. Linear regression lines are shown as solid black lines; shaded regions about each regression line represent the 95% confidence interval of the regression line; dashed lines of matching color represent the 95% prediction interval. . Curve fittings for open-cell and honeycomb models (Eq. 5) are shown as horizontal and vertical dashed lines, respectively. Coefficients and bulk material values of $\sigma_f^0$ are provided in Table 8. Plot created via the FreezeCasting.net open-data repository.

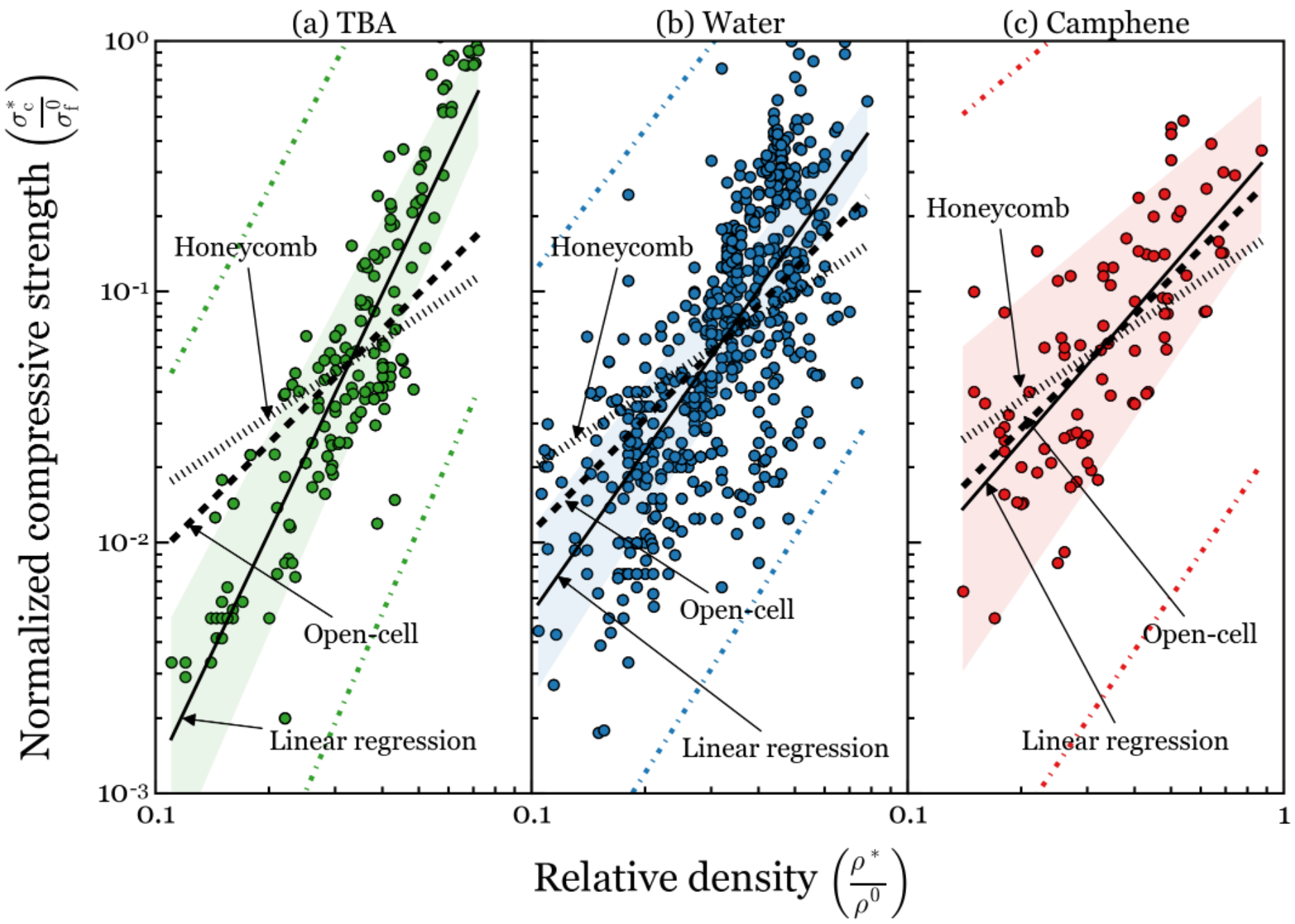

Fig. 30. Plots of normalized compressive strength of anisotropic freeze-cast ceramics as a function of relative density (1 - $\phi_p$), categorized by fluid type, including: (a) tert-butyl alcohol (*N*=171, [110, 224, 351, 492, 504, 647, 666, 678, 690, 700, 703, 704, 737, 738, 750]), (b) water (*N*=568, [112, 117, 120, 149, 196, 219, 224, 239, 240, 282, 283, 285, 286, 295, 296, 302, 312, 320, 324, 336, 339, 341, 342, 379, 384, 386, 390, 405, 414, 415, 429-431, 448, 460, 488, 505, 506, 514, 516, 563, 566, 569, 572, 579, 581, 584, 595, 617, 619, 620, 623, 624, 626-628, 633, 635, 637, 651, 652, 655, 662, 665, 670, 681-683, 688, 689, 705, 711, 712, 721, 742, 747, 792, 795]), (c) camphene (*N*=88, [262, 380, 381, 388, 410, 461, 575, 576, 606-608, 618, 640, 668, 672, 675, 691, 692, 697, 723, 726]). Solid lines are obtained using linear regression and shown as solid black lines; the shaded region about each regression line represents the 95% confidence interval, whereas the dashed line of similar color represents the 95% prediction region. Curve fittings for open-cell and honeycomb models (Eq. 5) are shown as horizontal and vertical dashed lines, respectively; coefficients and bulk material values of $\sigma_f^0$ are provided in Table 9. Plot created via the FreezeCasting.net open-data repository.

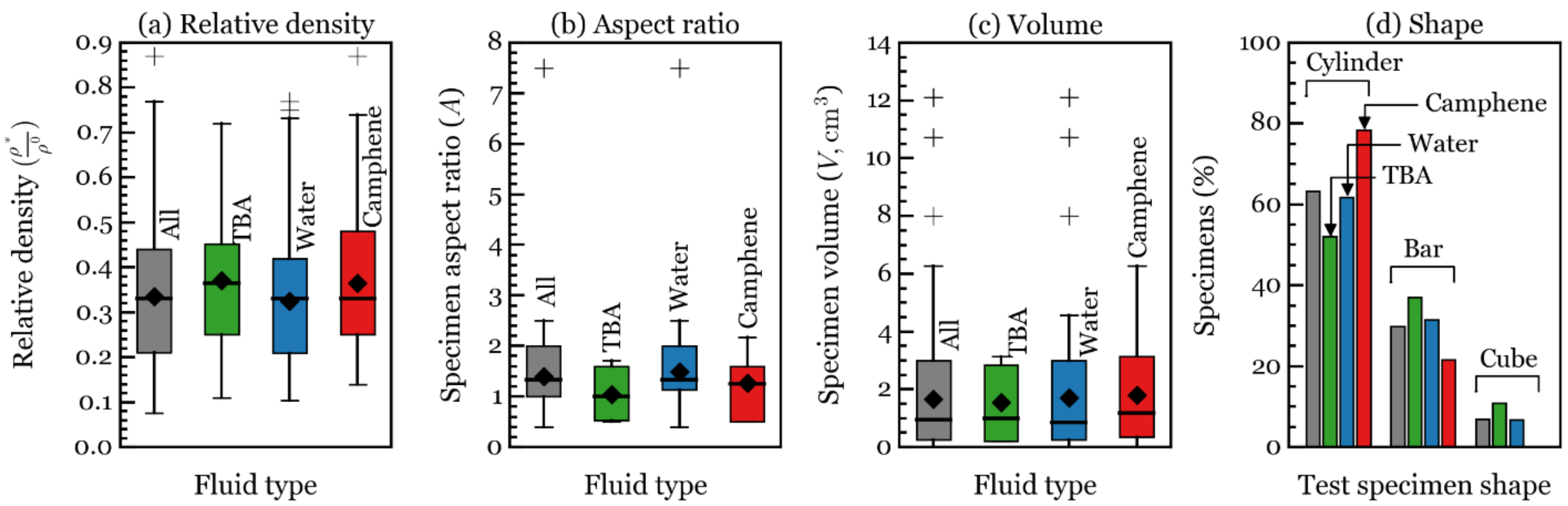

Fig. 31. Test specimen characteristics for data included in analysis of normalized compressive strength of anisotropic freeze-cast ceramics as a function of relative density (1 - $\phi_p$), specimen volume, and aspect ratio (Eq. 7), including: (a) relative density, (b) aspect ratio, (c) volume, and (d) shape. In all cases, a summary of all materials is shown in grey and *tert*-butyl alcohol (TBA), water, and camphene are shown in green, blue, and red, respectively. For (a-c), the mean for each condition is shown as a black diamond, the median as a black bar; outliers are shown as +, and error bars are derived using standard deviation. Refs: *tert*-butyl alcohol (*N*=116, [110, 224, 351, 492, 504, 647, 678, 690, 700, 703, 704]), (b) water (*N*=403, [112, 120, 149, 196, 219, 224, 239, 240, 282, 283, 285, 286, 295, 296, 302, 312, 320, 336, 339, 341, 390, 405, 414, 415, 429, 430, 448, 460, 514, 516, 563, 569, 579, 581, 595, 617, 619, 620, 623, 626-628, 633, 635, 651, 652, 670, 689, 705, 711, 747]), and (c) camphene (*N*=89, [262, 380, 381, 388, 410, 461, 575, 576, 606-608, 618, 640, 668, 672, 675, 691, 692, 697, 723, 726]). Plot created via the FreezeCasting.net open-data repository.

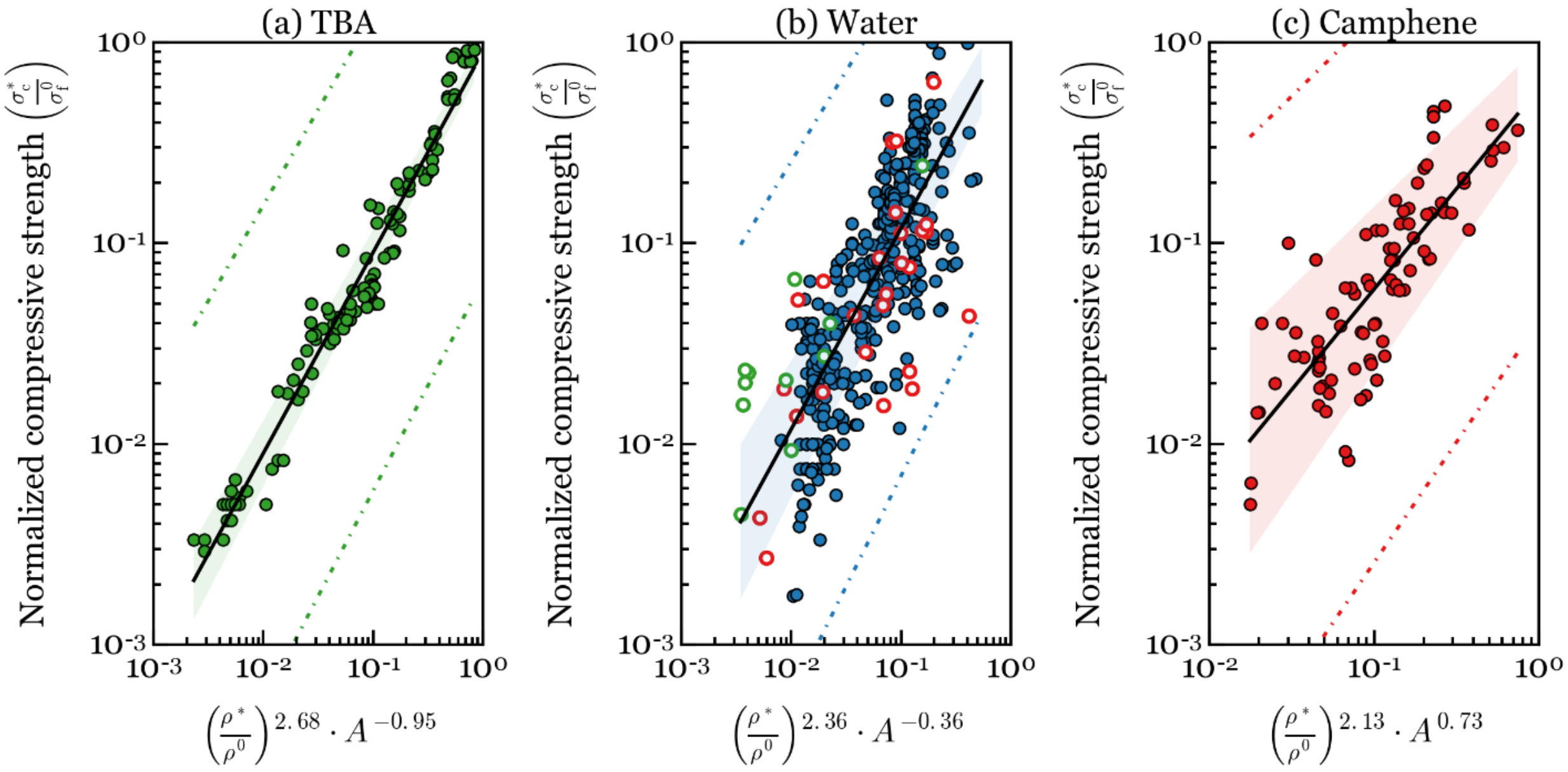

Fig. 32. Plots of normalized compressive strength of anisotropic freeze-cast ceramics as a function of relative density (1 - $\phi_p$), test specimen volume and aspect ratio (Eq. 8), categorized by fluid type, including: (a) *tert*-butyl alcohol (*N*=116, [110, 224, 351, 492, 504, 647, 678, 690, 700, 703, 704]), (b) water (*N*=403, [112, 120, 149, 196, 219, 224, 239, 240, 282, 283, 285, 286, 295, 296, 302, 312, 320, 336, 339, 341, 390, 405, 414, 415, 429, 430, 448, 460, 514, 516, 563, 569, 579, 581, 595, 617, 619, 620, 623, 626-628, 633, 635, 651, 652, 670, 689, 705, 711, 747]), and (c) camphene (*N*=89, [262, 380, 381, 388, 410, 461, 575, 576, 606-608, 618, 640, 668, 672, 675, 691, 692, 697, 723, 726]). For (b), lamellar structures are shown as blue circles; dendritic and honeycomb are shown in red and green, respectively. Solid lines are obtained using linear regression and shown as solid black lines; the shaded region about each regression line represents the 95% confidence interval, whereas the dashed line of similar color represents the 95% prediction region. Coefficients and bulk material values of $\sigma_f^0$ are provided in Table 10. Plot created via the FreezeCasting.net open-data repository.